\documentclass[twocolumn,showpacs,preprintnumbers,amsmath,amssymb]{revtex4}
\usepackage{epsfig}

\usepackage[table]{xcolor}
\usepackage{graphicx}
\usepackage{dcolumn}
\usepackage{bm}
\usepackage{esvect}

\usepackage{hyperref}
\hypersetup{colorlinks=true,linkcolor=magenta,anchorcolor=green,citecolor=cyan,filecolor=black,menucolor=black,urlcolor=brown}
\usepackage{slashed}

\newcommand{\be}{\begin{equation}}
\newcommand{\ee}{\end{equation}}
\newcommand{\ba}{\begin{eqnarray}}
\newcommand{\ea}{\end{eqnarray}}

\begin{document}

\title{Impact of kinetic and potential self-interactions on scalar dark matter}

\author{Philippe Brax}
\affiliation{Institut de Physique Th\'eorique, Universit\'e  Paris-Saclay, CEA, CNRS, F-91191 Gif-sur-Yvette Cedex, France}
\author{Jose A. R. Cembranos}
\affiliation{Departamento de  F\'{\i}sica Te\'orica and IPARCOS,\\ 
Universidad Complutense de Madrid, E-28040 Madrid, Spain}
\author{Patrick Valageas}
\affiliation{Institut de Physique Th\'eorique, Universit\'e  Paris-Saclay, CEA, CNRS, F-91191 Gif-sur-Yvette Cedex, France}

\begin{abstract}
We consider models of scalar dark matter with a generic interaction potential and non-canonical
kinetic terms of the K-essence type that are subleading with respect to the canonical term.
We analyze the low-energy regime and derive, in the nonrelativistic limit, the
effective equations of motions. In the fluid approximation they
reduce to the conservation of matter and to the Euler equation for the velocity field.
We focus on the case where the scalar field mass $10^{-21} \ll m \lesssim 10^{-4} \, {\rm eV}$
is much larger than for fuzzy dark matter, so that the quantum pressure is negligible
on cosmological and galactic scales, while the self-interaction potential and non-canonical
kinetic terms generate a significant repulsive pressure.
At the level of cosmological perturbations, this provides a dark-matter density-dependent
speed of sound. At the nonlinear level, the hydrostatic equilibrium obtained by balancing
the gravitational and scalar interactions imply that virialized structures have a solitonic
core of  finite size depending on the speed of sound of the dark matter fluid.
For the most relevant potential in $\lambda_4 \phi^4/4$ or K-essence with a $(\partial \phi)^4$ interaction, the size of such stable cores cannot
exceed 60 kpc. Structures  with a density contrast larger than $10^6$ can be accommodated
with a speed of sound $c_s\lesssim 10^{-6}$. We also consider the case of a cosine
self-interaction, as an example of bounded nonpolynomial self-interaction. This gives
similar results in low-mass and low-density halos whereas solitonic cores are shown to be
absent in massive halos.

\end{abstract}

\date{\today}

\maketitle


\section{Introduction}
\label{sec:introduction}

Astrophysical observations have collected a large amount of data over
the past decades. They have allowed cosmologists  to constrain with relatively good accuracy
cosmological scenarios and led to the $\Lambda$-CDM model.
This standard model of cosmology is based on the presence of a cold dark matter (CDM)
component whose origin is largely unknown. More often than not, it is assumed that
nonrelativistic collisionless particles form CDM and only interact
gravitationally with the other components of the standard model of particle physics.
However, there remains a long-standing debate about the nature of dark matter (DM)
and its behavior on small scales.
Indeed, there are  tensions between the predictions of the standard CDM model
and observations on galactic and subgalactic levels
\cite{Ostriker:2003qj,Weinberg:2013aya, Pontzen:2014lma}.
These discrepancies have been known as the ``{too big to fail}'' \cite{BoylanKolchin:2011de},
``{missing satellites}'' \cite{Moore:1999nt} and ``{core-cusp}'' problems \cite{deBlok:2009sp}.
The solution to these problems may come from baryonic effects, or from specific deviations of
DM with respect to the pure CDM paradigm on small scales.
Given the incompleteness of present galactic observations, it seems worth studying
new possibilities associated to new theoretical models.

One of these alternative DM models is based on coherent ultralight particles with
a de Broglie wavelength of the order of astrophysical scales \cite{Hu:2000ke}. This type of
coherent DM is associated with rapidly oscillating fields.
Indeed, massive scalars  \cite{Turner:1983he,Sahni:1999qe,Johnson:2008se} or vector fields
\cite{Cembranos:2012kk,Cembranos:2012ng,Cembranos:2013cba}
not only behave as CDM at the background level, but also at the perturbation level
\cite{Johnson:2008se, Hwang:2009js, Park:2012ru, Hlozek:2014lca, Cembranos:2015oya,Cembranos:2016ugq}
for distances larger than the associated Jeans scale. However, for shorter distances,
the matter power spectrum is characterized by a cutoff \cite{Hlozek:2014lca}.
It has also been shown that the formation of cusps can
be avoided for masses of order $m\sim 10^{-22}\;\text{eV}$  \cite{Schive:2014dra}.

A  well-motivated DM candidate associated to such coherent fields is the axion.
Originally, it was found in relation to the solutions of the strong $CP$ problem of QCD
\cite{Peccei:1977hh, Wilczek:1977pj, Weinberg:1977ma}.
However, these types of fields are ubiquitous in different beyond-standard-model scenarios,
as the ones motivated by string theory \cite{Marsh:2015xka, Hui:2016ltb}.
These Axion-Like-Particles (ALPs) may have a very large range of masses and present
a rich phenomenology
\cite{Khlopov:1985jw,Sarkar:2015dib,  Kobayashi:2017jcf, Abel:2017rtm,Banik:2017ygz, Hirano:2017bnu, Conlon:2017ofb,
Brito:2017wnc, Brito:2017zvb,Sarkar:2017vls, Diacoumis:2017hff}.
In particular, dark matter cores observed in several structures,
like galaxies or star clusters, could be explained by axion-like particles of different masses
\cite{Broadhurst:2018fei}.

In this work, we study the impact of anharmonic corrections to this type
of coherent DM.
Attractive quartic self-interactions have been studied in
\cite{Chavanis:2011zi,Cedeno:2017sou,Desjacques:2017fmf}, and the repulsive case in
\cite{Goodman:2000tg,Li:2013nal, Suarez:2016eez,
Suarez:2015fga, Suarez:2017mav, Chavanis:2018pkx}.
This yields an additional effective pressure which alleviates
the ``{core-cusp}'' problem \cite{Fan:2016rda} and
it could be the origin of vortices in galaxies \cite{RindlerDaller:2011kx}.
This deviation from the pure CDM behavior can be used to constrain the parameter space
of the model with CMB and large-scale structure (LSS) data \cite{Cembranos:2018ulm}.
The impact on the propagation of gravitational waves has been considered in
\cite{Dev:2016hxv} and on the inflationary gravitational wave background in
\cite{Li:2016mmc}.
Particle-physics models of such scenarios have been constructed in \cite{Fan:2016rda}.

In this paper, we consider models where the scalar potential and the non-canonical
kinetic terms can be generic.
We are, for instance, interested in models with a Lagrangian
\be
{\cal L}= X + \epsilon \frac{X^2}{\Lambda^4} - \frac{m^2}{2}\phi^2
\label{eq:Ke}
\ee
where $X=-\frac{(\partial \phi)^2}{2}$ and $\epsilon=\pm 1$.
When the term in $X^2$ is absent and the mass is larger than the Hubble rate at
matter-radiation equality, this describes the simplest model of scalar DM.
The effects of the new interaction in $X^2$ will be analyzed in this paper.
They can be summarized as follows. At low energy and in the nonrelativistic limit, this model
behaves like a DM model with an interaction potential term $\lambda_4 \phi^4$
with $ \lambda_4 = - \epsilon m^4 / \Lambda^4$, which is repulsive for $\epsilon=-1$ and
attractive when $\epsilon=1$. This model is valid up to matter-radiation equality as long as
$\Lambda \gtrsim  1 \, {\rm eV}$, which plays the role of the UV cutoff.
The bound on $\lambda_4$ from the collisionless nature of dark matter in the bullet cluster
implies $\lambda_4 \ll 1$. Interestingly, this is naturally achieved in these models as
the DM mass $m$ must be much lower than the cutoff scale $\Lambda$.
Moreover, the shift symmetry $\phi\to \phi+c$, which is softly broken by the mass term,
is restored when the mass vanishes. As a result, a small mass $m$ is technically natural
{\it \`a la} 't Hooft as the shift symmetry protects such a small mass from quantum corrections.
We will also consider models with a general K-essence term, and a generic potential,
and show that they are equivalent as DM models in the nonrelativistic limits.
The K-essence models of DM like (\ref{eq:Ke}) seem to be  more motivated from the particle
physics point of view as they could result from the physics of pseudo-Goldstone models.

 In this general setting,  we show that the nonrelativistic limit of the scalar models
can be described by a fluid where, on top of the well-known quantum pressure, a new potential
term arises in the Euler equation.
It characterizes both the self-interactions of the scalar field and the non-canonical kinetic
terms (when the latter are subleading with respect to the canonical term).
We focus on models with a scalar field mass $10^{-21} \ll m \lesssim 10^{-4} \, {\rm eV}$
that is much larger than for the well-studied fuzzy dark matter scenario,
so that the quantum pressure is negligible on cosmological and galactic scales,
while the self-interaction potential and non-canonical kinetic terms generate a significant
repulsive pressure.
At the perturbative level, this leads to a speed of sound of matter perturbations
that becomes density dependent. At the nonlinear level and when the self-interactions are
repulsive, we find that solitonic-like solutions of the hydrostatic equilibrium could describe the core
of virialized objects. For the particular case of a $\phi^4$ or $(\partial \phi)^4$ interactions, these
structures cannot exceed 60 kpc.
We also discuss the case of a cosine self-interaction, as an example of a bounded
self-interaction potential with physical consequences which differ largely from the polynomial models.

The paper is arranged as follows.
In section~\ref{sec:LG}, we describe the behavior of a generic model of scalar DM
with generic subleading potentials and kinetic terms beyond the leading quadratic terms.
We will call these models the Landau-Ginzburg models of DM as they possess generic potential and kinetic terms.
We derive the resummed effective potentials that appear in the nonrelativistic limit.
In section~\ref{sec:background}, we analyze how this model reproduces the features of
DM at the background level.
In section~\ref{sec:perturbations}, we study the cosmological perturbations in the subhorizon
regime and the quasistatic approximation. We obtain the speed of sound for these
Landau-Ginzburg models, focusing on the case of masses greater than for fuzzy dark matter,
where the quantum pressure is negligible.
In section~\ref{sec:small-scale}, we investigate the small-scale dynamics and the presence
of solitonic cores in hydrostatic equilibrium.
We also consider the stability of the resulting solitonic cores.
In section~\ref{sec:halos}, we match these cores to an outside NFW (Navarro-Frenk-White)
profile and study the impact on cosmological halos.
We then conclude in section~\ref{sec:conclusion}.

\section{Nonrelativistic Landau-Ginzburg Models}
\label{sec:LG}

\subsection{Scalar field dark matter}
\label{sec:scalarDM}

In this section, we recall how scalar fields can play the role of pressureless dark matter.
Let us consider the scalar-field action
\be
S_\phi = \int d^4x \sqrt{-g} \left[ K(X) - V(\phi) \right] ,
\ee
where $X=- \frac{1}{2} g^{\mu\nu} \partial_\mu\phi\partial_\nu\phi$ is the standard
kinetic term and $V(\phi)$ the potential. If $K(X)$ is nonlinear the scalar-field Lagrangian
shows a nonstandard kinetic term. This type of models could be the effective result at low energy of more fundamental theories. In particular the higher order interactions, whether polynomials or derivatives, could originate from integrating out massive field and keeping only $\phi$ as the only relevant degree of freedom at low energy in the DM sector.

Let us focus on power-law cases,
\be
K(X) = \frac{K_\star}{p} X^p , \;\;\; V(\phi) = \frac{V_\star}{n} \phi^ n ,
\ee
with $K_\star>0$, $V_\star>0$, and $n$ even.
Then, from the scalar-field energy-momentum tensor we can read the scalar-field
background density and pressure,
\ba
&& \rho_\phi = \frac{(2p-1) K_\star}{p} \left( \frac{\dot\phi^2}{2} \right)^p
+ \frac{V_\star}{n} \phi^n , \\
&& p_\phi = \frac{K_\star}{p} \left( \frac{\dot\phi^2}{2} \right)^p - \frac{V_\star}{n} \phi^n .
\ea
In the limit of fast oscillations in the potential well, we can neglect the Hubble expansion
and the equation of motion of the background scalar field reads
\be
(2p-1) K_\star \left( \frac{\dot\phi^2}{2} \right)^{p-1} \ddot\phi + V_{\star} \phi^{n-1} = 0 .
\label{eq:phi-p-n}
\ee
Multiplying by $\dot\phi$, we obtain as a first integral of motion that $\rho_\phi$ is constant.
If $p>1$ or $n>2$ the oscillations are not harmonic and can be integrated as
\be
dt = d\phi \left[ \frac{2^p p}{(2p-1) K_\star} \left( \rho_\phi - \frac{V_\star}{n} \phi^n \right) \right]^{-1/(2p)} .
\ee
On the other hand, defining $\langle \dots \rangle$ as the average over one oscillation period,
we have
\be
\langle \frac{d}{dt} \left( \phi \dot\phi^{2p-1} \right) \rangle = 0 ,
\ee
and hence
\be
\langle \dot\phi^{2p} \rangle = - (2p-1) \langle \phi \dot\phi^{2p-2} \ddot\phi \rangle .
\ee
Combining with the equation of motion (\ref{eq:phi-p-n}), we obtain
\be
\langle \dot\phi^{2p} \rangle = \frac{V_\star}{K_\star} 2^{p-1} \langle \phi^n \rangle .
\ee
This yields for the averaged density and pressure
\ba
&& \langle \rho_\phi \rangle = \left( \frac{2p-1}{2p} + \frac{1}{n} \right) V_\star
\langle \phi^n \rangle , \\
&& \langle p_\phi \rangle = \left( \frac{1}{2p} - \frac{1}{n} \right) V_\star
\langle \phi^n \rangle ,
\ea
which gives the averaged equation of state parameter
\be
w = \frac{\langle p_\phi \rangle}{\langle \rho_\phi \rangle} = \frac{n-2p}{n(2p-1)+2p} .
\label{eq:w-def}
\ee
Thus, the scalar field behaves as pressureless cold dark matter if $n=2p$,
\be
w=0 \;\;\; \mbox{if} \;\;\; n = 2 p .
\label{eq:w-n-2p}
\ee
This includes in particular the standard case of the massive free scalar field,
with $p=1$ and $n=2$, with a standard kinetic term and a quadratic potential.
In this paper, we shall focus on this standard harmonic case, with subleading higher-order
corrections to both the kinetic and potential terms.
We require that these anharmonic corrections are sufficiently small for the scalar field
to behave as pressureless cold dark matter at the level of the cosmological background,
through the whole matter era. However, they could play a significant role on small galactic
scale and give rise to an effective pressure that could support galactic dark matter halos
against Newtonian gravity.

According to Eq.(\ref{eq:w-n-2p}), it should be possible to generalize this scenario
to strongly nonlinear models, with $p>1$ and $n=2p$. However, we do not investigate
this case further in this paper.

\subsection{The Landau-Ginzburg models}
\label{sec:the-model}

\subsubsection{Scalar-field action}
\label{sec:scalar-field-action}

We are interested in an effective model of scalar dark matter valid below a cutoff energy scale $\Lambda$. The theory describing
the Universe beyond this energy scale is left unspecified. As an effective theory, we only assume that the action is local and
described by an interaction potential $V_{\rm I}$ for a particle of mass $m$ and that it only involves first derivatives of the scalar fields.
In general, higher-order Lagrangians lead to ghostlike behaviors, although as Horndeski shows, their functional forms can be tuned in order to avoid such problems.
We do not assume this here, and simply consider that all the higher derivative terms correspond to the propagation of extra degrees of freedom with a mass larger or equal to the cutoff $\Lambda$. As a result, their effects can be neglected at energies below $\Lambda$.
Thus, we consider the scalar model of a light particle of mass $m$ subject
to self-interactions, defined by the Lagrangian
\be
{\cal L}_\phi = X + K_{\rm I}(X) - \frac{m^2}{2} \phi^2 - V_{\rm I}(\phi) ,
\label{eq:L-phi}
\ee
where $X=- \frac{1}{2} g^{\mu\nu} \partial_\mu\phi\partial_\nu\phi$ is the standard
kinetic term, $V_{\rm I}$ the self-interaction potential
\be
V_{\rm I}(\phi) = \Lambda^4 \sum_{p\ge 3} \frac{\lambda_p}{p} \frac{\phi^p}{\Lambda^p} ,
\label{eq:V-I-def}
\ee
and $K_{\rm I}$ the nonstandard kinetic term,
\be
K_{\rm I}(X)= \Lambda^4 \sum_{n \geq 2} \frac{k_n}{n} \frac{X^n}{\Lambda^{4n}} .
\label{eq:K-I-def}
\ee
We assume that this corresponds to an effective theory, on scales larger than a cutoff
$\Lambda^{-1}$, which can be taken for instance from a fraction of millimeters to a few kpc's,
or to a fully nonlinear theory defined by the resummed potential $V_{\rm I}$
and kinetic term $K_{\rm I}$, which can be nonpolynomial.

The total action of the system is
\be
S = S_{\rm EH} + S_{\phi} + S_{\rm m} ,
\ee
where $S_{\rm EH}$ is the Einstein-Hilbert action of General Relativity,
$S_\phi= \int d^4 x \sqrt{-g} {\cal L}_\phi$ the scalar field action, and $S_{\rm m}$ the action
of the standard model particles (baryons, photons) and possible dark energy components.
In this work we work in the Newtonian gauge, around the
Friedmann-Lemaitre-Robertson-Walker (FLRW) background,
\be
ds^2 =  - (1+2\Phi) dt^2 + a^2(t) (1-2\Psi) d\vec x^{\,2} ,
\ee
where $a(t)$ is the cosmological scale factor, $\vec x$ the comoving spatial coordinate,
and $\Phi$ and $\Psi$ the Newtonian metric potentials.

\subsubsection{Small-amplitude nonlinear corrections}
\label{sec:small-amplitude}

As we have seen in section~\ref{sec:scalarDM}, when the potential is very close to harmonic
and the kinetic terms are close to canonical, the scalar field $\phi$ can play the role
of the dark matter. Indeed, in the regime where it oscillates very fast as compared with
the Hubble expansion rate, the harmonic oscillator shows equipartition between the
kinetic and potential energy and the averaged pressure is zero,
$\langle p_\phi \rangle = 0$.
Then, the scalar field behaves as pressureless cold dark matter.
This balance is modified by anharmonic corrections, which give rise to a nonzero pressure.
Therefore, we require that the self-interaction potential and the higher-order kinetic terms
be small, from the matter-radiation equality until now,
\be
V_{\rm I} \ll \frac{m^2}{2} \phi^2 , \;\;\; K_I \ll X .
\label{eq:V-I-small}
\ee
This typically corresponds to cases where the Lagrangian receives contributions from
different terms arising from a more complete theory.
For instance, let us consider a potential that can be written as the sum
$V(\phi) = M_1^4 V_1(\phi/\Lambda_1) +  M_2^4 V_2(\phi/\Lambda_2)$,
with $\Lambda_1 \gg \Lambda_2$ and $M_1 \gg M_2$.
Then, if $(\Lambda_2/\Lambda_1)^{3/4} \ll M_2/M_1 \ll (\Lambda_2/\Lambda_1)^{1/2}$,
the first potential gives a leading contribution that can be approximated by its quadratic term
while the second potential gives a subleading contribution that enters its nonlinear regime
before the cubic term $M_1^4 (\phi/\Lambda_1)^3$ from $V_1$ becomes relevant.
This gives for instance a total potential that is parabolic at zeroth order
but shows smaller-amplitude and higher-frequency structures on top of this mean parabola.
The other natural setting is to have $V_{\rm I}$ and $K_{\rm I}$ be the first-order corrections
to the free massive scalar field, in which case we expect $V_{\rm I}$ and $K_{\rm I}$
to be governed by the first terms $\phi^4$ and $X^2$ in this perturbative regime.

\begin{figure}
\begin{center}
\epsfxsize=8. cm \epsfysize=5. cm {\epsfbox{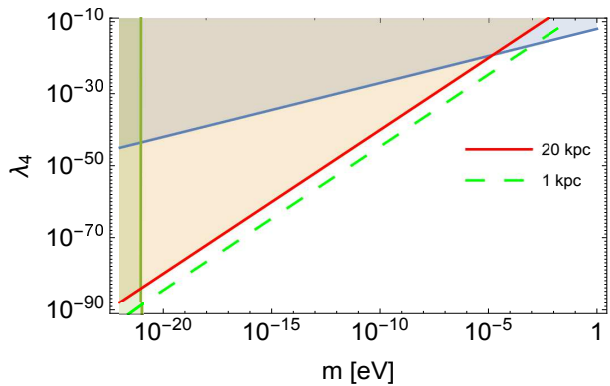}}
\end{center}
\caption{Range of interest in the plane $(m,\lambda_4)$. The left and upper exclusion regions
correspond to Eqs.(\ref{eq:no-quantum-pressure}), (\ref{eq:lambda4-cross-section})
and (\ref{eq:lambda4-DM}). The diagonal lines labeled ``20 kpc'' and ``1 kpc'' correspond
to Eq.(\ref{eq:lambda4-ra}), with $r_a=20 \, {\rm kpc}$ and $r_a=1 \, {\rm kpc}$.}
\label{fig_m-lambda4}
\end{figure}

To be more explicit, we now describe the range of parameters
$(m,\lambda_4)$ that we consider in this paper, neglecting at this stage
higher-order corrections, i.e. we focus first on the $\phi^4$ theory with canonical kinetic terms.
{ As we will see shortly, this also corresponds to the quartic K-essence model (\ref{eq:Ke}), which can be obtained as a pseudo-Goldstone model of DM.}
This is shown by the non-shaded region in Fig.~\ref{fig_m-lambda4}.
First, the condition of fast oscillations from the matter-radiation equality implies
\be
m \gg H_{\rm eq} \sim 10^{-28} \, {\rm eV} ,
\label{eq:m-Heq}
\ee
where $H_{\rm eq}$ is the Hubble expansion rate at the matter-radiation equality.
In this paper, we focus on scenarios where $m$ is also much greater than
cosmological wave numbers $k/a$, down to distances of the order of the kpc,
\be
\frac{k}{am} \ll 1 , \;\;\; \mbox{with} \;\;\;
\frac{1}{m} = 6.4 \times 10^{-27}
\left( \frac{m}{1\,{\rm eV}} \right)^{-1} \, {\rm kpc} .
\label{eq:m-k}
\ee
This corresponds to masses much greater than $10^{-27} {\rm eV}$,
\be
m \gg 10^{-27} \, {\rm eV} .
\ee
Masses of order $10^{-22}\,{\rm eV}$ correspond to the Fuzzy Dark Matter scenario,
where the so-called quantum pressure associated with interference patterns of the
scalar field can balance gravity on galactic scales. This may cure some of the small-scale
problems of the standard cold dark matter scenario, but also faces severe constraints
from Lyman-$\alpha$ forest statistics.
The impact of small attractive self-interactions in such scenarios has been investigated
in \cite{Desjacques:2017fmf}.
In this paper, we instead focus on more massive scalar fields with repulsive self-interactions
that may balance gravity on galactic scales.
Thus, we consider larger masses, where the quantum pressure is negligible from
cosmological to galactic scales. As recalled in Eq.(\ref{eq:rQ-def}) below, this corresponds to
\be
\mbox{negligible quantum pressure:} \;\;\; m \gg 10^{-21} \, {\rm eV} .
\label{eq:no-quantum-pressure}
\ee
This lower bound is shown by the vertical shaded area on the left part of
Fig.~\ref{fig_m-lambda4}.

Taking into account the quartic interaction only, the cross section per unit mass is
\cite{Desjacques:2017fmf,Tulin:2017ara}
\be
\frac{\sigma}{m} = \frac{9\lambda_4^2}{8\pi m^3} \sim \lambda_4^2
\left( \frac{m}{\rm 1 \, eV} \right)^{-3} 10^{23} \, {\rm cm^2/g} .
\label{eq:cross-section}
\ee
On the other hand, observations of the merging of clusters provide the upper bound
$\sigma/m \lesssim 1 \, {\rm cm^2/g}$ on self-interacting dark matter \cite{Randall:2007ph}.
This gives
\be
\lambda_4 \lesssim 10^{-12} \left( \frac{m}{\rm 1 \, eV} \right)^{3/2} .
\label{eq:lambda4-cross-section}
\ee
From Eq.(\ref{eq:L-phi}) we have at leading order $\rho \sim m^2\phi^2$ and
$V_{\rm I} \sim \lambda_4 \phi^4$. As $\phi$ decreases with cosmic time, along
with the scalar-field dark matter density $\rho \sim m^2\phi^2$, the constraint
$V_{\rm I} \lesssim \rho$ from the matter-radiation equality until now is set
by the condition at $z_{\rm eq}$ and it implies
\be
V_{\rm I} \lesssim \rho \;\;\; \mbox{for} \;\;\; z \leq z_{\rm eq} : \;\;\;
\lambda_4 \lesssim  \left( \frac{m}{\rm 1 \, eV} \right)^4 .
\label{eq:lambda4-DM}
\ee
The upper boundaries (\ref{eq:lambda4-cross-section}) and (\ref{eq:lambda4-DM})
are shown by the two upper shaded areas in Fig.~\ref{fig_m-lambda4}.
They cross at $m \sim 10^{-5} \, {\rm eV}$.

Finally, as we shall see below in Eqs.(\ref{eq:rJ-quartic}) and (\ref{eq:ra-def}), the repulsive
self-interaction gives rise to a characteristic length $r_a$, which corresponds to both
the Jeans length, beyond which gravitational instability amplifies primordial density
fluctuations, and to the scale of solitonic cores in collapsed halos.
It also reads as \cite{Goodman:2000tg,Arbey:2003sj}
\be
r_a = \sqrt{\frac{3 \lambda_4}{2}} \frac{M_{\rm Pl}}{m^2} ,
\label{eq:ra-lambda4}
\ee
which gives
\be
\lambda_4 = \left( \frac{r_a}{20 \, {\rm kpc}} \right)^2  \left( \frac{m}{\rm 1 \, eV} \right)^4 .
\label{eq:lambda4-ra}
\ee
This is shown for $r_a=20 \, {\rm kpc}$ and $1 \, {\rm kpc}$ by the diagonal solid and
dashed lines  in Fig.~\ref{fig_m-lambda4}.
The line $r_a=20 \, {\rm kpc}$ coincides with the upper boundary (\ref{eq:lambda4-DM}).
Because we require galaxies and Lyman-$\alpha$ clouds to form by gravitational
instability, as in the standard CDM scenario, we have the upper bound
$r_a \lesssim 20 \, {\rm kpc}$. This gives the same upper bound as in Eq.(\ref{eq:lambda4-DM}).
This coincidence means that there is some tension \cite{Goodman:2000tg,Peebles:2000yy}
between the wish to have a large
radius $r_a$ that can play a role on galactic scales and the constraints derived
from the cosmological background (the scalar field density must be small during
the radiation era, especially at the BBN, and it must behave like pressureless dark matter
up to high redshifts). However, this tension can be relaxed if the scalar-field potential
is not quartic up to high energies. In particular, a bounded potential such as the cosine
model we study in this paper ensures that we recover the standard cosmological background
to a high accuracy at high redshifts.

One may also derive constraints on scalar-field dark matter models from
the BBN \cite{Li:2013nal}, by requiring that the Hubble expansion rate does not deviate
too much from the standard cosmology.
Indeed, let us recall that the behavior of scalar field dark matter typically shows three
phases \cite{Li:2013nal}.
Looking backward in time, we have seen that, at least up to $z_{\rm eq}$,
we require $m \gg H$ and $V_{\rm I} \ll m^2\phi^2$, so that the scalar
field behaves as pressureless dark matter, $w=0$. At earlier times,
we still have $m \gg H$ but the self-interactions may become important.
Then, for the quartic model Eq.(\ref{eq:w-def}), this gives $w=1/3$ and the scalar field
behaves like radiation, but with a negligible relative density because
$\rho_\phi \ll \rho_\gamma$ as this transition occurs at a higher redshift than $z_{\rm eq}$.
At even earlier times, the Hubble expansion rate becomes greater than the
oscillation frequency. This corresponds to a stiff phase ($w=1$) where the scalar field
density and pressure are dominated by the kinetic term and one obtains
\cite{Li:2013nal}
$\rho_\phi \propto a^{-6}$, $a \propto t^{1/3}$, $\dot\phi \propto a^{-3}$ and $\phi \propto \ln a$.
Then, the scalar field might dominate at earlier times.
However, if $m \gg H_{\rm BBN} \sim 10^{-13} \, {\rm eV}$ this stiff phase occurs
before the BBN and the latter follows the standard cosmological model.
On the other hand, the quartic model may not extend up to these redshifts.
Therefore, we do not show this constraint in Fig.~\ref{fig_m-lambda4},
as we focus on lower redshifts and assume that in case $m \lesssim 10^{-13} \, {\rm eV}$
the potential differs from the quartic model at earlier times, as for the cosine
model described in section~\ref{sec:cosine-potential}.
We implicitly assume that the initial conditions are such that the scalar field density at late
times corresponds to the observed dark matter density.

Thus, in this article we focus on models where the parameters $(m,\lambda_4)$ fall
between the two diagonal lines of Fig.~\ref{fig_m-lambda4}, with
$1 \lesssim r_a \lesssim 20 \, {\rm kpc}$, and
$10^{-21} \lesssim m \lesssim 10^{-4} \, {\rm eV}$.
Smaller masses correspond to Fuzzy Dark Matter models, where quantum pressure is important.
Lower values of $\lambda_4$, or higher values of $m$, correspond to models that
behave like standard Cold Dark Matter particles down to galactic scales, as the
scale $r_a$ decreases. They are consistent with observations but do not show
new properties as compared with CDM on cosmological scales.

\subsection{Nonrelativistic limit}
\label{sec:nonrelativistic}

In quantum field theory, the scalar field model describes the behavior of massive scalars.
In canonical quantization, states $\vert \vec k \, \rangle$ correspond to single particle states
with energies
\be
E_k= \sqrt {k^2 +m^2} ,
\ee
where we have neglected the interactions.
When $\vert \vec k \vert \ll m$, the energy becomes
\be
E_k= m + \frac{k^2}{2m} ,
\ee
corresponding to nonrelativistic excitations. At these energies, antiparticles cannot be created
so that the particle number is conserved and the quantum field theory reduces to quantum
mechanics. The field $\phi$ can be decomposed as
\be
\phi= \frac{1}{\sqrt{2m}}( e^{-imt} \psi + e^{imt} \psi^\star) ,
\label{eq:psi-def}
\ee
where $\psi$ is now a complex scalar field. This corresponds to a Bose-Einstein condensate,
where all particles have negligible momentum, $\vert \vec k \vert \ll m$,
and their number is conserved. Then, the system is described by the classical
complex field $\psi$.
We can note that $\phi$ is invariant under the transformation
\be
t \to t+\alpha/m , \;\;\; \psi \to e^{i\alpha} \psi , \;\;\; \phi \to \phi .
\label{eq:t-alpha-psi}
\ee
As we shall see below, this will lead to a { global} U(1) symmetry for $\psi$ in the
nonrelativistic limit. { This global symmetry is the remnant of the time translation
invariance $t\to t-\xi^0(t,\vec x)$, which is a part of the diffeomorphism invariance of the theory.}

\subsection{Fluid picture}
\label{sec:fluid}

The complex scalar field $\psi$ can be mapped to a hydrodynamical system,
using the Madelung transformation \cite{Madelung_1927}
\be
\psi=\sqrt{\frac{\rho}{m}} e^{iS} ,
\label{eq:rho-S-def}
\ee
which defines the amplitude $\sqrt{\rho/m}$ and the phase $S$.
As is well known, the equations of motion will take the form of the hydrodynamical continuity
and Euler equations, where $\rho$ is interpreted as the density and $\vec v$ as the fluid
velocity defined by
\be
\vec v = \frac{\nabla S}{m a} .
\label{eq:v-S-def}
\ee
The symmetry (\ref{eq:t-alpha-psi}) now reads as the shift symmetry
\be
t \to t+\alpha/m , \;\;\; S \to S+\alpha .
\label{eq:t-alpha-S}
\ee
{ As a result of this global symmetry, there always exists a Noether current whose conservation
is guaranteed. We will identify it as the matter density current implying that matter is always
conserved in the fluid description. Moreover, the $S$ field can be seen as the St\"uckelberg
field which restores the local time invariance symmetry as
\be
t\to t+ \xi^0, \ \ \  S \to S + m\xi^0 .
\ee
This local invariance is broken by the background geometry and its scale factor
$a(t)$, with an explicit time dependence which is not compensated by $S$.}

\subsection{Weak gravity regime and effective actions}
\label{sec:weak gravity}

\subsubsection{Einstein-Hilbert action}
\label{sec:S-EH-weak}

On cosmological and galactic scales, the Newtonian potentials $\Phi$ and $\Psi$
are small, typically of order $10^{-6}$ to $10^{-5}$. Therefore, we can expand
the action over $\Phi$ and $\Psi$.
As  recalled in Appendix~\ref{app:EH},
up to quadratic order, the Einstein-Hilbert action is given by the expression
(\ref{eq:S-EH-2}), while up to linear order the Einstein tensor $G^\mu_\nu$ is given
by Eqs.(\ref{eq:G00})-(\ref{eq:Gij}).
At this order, they are related by
\be
\frac{\delta S_{\rm EH}}{\delta\Phi} = - M_{\rm Pl}^2 a^3 G^0_0 , \;\;\;
\frac{\delta S_{\rm EH}}{\delta\Psi} = M_{\rm Pl}^2 a^3 G^i_i ,
\label{eq:S-EH-G}
\ee
where in the last expression we sum over the index $i$ as in the Einstein convention.
By working in the Newtonian gauge and restricting the study to the
scalar perturbations, we obtain two equations of motion
for  $\Phi$ and $\Psi$.
They correspond to the
(00) component and to the spatial trace part of the full set of Einstein equations.

\subsubsection{Scalar-field action}
\label{sec:S-phi-weak}

Because $\Phi$ and $\Psi$ are very small, below $10^{-5}$, we only need to expand
the scalar-field action up to linear order in $\Phi$ and $\Psi$ yielding
\ba
S_\phi = && \int d^4x \; a^3 \biggl [ \frac{1-\Phi-3\Psi}{2} \dot\phi^2 - \frac{1+\Phi-\Psi}{2 a^2}
(\nabla\phi)^2 \nonumber \\
&& -  \frac{1+\Phi-3\Psi}{2} m^2 \phi^2 + K_{\rm I}(X) - V_{\rm I}(\phi) \biggl ] .
\label{eq:S-phi-2}
\ea
At linear order in $\Phi$ and $\Psi$, we actually get a term
$(1+\Phi-3\Psi) (K_{\rm I} - V_{\rm I})$
instead of $(K_{\rm I}-V_{\rm I})$ in the action. However, because of the constraint
(\ref{eq:V-I-small}), $K_{\rm I}-V_{\rm I}$ is subdominant as compared with the
scalar-field energy density $\rho$, $| K_{\rm I} - V_{\rm I} | \ll \rho$.
We shall see below that $| K_{\rm I} - V_{\rm I} | \lesssim \rho | \Phi |$, and therefore
we can drop the term $(\Phi-3\Psi) (K_{\rm I} - V_{\rm I})$ in the action.
As in (\ref{eq:S-EH-G}), the derivatives of the scalar-field action with respect to the metric
potentials $\Phi$ and $\Psi$ are related to the scalar-field energy-momentum tensor
$T^\mu_\nu$ by
\be
\frac{\delta S_{\phi}}{\delta\Phi} = a^3 T^0_0 , \;\;\;
\frac{\delta S_{\phi}}{\delta\Psi} = - a^3 T^i_i ,
\label{eq:S-phi-T}
\ee
where we have defined
$T^{\mu\nu}= \frac{2}{\sqrt{-g}} \frac{\delta S_\phi}{\delta g_{\mu\nu}}$.
At the level of the scalar field action, the invariance under diffeomorphisms,
which locally take the form $x^\mu \to x^\mu - \xi^\mu$ and
$g_{\mu\nu} \to g_{\mu\nu} + \nabla_\mu \xi_\nu +\nabla_\nu \xi_\mu$, implies
$\int d^4 x \sqrt {-g} T^{\mu\nu} (\nabla_\mu \xi_\nu + \nabla_\nu \xi_\mu) =0$,
when the equations of motion are satisfied.
This gives the usual conservation equations
\be
\nabla_\mu T^{\mu\nu} = 0 .
\ee
The case $\nu=0$, associated with time diffeomorphisms, gives the continuity
equation for the energy density,
\ba
&& \partial_t \left[ a^3 \left( \frac{\dot\phi^2}{2} + \frac{(\nabla\phi)^2}{2a^2}
+ \frac{m^2\phi^2}{2} \right) \right] - \nabla ( a \dot\phi \nabla\phi ) \nonumber \\
&& + 3 H a^3 \left( \frac{\dot\phi^2}{2} - \frac{(\nabla\phi)^2}{6a^2}
- \frac{m^2\phi^2}{2} \right) = 0 ,
\label{eq:continuity-GR}
\ea
where we used $\Phi \sim \Psi \ll 1$ and
$\dot\Phi \sim \dot\Psi \sim H \Phi \ll H$.
At the level of the background, this corresponds to the usual relativistic
continuity equation $\dot\rho+3H(\rho+p)=0$.

\subsubsection{Nonrelativistic limit}
\label{sec:S-psi-weak}

We can obtain the nonrelativistic limit by substituting the expression
(\ref{eq:psi-def}) into the scalar-field action (\ref{eq:S-phi-2}).
The full expression is given in Eq.(\ref{eq:S-psi-full-2}) in the
appendix~\ref{app:complex-scalar}.
In the nonrelativistic regime we neglect the fast oscillatory terms and the action
simplifies as
\ba
&& S_\phi = \int d^4x \; a^3 \biggl \lbrace \frac{1-\Phi-3\Psi}{2m}
( i m \dot\psi \psi^\star - i m \psi \dot\psi^\star)
\nonumber \\
&& - \frac{1+\Phi-\Psi}{2 m a^2} (\nabla\psi)\cdot(\nabla\psi^\star) - m \Phi \psi \psi^\star
\nonumber \\
&& + {\cal K}_{\rm I}(\psi,\psi^\star) - {\cal V}_{\rm I}(\psi,\psi^\star) \biggl \rbrace .
\label{eq:S-psi-2}
\ea
Here we neglected the term $\dot\psi\dot\psi^\star$ in the standard kinetic part, because
it is negligible as compared with $m \dot\psi \psi^\star$, as $\psi$ evolves on
cosmological or galactic timescales, which are much larger than the fast oscillatory
period $2\pi/m$.

The self-interaction potential ${\cal V}_{\rm I}(\psi,\psi^\star)$ is obtained from $V_{\rm I}(\phi)$
by substituting the decomposition (\ref{eq:psi-def}). As for the other terms of the nonrelativistic
action, we only keep the non-oscillatory terms. This means that in the
series expansion (\ref{eq:V-I-def}) we only keep the even order terms $\phi^{2n}$,
where we pair $n$ factors $e^{-imt}$ with $n$ factors $e^{imt}$.
This gives
\be
{\cal V}_{\rm I}(\psi,\psi^\star) = \Lambda^4 \sum_{n=2}^{\infty} \frac{\lambda_{2n}}{2n}
\frac{(2n)!}{(n!)^2} \left( \frac{\psi \psi^\star}{2 m \Lambda^2} \right)^n .
\label{eq:V-I-psi}
\ee
The combinatorial factor is the number of ways of choosing $n$ factors
$e^{-imt}$ (and the remaining $n$ factors $e^{imt}$) among the $2n$ factors $\phi$.
In a similar fashion, from the series expansion (\ref{eq:K-I-def}) the remaining contribution
from the nonstandard kinetic term reads
\be
{\cal K}_{\rm I}(\psi,\psi^\star) = \Lambda^4 \sum_{n=2}^{\infty} \frac{k_n}{n}
\frac{(2n)!}{(n!)^2} \left( \frac{m\psi \psi^\star}{4 \Lambda^4} \right)^n ,
\label{eq:K-I-psi}
\ee
where we used the conditions $m \gg H$ and $m \gg k/a$,
from Eq.(\ref {eq:m-k}), to  keep only the leading terms.
Thus, in the nonrelativistic limit that we consider in this paper, with weak interactions
and nonstandard kinetic terms, the latter are equivalent to a potential term
and only lead to a correction to the nonrelativistic
potential ${\cal V}_{\rm I}$. This yields an effective nonrelativistic potential
\be
{\cal V}_{\rm I}^{\rm eff} = {\cal V}_{\rm I} - {\cal K}_{\rm I} ,
\label{eq:V-I-eff}
\ee
with the effective coefficients
\be
\lambda^{\rm eff}_{2n} = \lambda_{2n} - 2 k_n \left( \frac{m^2}{2\Lambda^2} \right)^n .
\label{eq:lambda-2n-eff}
\ee
The expression (\ref{eq:S-psi-2}) provides the effective scalar-field action that governs
the weak-gravity nonrelativistic regime.
{ Notice that the K-essence models of DM, such as (\ref{eq:Ke}),
with their shift symmetry $\phi\to \phi +c$ softly broken by the mass term,
behave in the same fashion as DM models with a potential term.}

At leading order over $m$, and for wave numbers $k/a\ll m$ as in Eq.(\ref{eq:m-k}),
the continuity equation (\ref{eq:continuity-GR}) simplifies as
\be
\partial_t \left[ a^3 m \psi \psi^\star \right] + \nabla \cdot \left[
\frac{ia}{2} ( \psi \vec\nabla \psi^\star - \psi^\star \vec\nabla \psi ) \right] = 0 .
\label{eq:continuity-NR}
\ee
We recover the fact that the pressure term associated with the last term
in Eq.(\ref{eq:continuity-GR}) is negligible, as the scalar field behaves
as pressureless dark matter.

On the other hand, the nonrelativistic action (\ref{eq:S-psi-2}) satisfies
the U(1) symmetry $\psi \to e^{i\alpha} \psi$,
$\psi^\star \to e^{-i\alpha} \psi^\star$.
This follows directly from the symmetry (\ref{eq:t-alpha-psi}), associated with the
definition of $\psi$, and the fact that the action (\ref{eq:S-psi-2}) does not
explicitly depend on time [because we discarded the oscillatory terms
$e^{\pm 2 imt}$ of Eq.(\ref {eq:S-psi-full-2})].
To this symmetry is associated the Noether current $J^\mu$, with
\be
\partial_\mu J^\mu = 0 ,
\label{eq:Jmu-Noether}
\ee
and
\be
J^0= a^3 m \psi \psi^\star , \;\;\;
\vec J = \frac{ia}{2} \left( \psi \vec\nabla \psi^\star
- \psi^\star \vec\nabla \psi\right ),
\label{eq:Jmu-def}
\ee
where we used again $\Phi \sim \Psi \ll 1$.
We recognize the standard conserved current of nonrelativistic quantum mechanics.
In the nonrelativistic limit, this corresponds to the conservation of the matter
density.
We note that Eq.(\ref{eq:Jmu-Noether}) is identical to Eq.(\ref{eq:continuity-NR}).
This can be understood from the fact that the U(1) symmetry leading to
Eq.(\ref{eq:Jmu-Noether}) is a consequence of the transformation (\ref{eq:t-alpha-psi}),
which is related to uniform translations over time, while the conservation equation
(\ref{eq:continuity-NR}) is related to the invariance with respect to time diffeomorphisms,
described by Eq.(\ref{eq:continuity-GR}).
Therefore, they are closely related and lead to the same conservation equation.

\subsection{Fluid picture}
\label{sec:fluid-weak}

We can obtain the hydrodynamical action by substituting the expression
(\ref{eq:rho-S-def})
into the action (\ref{eq:S-psi-full-2}) of the complex scalar field $\psi$.
The full expression is given in Eq.(\ref{eq:S-rho-full-2}) in the
appendix~\ref{app:fluid}. Again, in the nonrelativistic regime, we neglect the fast oscillatory
terms and the action simplifies to
\ba
&& S_\phi = \int d^4x \; a^3 \biggl \lbrace \frac{1-\Phi-3\Psi}{2m^2} ( - 2 m \rho \dot{S} )
\nonumber \\
&& - \frac{1+\Phi-\Psi}{2 m^2 a^2} \biggl( \frac{(\nabla\rho)^2}{4\rho} + \rho (\nabla S)^2 \biggl)
- \rho \Phi - {\cal V}_{\rm I}^{\rm eff}(\rho) \biggl \rbrace . \hspace{0.7cm}
\label{eq:S-rho-2}
\ea
As in Eq.(\ref{eq:S-psi-2}), we only keep the leading term with the highest power of $m$
in the standard time-derivative kinetic term.
The self-interaction potential is obtained by substituting the expression (\ref{eq:rho-S-def})
in the potential (\ref{eq:V-I-psi}), with the correction (\ref{eq:lambda-2n-eff}) in the case
of nonstandard kinetic terms. It gives
\be
{\cal V}_{\rm I}^{\rm eff}(\rho) = \Lambda^4 \sum_{n=2}^{\infty} \frac{\lambda_{2n}^{\rm eff}}{2n}
\frac{(2n)!}{(n!)^2} \left( \frac{\rho}{2 m^2 \Lambda^2} \right)^n ,
\label{eq:V-I-rho}
\ee
which does not depend on $S$.
We can also identify the conserved current (\ref{eq:Jmu-def}) with
\be
J^0 = a^3 \rho , \;\;\;
\vec J = \frac{a \rho \vec \nabla S}{m} = a^2 \rho {\vec v} .
\label{eq:Jmu-fluid}
\ee

\subsection{Nonrelativistic effective potential}
\label{sec:NR-potential}

\subsubsection{Resummation}
\label{sec:resum}

Thus, in the nonrelativistic limit the self-interactions and the nonstandard kinetic terms
appear through the transformed potential ${\cal V}_{\rm I}^{\rm eff}$.
In practice, the equation of motion will involve the derivative
$d{\cal V}_{\rm I}^{\rm eff}/d\rho$. We can obtain a more explicit relationship with $V_{\rm I}$
an $K_{\rm I}$ by resumming the series (\ref{eq:V-I-rho}). First, we can extract the even terms
of $V_{\rm I}(\phi)$ by defining the function
\ba
&& x> 0 : \;\;\;   U_{\rm I}(x) = \sum_{n=2}^{\infty} \lambda_{2n}^{\rm eff} x^{n-1}
\hspace{3cm} \nonumber \\
&& \hspace{0.3cm} = \frac{V'_{\rm I}(\sqrt{x} \Lambda) - V'_{\rm I}(-\sqrt{x}\Lambda)}
{2\Lambda^3 \sqrt{x}} - \frac{m^2}{\Lambda^2}
K'_{\rm I}\left(x \frac{m^2\Lambda^2}{2} \right) , \hspace{0.7cm}
\label{eq:U-I-def}
\ea
where $V_{\rm I}'(\phi) = dV_{\rm I}/d\phi$ and
$K'_{\rm I}(X)=dK_{\rm I}/dX$.
Then, using the integral representation of the beta function \cite{Gradshteyn1965},
\be
\frac{(2n)!}{(n!)^2} = \frac{2}{\pi} \int_0^1 \frac{du}{\sqrt{1-u^2}} \; [ 4 (1-u^2) ]^n ,
\ee
we obtain for the transformed function ${\cal U}_{\rm I}(x)$,
\ba
{\cal U}_{\rm I}(x) & = & \sum_{n=2}^{\infty} \lambda_{2n}^{\rm eff} \frac{(2n)!}{(n!)^2} x^{n-1}
\label{eq:cU_I-series} \\
& = & \frac{8}{\pi} \int_0^1 du \; \sqrt{1-u^2} \; U_{\rm I}[ 4(1-u^2)x ] ,
\label{eq:cUI-UI}
\ea
and the first derivative of the nonrelativistic potential ${\cal V}_{\rm I}^{\rm eff}$ reads
\be
\Phi_{\rm I}^{\rm eff}(\rho) = \frac{d{\cal V}_{\rm I}^{\rm eff}}{d\rho}
= \frac{\Lambda^2}{4m^2} \, {\cal U}_{\rm I}\left( \frac{\rho}{2m^2\Lambda^2}\right) .
\label{eq:Phi-I-def}
\ee
The relation (\ref{eq:cUI-UI}) can be inverted in a similar fashion. Defining the function
\ba
&& x> 0 : \;\;\;   W_{\rm I}(x) = \sum_{n=2}^{\infty} \frac{\lambda_{2n}^{\rm eff}}{2n} x^{n-1}
\hspace{2cm} \nonumber \\
&& \hspace{1.cm} = \frac{V_{\rm I}(\sqrt{x} \Lambda) + V_{\rm I}(-\sqrt{x}\Lambda)}
{2\Lambda^4 x} - \frac{K_{\rm I}\left(x \frac{m^2\Lambda^2}{2} \right)}{\Lambda^4 x} ,
\hspace{0.5cm}
\ea
which describes the even part of the initial potential $V_{\rm I}(\phi)$, we obtain
\be
W_{\rm I}(x) = \frac{1}{2} \int_0^1 du \; (1-u) \; {\cal U}_{\rm I}[u (1-u) x] .
\ee
These expressions assume that the self-interaction potential and the nonstandard kinetic term
are given by their series expansion (\ref{eq:V-I-def}) and (\ref{eq:K-I-def}) over the range
of interest.

\subsubsection{Power-law potentials or kinetic terms}
\label{sec:power-law-potentiall}

If the kinetic term is canonical and the self-interaction potential is a monomial,
\be
K_{\rm I}(X) = 0 , \;\;\;
V_{\rm I}(\phi) = \Lambda^4 \frac{\lambda_{2n}}{2n} \frac{\phi^{2n}}{\Lambda^{2n}} ,
\label{eq:K-I-V-I-quartic}
\ee
or if the nonlinear kinetic term is a monomial and the self-interaction potential vanishes,
\be
V_{\rm I}(\phi) = 0 , \;\;\;
K_{\rm I}(X) = \Lambda^4 \frac{k_n}{n} \frac{X^n}{\Lambda^{4n}} .
\label{eq:K-I-V-I-quartic-K-essence}
\ee
The nonrelativistic self-interaction potential $\Phi_{\rm I}$ is also a power law,
\be
\Phi_{\rm I}(\rho) = \left( \frac{\rho}{\rho_a} \right)^{n-1} ,
\label{eq:Phi-rho-power-law}
\ee
with
\be
\rho_a = \left( \frac{\lambda_{2n} \Lambda^2}{4 m^2} \frac{(2n)!}{(n!)^2} \right)^{-1/(n-1)}
2 m^2 \Lambda^2
\label{eq:rhoa-def}
\ee
for the potential case (\ref{eq:K-I-V-I-quartic}),
{
and
\be
\rho_a = \left( - \frac{k_n}{4} \frac{(2n)!}{(n!)^2} \right)^{-1/(n-1)} 4 \Lambda^4
\label{eq:rhoa-def-kinetic}
\ee
for the kinetic case (\ref{eq:K-I-V-I-quartic-K-essence}).
Here we focus on the cases $\lambda_{2n} > 0$ or $k_n<0$, where the potential
$\Phi_{\rm I}$ gives a repulsive force.
}

To ensure that the background scalar field behaves like pressureless dark matter,
at least from the time of radiation-matter equality until now, we must satisfy the constraint
(\ref{eq:V-I-small}). This implies $\Phi_{\rm I}(\bar\rho_{\rm eq}) \lesssim 1$, hence
\be
\bar{{\cal V}}^{\rm eff}_{\rm I} \lesssim \bar\rho : \;\;\;
\rho_a \gtrsim \bar\rho_{\rm eq} \sim 10^{11} \bar\rho_0 \sim 10^{-36} \, {\rm GeV}^4  .
\label{eq:r-I-20}
\ee
{
In the kinetic case (\ref{eq:K-I-V-I-quartic-K-essence}), this implies for coefficients
$k_n$ of order unity that the cutoff $\Lambda$ must be above $1 \, {\rm eV}$,
\be
\mbox{if } k_n \sim 1 : \;\;\;  \Lambda \gtrsim 1 \, {\rm eV} .
\ee
}

\subsubsection{Cosine potential}
\label{sec:cosine-potential}

For illustrative purposes, let us consider a bounded potential such as a cosine,
with a standard kinetic term.
As explained above, this could also correspond to a bounded nonlinear correction
to the kinetic term.
Following the two-scale scenario discussed below Eq.(\ref{eq:V-I-small}),
we write the full scalar-field potential as the sum of a leading quadratic term
and a subleading nonlinear potential, taken to be a cosine,
\be
V(\phi) = \frac{m_0^2}{2} \phi^2 + M_I^4 \left[ \cos(\phi/\Lambda) - 1 \right] ,
\;\;\; \frac{M_{\rm I}^4}{\Lambda^2} \ll m_0^ 2 .
\ee
We can absorb the quadratic part of the cosine into the mass term
and write $V(\phi) = \frac{m^2}{2} \phi^2 + V_{\rm I}(\phi)$, with
\ba
&& m^2 = m_0^2 - \frac{M_{\rm I}^4}{\Lambda^2} \simeq m_0^2 , \\
&& V_{\rm I}(\phi) = M_{\rm I}^4 \left[ \cos(\phi/\Lambda) - 1
+ \frac{\phi^2}{2\Lambda^2} \right]  .
\ea
For $\phi \ll \Lambda$ we recover a quartic potential, with
$\lambda_4 = M_{\rm I}^4/(6\Lambda^4)$.
Using the resummation described in section~\ref{sec:resum},
the function $U_{\rm I}$ defined in Eq.(\ref{eq:U-I-def}) reads
\be
U_{\rm I}(x) = \frac{M_{\rm I}^4}{\Lambda^4} \left[ 1 - \frac{\sin\sqrt{x}}{\sqrt{x}} \right ] ,
\ee
and the function ${\cal U}_{\rm I}(x)$ defined in Eq.(\ref{eq:cU_I-series}) reads
\be
{\cal U}_{\rm I}(x) = \frac{2M_{\rm I}^4}{\Lambda^4} \left[ 1 - \frac{J_1(2\sqrt{x})}{\sqrt{x}}
\right] .
\ee
This yields for the nonrelativistic self-interaction potential $\Phi_{\rm I}(\rho)$,
\be
\Phi_{\rm I}(\rho) = \frac{8\rho_b}{\rho_a} \left[ 1 - \frac{2J_1(\sqrt{\rho/\rho_b})}{\sqrt{\rho/\rho_b}} \right] ,
\label{eq:Phi-I-J1}
\ee
with
\be
\rho_a = \frac{8 m^4 \Lambda^4}{M_{\rm I}^4} , \;\;\; \rho_b = \frac{m^2\Lambda^2}{2} ,
\;\;\; \rho_b \ll \rho_a .
\ee
At low densities we again recover the case of the quartic potential,
while at high densities the self-interaction potential converges to a finite value,
\ba
\rho \ll \rho_b : && \Phi_{\rm I}(\rho) = \frac{\rho}{\rho_a} + ... \\
\rho \gg \rho_b : && \Phi_{\rm I}(\rho) = \frac{8\rho_b}{\rho_a}  \ll 1 .
\ea
The resummation (\ref{eq:Phi-I-J1}) is justified because the series expansions of
$V_{\rm I}$, $U_{\rm I}$ and ${\cal U}_{\rm I}$ converge over the full positive real axis.
Independently of the details of the scalar-field potential, the generic consequence
of a bounded $V_{\rm I}(\phi)$ is a bounded nonrelativistic potential $\Phi_{\rm I}(\rho)$.

Because the potential $\Phi_{\rm I}$ now satisfies a small upper bound,
we automatically verify the pressureless condition
(\ref{eq:V-I-small}) for the background at all redshifts. This no longer constrains
$\rho_a$ to be larger than $\bar\rho_{\rm eq}$, or the first expansion coefficient
$\lambda_4$ to obey Eq.(\ref{eq:lambda4-DM}), as long as $\rho_b \ll \rho_a$
and $\rho_b < \bar\rho_{\rm eq}$.
However, the constraints (\ref{eq:lambda4-DM}) and (\ref{eq:r-I-20}) still apply,
for the other reason described in Eq.(\ref{eq:ra-lambda4})
and section~\ref{sec:Jeans-quartic} below, associated with the formation of large-scale
structures.
Indeed, the Jeans length set by the repulsive self-interaction, given by
Eqs.(\ref{eq:cs-rho})-(\ref{eq:rJ}), must remain below $20 \, {\rm kpc}$ to ensure that
Lyman-$\alpha$ clouds and galaxies can form
(we assume that we are in the low-density regime $\bar\rho \ll \rho_b$
for $z \lesssim 6$).

\section{Cosmological background}
\label{sec:background}

\subsection{Real scalar field $\phi$}
\label{sec:background-phi}

For the cosmological background, the Einstein equations, or equivalently the derivatives
of the full action with respect to $\Phi$ and $\Psi$, give the Friedmann equations
\ba
&& 3 M_{\rm Pl}^2 H^2 = \frac{1}{2} \dot{\bar\phi}^2 + \frac{1}{2} m^2 \bar\phi^2
+ \bar\rho_{\rm de} + \bar\rho_\gamma , \\
&& M_{\rm Pl}^2 (3H^2 +2 \dot{H} ) = - \frac{1}{2} \dot{\bar\phi}^2 + \frac{1}{2} m^2 \bar\phi^2
+ \bar\rho_{\rm de} - \frac{\bar\rho_\gamma}{3} , \hspace{1cm}
\ea
where we included the additional dark energy and radiation components.
The derivative of the action with respect to $\phi$ gives the equation of motion
\be
(1+K'_{\rm I}+2 \bar{X} K''_{\rm I}) \ddot{\bar\phi} + 3 H (1+K'_{\rm I}) \dot{\bar\phi}
+  m^2 \bar\phi + \frac{dV_{\rm I}}{d\phi} = 0 .
\label{eq:phi-KG}
\ee
Throughout this article we consider the regime where $m$ is much larger than other
energy scales, as in Eq.(\ref{eq:m-k}).
Thus, we look for asymptotic solutions of the dynamics in the limit
$m\to\infty$. Then, at leading order the equation of motion simplifies as
$\ddot\phi+m^2\phi=0$, with the solutions $e^{\pm imt}$. This is the basis for the decomposition
(\ref{eq:psi-def}), which is only relevant in this nonrelativistic regime for
$m \gg H$ and $m \gg |\vec k|$.
In this regime, we can obtain the asymptotic solution of the equation of motion
(\ref{eq:phi-KG}) by a standard periodic averaging method \cite{Verhulst2005},
which is also related to the ``variation of constants'' method introduced by Lagrange.
Thus, starting from the unperturbed equation of motion, $\ddot\phi+m^2\phi=0$,
we look for solutions of the form
\be
\bar\phi(t) = \bar\varphi(t) \cos(mt-\bar{S}(t)) ,
\label{eq:phi-cos}
\ee
such that
\be
\dot{\bar\phi} = - m \bar\varphi(t) \sin(mt-\bar{S}(t)) .
\label{eq:dphi-sin}
\ee
These two equations implicitly define the two functions $\bar\varphi$ and $\bar S$
from $\bar\phi$. The second equation (\ref{eq:dphi-sin}) for the time derivative of
$\bar\phi$ also implies
\be
\dot{\bar\varphi} \cos(mt-\bar S) + \bar\varphi \dot{\bar S} \sin(mt-\bar S) = 0 .
\label{eq:dphi-consistency}
\ee
For the unperturbed solution $e^{\pm imt}$, $\bar\varphi$ and $\bar S$ are constant.
In the perturbed case, they slowly vary with time and modulate the amplitude and phase
of the fast oscillations of the unperturbed solution $\cos(mt)$.
Substituting into the equation of motion (\ref{eq:phi-KG}), we obtain
\ba
&& \dot{\bar\varphi} \sin(mt-\bar{S}) - \bar\varphi \dot{\bar S} \cos(mt-\bar S)
= \nonumber \\
&& - 3 H \bar\varphi \sin(mt-\bar{S}) - ( K'_{\rm I} + 2 \bar{X} K''_{\rm I} ) m \bar\varphi
\cos(mt-\bar S) \nonumber \\
&& + \frac{1}{m} \frac{dV_{\rm I}}{d\bar\varphi} ,
\label{eq:phi-KG-cos}
\ea
where we only kept the leading terms in $K_{\rm I}$, using
\be
K'_{\rm I} \ll 1, \;\;\; X K''_{\rm I} \ll 1,
\ee
in agreement with Eq.(\ref{eq:V-I-small}).
Combining Eqs.(\ref{eq:dphi-consistency}) and (\ref{eq:phi-KG-cos}) gives
\ba
&& \dot{\bar\varphi} = - 3 H \bar\varphi \sin^2(mt-\bar{S}) - ( K'_{\rm I} + 2 \bar{X} K''_{\rm I} )
m \bar\varphi  \nonumber \\
&& \times  \cos(mt-\bar S)  \sin(mt-\bar S) + \frac{1}{m} \sin(mt-\bar{S})
\frac{dV_{\rm I}}{d\bar\varphi} , \hspace{1cm}
\ea
and
\ba
&& \dot{\bar S} =  3 H \cos(mt-\bar S) \sin(mt-\bar{S}) +  ( K'_{\rm I} + 2 \bar{X} K''_{\rm I} ) m
\nonumber \\
&& \times \cos^2(mt-\bar S) - \frac{1}{m\bar\varphi} \cos(mt-\bar S)
\frac{dV_{\rm I}}{d\bar\varphi} .
\ea
In agreement with the role of $\bar\varphi$ and $\bar{S}$ as slow variables, we can
check that $\dot{\bar\varphi}$ and $\dot{\bar{S}}$ do not show high-frequency factors $m$
(the only factors $m$ that appear in the right-hand sides are multiplied by the small
quantities $K'_{\rm I}$ and $X K''_{\rm I}$).
Then, the idea of the method is to average the right-hand sides over the fast-oscillation period
$2\pi/m$.
Using the expansions (\ref{eq:V-I-def}) and (\ref{eq:K-I-def}), this gives for the averaged quantities,
\ba
&& \dot{\bar\varphi} = - \frac{3}{2} H \bar\varphi , \\
&& \dot{\bar S} = -  \frac{\Lambda^2}{4 m} \sum_{n=2}^{\infty} \lambda_{2n}^{\rm eff}
\frac{(2n)!}{(n!)^2} \left( \frac{\bar\varphi^2}{4\Lambda^2}\right)^{n-1} ,
\ea
where the coefficients $\lambda_{2n}^{\rm eff}$ are given by Eq.(\ref{eq:lambda-2n-eff}).
This gives the solutions
\be
\bar\varphi = \bar\varphi_0 \, a^{-3/2}
\label{eq:psi0}
\ee
and
\be
\dot{\bar S} = -  \frac{\Lambda^2}{4 m}
\sum_{n=2}^{\infty} \lambda_{2n}^{\rm eff} \frac{(2n)!}{(n!)^2}
\left( \frac{\bar\varphi_0^2}{4\Lambda^2a^3}\right)^{n-1} ,
\label{eq:Sdot-series}
\ee
which yields
\be
\bar S(t) = \bar S_0 - \int_{t_0}^t dt \; \frac{\Lambda^2}{4 m} \;
{\cal U}_{\rm I}\left( \frac{\bar\varphi_0^2}{4\Lambda^2a^3}\right) ,
\label{eq:S-S0}
\ee
where $\bar\varphi_0$ and $\bar{S}_0$ are integration constants and we recognized
the function ${\cal U}_{\rm I}$ defined in Eq.(\ref{eq:cU_I-series}).
Thus, the fast oscillations remove the contributions from the
odd terms $\lambda_{2n+1} \phi^{2n+1}$ of the scalar-field self-interaction potential
and give rise to the factor $(2n)!/(n!)^2$, associated with the nonrelativistic potential
(\ref{eq:V-I-psi}), which here does not appear from combinatorics but from
averages over powers of trigonometric functions.

We also recover the property that the nonstandard kinetic terms only give rise to
an additional contribution to the nonrelativistic potential, which agrees with
Eqs.(\ref{eq:K-I-psi}) and (\ref{eq:V-I-eff}).
Therefore, in the following we no longer explicitly consider the nonstandard kinetic
contribution ${\cal K}_{\rm I}$, as it is understood it is included in the potential
${\cal V}_{\rm I}^{\rm eff}$, and we omit he superscript ``eff'' to simplify notations.

As is well known, this integration method ensures that secular terms are absent.
Thus, the amplitude decays as $a^{-3/2}$, as for the harmonic case,
and the small anharmonic correction only generates a nonzero phase shift.
This means that the scalar field energy density,
$\bar\rho_\phi = \dot{\bar\phi}^2/2 + m^2 \bar\phi^2/2$, decays as $a^{-3}$
and can play the role of the cold dark matter.

At this leading order for the real scalar field $\bar\phi(t)$, its energy density and pressure are
\be
\bar\rho_\phi = \frac{m^2\bar\varphi_0^2}{2a^3} , \;\;\; \bar{p}_\phi = 0 ,
\ee
where we averaged over the fast oscillations.
Therefore, $\bar\varphi_0$ is set by the scalar-field energy density, which can be set equal
to the dark matter density if there is no other dark matter component.

At a subleading order, the contribution of the self-interactions $V_{\rm I}$ generate a nonzero
pressure. However, as in the action (\ref{eq:S-phi-2}) (where it would correspond to the term
$\Psi V_{\rm I}$), we neglect this contribution to the background dynamics, in agreement
with the constraint (\ref{eq:V-I-small}).
Clearly, if $V_{\rm I}$ is of the same order as $\rho_\phi$, it is not a small correction to the
Klein-Gordon equation of motion (\ref{eq:phi-KG}). Then, it will modify the period and the
amplitude of the oscillations, which also show higher-order harmonics, and the
asymptotic solution (\ref{eq:phi-cos}) is no longer valid.
Then, the nonrelativistic decomposition (\ref{eq:psi-def}) is no longer relevant.
The same constraints apply to the nonstandard kinetic term $K_{\rm I}$.

\subsection{Nonrelativistic limit}
\label{sec:background-psi}

Comparing the solution (\ref{eq:phi-cos}) with the decomposition (\ref{eq:psi-def}),
we can see that at the background level, the complex scalar field $\psi$
defined by Eq.(\ref{eq:psi-def}) is
\be
\bar\psi(t) = \bar\psi_0 \, a^{-3/2} e^{i\bar{S}} ,
\;\;\; \mbox{with} \;\;\; \bar\psi_0 = \sqrt{\frac{m}{2}} \bar\varphi_0 = \sqrt{\frac{\bar\rho_0}{m}} .
\label{eq:psi-background}
\ee
Then, we can check that (\ref{eq:psi-background}) is indeed the solution of the equation of
motion derived from the nonrelativistic action (\ref{eq:S-psi-2}), which reads 
\be
i \left( \dot{\bar\psi} + \frac{3}{2} H \bar\psi \right) = \frac{\partial {\cal V}_{\rm I}}{\partial\psi^\star} .
\ee
Substituting the expression (\ref{eq:psi-background}) and using Eq.(\ref{eq:V-I-psi})
for the self-interaction potential ${\cal V}_{\rm I}(\psi,\psi^\star)$, we recover the equation of
motion (\ref{eq:Sdot-series}) for $\dot{\bar S}$.
It provides the explicit connection, at the background level, between the periodic averaged
asymptotic solution (\ref{eq:phi-cos}) of the real scalar field $\phi$, in the limit where
$m$ is the largest energy scale of the system, the nonrelativistic decomposition
(\ref{eq:psi-def}), and the nonrelativistic action (\ref{eq:S-psi-2}).

\subsection{Fluid picture}
\label{sec:background-fluid}

Comparing the solution (\ref{eq:psi-background}) with the decomposition (\ref{eq:rho-S-def}),
we find that at the background level, $\bar\rho$ and $\bar{S}$ can be identified with the quantities
introduced in the solutions for the real scalar field $\phi$ and the complex scalar field $\psi$.
We can check that the solution defined by $\bar\rho= \bar\rho_0/a^3$ and $\bar S$ given by
Eq.(\ref{eq:Sdot-series}), which also can be written as
\be
\dot{\bar S} = -  \frac{m\Lambda^4 a^3}{2\bar\rho_0}
\sum_{n=2}^{\infty} \lambda_{2n} \frac{(2n)!}{(n!)^2}
\left( \frac{\bar\rho_0}{2m^2\Lambda^2a^3}\right)^n ,
\label{eq:Sdot-series-rho}
\ee
or in its integrated form as
\be
\bar S(t) = \bar S_0 - \int_{t_0}^t dt  \; \frac{\Lambda^2}{4m} \;
{\cal U}_{\rm I} \left( \frac{\rho_0}{2m^2\Lambda^2a^3} \right) ,
\label{eq:S-S0-rho}
\ee
is indeed the solution of the equations of motion derived from the hydrodynamical action
(\ref{eq:S-rho-2}), which read
\ba
&& \dot{\bar S} = - m \frac{d {\cal V}_{\rm I}}{d\rho} ,
\label{eq:dot-S-rho} \\
&& \dot{\bar\rho} + 3 H \bar\rho = 0 .
\ea
From the expression (\ref{eq:V-I-rho}), we can see that Eq.(\ref{eq:dot-S-rho}) coincides
with Eq.(\ref{eq:Sdot-series-rho}).
It provides the explicit link, at the background level, between the periodic averaged
asymptotic solution (\ref{eq:phi-cos}) of the real scalar field $\phi$ and the nonrelativistic
fluid picture.

\section{Perturbations}
\label{sec:perturbations}

Whereas for the study of the cosmological background we started from the scalar field
$\phi$ to  make the connection with the nonrelativistic complex field $\psi$ and the
hydrodynamical fields $(\rho,S)$, for the perturbations it is more convenient to start
from the fluid picture, which is similar to the standard treatment of the CDM scenario,
and next make the connection with the fields $\psi$ and $\phi$.

\subsection{Fluid picture}
\label{sec:perturbations-fluid}

Within the fluid picture, the metric and scalar-field perturbations are governed by the action
(\ref{eq:S-rho-2}). As the latter only depends on derivatives of $S$, it is
invariant under the global shift symmetry
\be
S\to S +\alpha ,
\label{eq:shift}
\ee
which is again a nonrelativistic consequence of the symmetry (\ref{eq:t-alpha-S}),
and the associated Noether current
\be
J^\mu = -m \frac{\delta{\cal L}}{\delta(\partial_\mu S)}
\ee
is conserved.
This corresponds to the U(1) symmetry for $\psi$ and to the conserved
current $J^\mu$ of Eq.(\ref{eq:Jmu-Noether}), given by
Eqs.(\ref{eq:Jmu-def}) and (\ref{eq:Jmu-fluid}).
The conservation law $\dot{J}^0 + \nabla \cdot \vec J = 0$
reads
\be
\dot \rho + 3 H \rho + \frac{1}{m a^2} \nabla (\rho \nabla S) = 0 ,
\label{eq:continuity}
\ee
and it is equivalent to the equation of motion for $S$.
Defining the velocity field by Eq.(\ref{eq:v-S-def}), it takes the form of the hydrodynamical
continuity equation
\be
\dot \rho + 3 H \rho + \frac{1}{a} \nabla (\rho \vec v) = 0 .
\label{eq:continuity-cosmo}
\ee
We can see that the self-interactions do not modify this continuity equation.
The Euler-Lagrange equation for $\rho$ provides the second equation
of motion,
\be
\dot S + \frac{(\nabla S)^2}{2ma^2} = - m \Phi - m \frac{d{\cal V}_{\rm I}}{d\rho}
+ \frac{1}{2m a^2} \frac{\nabla^2 \sqrt{\rho}}{\sqrt{\rho}} .
\label{eq:S-pert}
\ee
Taking the gradient of this equation gives the hydrodynamical Euler equation,
\be
\dot {\vec v} +  H \vec v + \frac{1}{a} (\vec v \cdot \nabla) \vec v = - \frac{1}{a}
\nabla (\Phi + \Phi_{\rm I} + \Phi_{\rm Q} ) ,
\label{eq:Euler}
\ee
where we used $\nabla(\vec v^{\,2}) = 2 (\vec v \cdot \nabla) \vec v$ because
$\nabla\times\vec v=0$.
The self-interaction potential $\Phi_{\rm I}$ is defined in
Eq.(\ref{eq:Phi-I-def}) and we have introduced the ``quantum pressure'' term
\be
\Phi_{\rm Q} = - \frac{\nabla^2 \sqrt \rho}{2m^2a^2 \sqrt \rho} .
\label{eq:Phi-Q}
\ee
Thus, we recover the dynamics of the standard cold dark matter scenario,
with density $\rho$ and fluid velocity $\vec v$, with the additional interaction potential
and quantum potential $\Phi_{\rm I}$ and $\Phi_{\rm Q}$
\cite{Marsh:2015daa}.
These new terms must remain small on large scales, so as to match observational
data, but they could lead to significant effects up to galactic scales, where
the CDM scenario shows several tensions with observations.

Using the expressions of the Einstein tensor recalled in appendix~\ref{app:weak-gravity},
the $\Phi$ and $\Psi$ equations read
\ba
&& M_{\rm Pl}^2 [ - 3 H^2 + 6 H^2 \Phi + 6 H \dot\Psi - 2 a^{-2} \nabla^2 \Psi ] =
\frac{\rho\dot{S}}{m} \nonumber \\
&& - \frac{1}{2m^2a^2} \left( \frac{(\nabla\rho)^2}{4\rho} + \rho
(\nabla S)^2 \right) - \rho -\bar\rho_{\rm de} - \bar\rho_{\gamma} , \hspace{1cm}
\label{eq:E00}
\ea
and
\ba
&& M_{\rm Pl}^2   \left[ - 3 H^2 - 2 \dot{H} + 2 \ddot\Psi + 6 H \dot\Psi
+ 2 H \dot\Phi \right. \nonumber \\
&& \left. + (6H^2+4\dot{H}) \Phi + a^{-2} \frac{2}{3}\nabla^2 (\Phi-\Psi)  \right]
 \nonumber \\
&& =   \left[ - \frac{\rho\dot{S}}{m} - \frac{1}{6m^2a^2} \left( \frac{(\nabla\rho)^2}{4\rho}
+ \rho (\nabla S)^2 \right) - \bar\rho_{\rm de} + \frac{\bar\rho_\gamma}{3} \right] \nonumber \\
&&  ,
 \label{eq:Eij}
\ea
where we neglect the perturbations of the radiation and dark energy component
(which may be a cosmological constant).
As noticed in section~\ref{sec:S-EH-weak} from Eq.(\ref{eq:S-EH-G}),
the first equation (\ref{eq:E00}) and the trace of the equations (\ref{eq:Eij}) can also be obtained
from derivatives of the full action with respect to $\Phi$ and $\Psi$.

In the nonrelativistic regime that we consider in our analysis, most terms in these Einstein
equations can actually be neglected. For the study of perturbations, we are interested
in small subhorizon scales with wave numbers
\be
\frac{k}{a} \gg H .
\ee
On the other hand, we focus on scenarios where $1/m$ is much smaller than
cosmological length scales, $m \gg 10^{-27} {\rm eV}$,
\be
\frac{k}{am} \ll 1 , \;\;\; \mbox{with} \;\;\;
\frac{1}{m} = 6.4 \times 10^{-27}
\left( \frac{m}{1\,{\rm eV}} \right)^{-1} \, {\rm kpc} .
\ee
The definition (\ref{eq:v-S-def}) gives
\be
\delta S \sim \frac{a m v}{k} , \;\;\; \delta \dot{S} \sim \frac{H a m v}{k} , \;\;\;
\dot{\bar S} = - m \bar{\Phi}_{\rm I}
\ee
where $\delta S = S - \bar{S}$, we used the fact that $\vec v$ varies on the Hubble
timescale for cosmological structures, and we used Eq.(\ref{eq:dot-S-rho})
with the definition (\ref {eq:Phi-I-def}).
Then, using $v^2 \ll 1$, as cosmological and galactic structures have
$v^2 \sim \Phi \lesssim 10^{-5}$, we find that  (\ref{eq:E00})
simplifies to
\be
\frac{\nabla^2 \Psi}{a^2} = \frac{\rho-\bar\rho}{2 M_{\rm Pl}^2} , \;\;\;
\mbox{hence} \;\;\; \Psi \sim \frac{a^2 H^2 \delta}{k^2} ,
\ee
with $\delta = (\rho-\bar\rho)/\bar\rho$.
and  Eq.(\ref{eq:Eij}) gives
\be
\Psi-\Phi \sim \frac{\delta^2 H^2}{m^2} + \frac{a^2 H^2 (1+\delta) }{k^2} v^2 .
\ee
Therefore, $|\Psi-\Phi| \ll \Psi$ as $v \ll 1$ and $m^2 \gg k^2 \delta /a^2$, implying 
therefore
\be
\Phi \simeq \Psi ,
\ee
at first order over $\Phi$ and $\Psi$.
We also have the scaling $v \sim Ha/k$, which simply states that cosmological
structures evolve on Hubble timescales.
Therefore, we recover the standard Poisson equation in the nonrelativistic regime,
\be
\frac{\nabla^2 \Phi}{a^2} = 4\pi {\cal G} \bar\rho \delta .
\label{eq:Poisson}
\ee
On the other hand, the Euler equation (\ref{eq:Euler}) implies
\be
\frac{d {\cal V}_{\rm I}}{d\rho} = \Phi_{\rm I} \lesssim \Phi .
\ee
Indeed, in realistic scenarios the departure from CDM must be small on large 
cosmological scales, beyond clusters scales, where the linear power spectrum
and growth rate are measured to better than $10\%$. This implies from
Eq.(\ref{eq:Euler}) that $\Phi_{\rm I} \ll \Phi$ on these linear scales. 
On the other hand, on galactic scales we can authorize effects of order unity
(but not much greater, otherwise this would generate large velocities according
to Eq.(\ref{eq:Euler})).
Therefore, in realistic scenarios we have ${\cal V}_{\rm I} \lesssim \rho \Phi \ll \rho$, 
from subgalactic to Hubble scales.
This agrees with the constraint
(\ref{eq:V-I-small}) and confirms that we could drop the term $(\Phi-3\Psi) V_{\rm I}$
in the action (\ref{eq:S-phi-2}), as it is a higher-order term at most of order $\rho \Phi^2$.

\subsection{Complex scalar field $\psi$}
\label{sec:perturbations-complex-scalar}

The equation of motion of the nonrelativistic complex scalar field $\psi$ is obtained
from the action (\ref {eq:S-psi-2}),
\be
i \left( \dot{\psi} + \frac{3}{2} H \psi \right) = - \frac{\nabla^2 \psi}{2 m a^2} + m \Phi \psi
+ \frac{\partial {\cal V}_{\rm I}}{\partial\psi^\star} .
\label{eq:psi-NR-motion}
\ee
This nonlinear Schrodinger equation is a generalized Gross-Pitaevskii equation
if ${\cal V}_{\rm I}$ is not quartic.
We recover Eqs.(\ref{eq:continuity}) and (\ref{eq:S-pert}) when we make the substitution
(\ref{eq:rho-S-def}).
If we consider linear perturbations in $\rho$ and $S$ defined by
\be
\rho = \bar\rho (1+\delta) , \;\;\; S= \bar{S} + \delta S ,
\label{eq:linear-rho-S}
\ee
the linear perturbation over $\psi$ reads
\be
\psi = \bar\psi + \delta\psi \;\;\; \mbox{with} \;\;\;
\delta\psi = \bar\psi \left( \frac{\delta}{2} + i \delta S \right) .
\ee

\subsection{Real scalar field $\phi$}
\label{sec:perturbations-real-scalar}

The equation of motion of the real scalar field $\phi$ is obtained
from the action (\ref {eq:S-phi-2}),
\be
\ddot{\phi} + 3 H \dot{\phi} - \frac{1}{a^2} \nabla^2\phi
+ m^2 \phi + \frac{dV_{\rm I}}{d\phi} = 0 .
\label{eq:phi-pert}
\ee
The linear perturbations defined by Eq.(\ref{eq:linear-rho-S}) correspond to
$\phi = \bar\phi+\delta\phi$ with
\be
\delta\phi = \frac{\sqrt{2\bar\rho}}{m} \left[ \frac{\delta}{2} \cos(mt-\bar{S})
+ \delta S \sin(mt-\bar{S}) \right] .
\ee
This can also be written as
\be
\phi = \frac{\sqrt{2\bar\rho}}{m} \left( 1+\frac{\delta}{2} \right) \cos(mt-\bar{S}-\delta S) .
\ee
As expected and as for the background solution (\ref{eq:phi-cos}), in the nonrelativistic
limit the scalar field $\phi$ can be written as a modulation of the fast oscillations
$\cos(mt)$ by slow background and perturbation corrections.

\subsection{Gravitational instability}
\label{sec:grav-instab}

For small perturbations with respect to the FLRW background, we can linearize the
equations of motion. It is convenient to work with the fluid approach, defining
the linear density contrast $\delta$ and the divergence $\theta$ of the fluid velocity,
\be
\delta = \frac{\rho-\bar\rho}{\bar\rho} , \;\;\;   \theta = \frac{\nabla\cdot\vec{v}}{a} .
\ee
The continuity equation (\ref{eq:continuity}) gives
\be
\theta = -\dot\delta ,
\ee
and the Euler equation (\ref{eq:Euler}) gives
\be
\dot\theta + 2 H \theta = - \frac{1}{a^2} \nabla^2 (\Phi + \Phi_{\rm I} + \Phi_Q ) .
\ee
Combining these two equations, using the Poisson equation (\ref{eq:Poisson})
and the expression (\ref{eq:Phi-Q}) of the quantum potential,
we obtain in Fourier space the modified growth equation \cite{Chavanis:2011uv}
\be
\ddot\delta + 2 H \dot\delta + \left( c_s^2 \frac{k^2}{a^2} - 4 \pi {\cal G} \bar\rho \right) \delta
= 0 ,
\label{eq:delta-cosmo}
\ee
where we introduced the speed of sound $c_s$ as
\be
c_s^2 = \frac{k^2}{4 a^2 m^2} + \bar\rho \frac{d\bar\Phi_{\rm I}}{d\bar\rho}
=  \frac{k^2}{4 a^2 m^2} + \bar\rho \frac{d^2\bar{\cal V}_{\rm I}}{d\bar\rho^2} .
\ee
The first term comes from the quantum potential and only plays a role at short distances.
As dynamical timescales are of the order of the Hubble time for galactic and cosmological
structures, the nonzero speed of sound plays a role up to the acoustic length scale
$r_s \sim c_s/H$. This implies that the quantum pressure becomes important at the
acoustic quantum scale $r_{\rm Q}$, with
\be
r_{\rm Q} = \frac{1}{\sqrt{Hm}} \simeq 1.7 \times 10^{-10}
\left( \frac{m}{1\,{\rm eV}} \right)^{-1/2} \, {\rm kpc} ,
\label{eq:rQ-def}
\ee
at redshift $z=0$.
This regime has already been studied in detail in the literature and corresponds to light
scalar fields, with $m \lesssim 10^{-21} {\rm eV}$, associated with fuzzy dark matter.
In this work, we focus on models with $m \gg 10^{-21} {\rm eV}$, where the quantum
pressure is negligible, and the speed of sound is set by the self-interactions.
Then, it becomes scale independent and only depends on the density,
\be
c_s^2(\bar\rho) =  \bar\rho \frac{d\Phi_{\rm I}}{d\bar\rho} .
\label{eq:cs-rho}
\ee
From Eq.(\ref{eq:delta-cosmo}), we recover the usual gravitational instability on large scales,
$k/a<r_J^{-1}$, beyond the Jeans length $r_J$, and acoustic oscillations on smaller scales,
with
\be
\frac{k_J}{a} = \frac{1}{r_J} \;\;\; \mbox{with} \;\;\;
r_J = \frac{c_s}{\sqrt{4\pi {\cal G} \bar\rho}} .
\label{eq:rJ}
\ee

\subsection{The quartic model}
\label{sec:Jeans-quartic}

{
The quartic model corresponds to $V_{\rm I}(\phi) \propto \phi^4$, that is, $n=2$ in
Eq.(\ref{eq:K-I-V-I-quartic}), or to $K_{\rm I}(X) \propto X^2$, that is, $n=2$ in
Eq.(\ref{eq:K-I-V-I-quartic-K-essence}).
In both cases we obtain
\be
\rho_a = \frac{4m^4}{3\lambda_4} , \;\;\; \Phi_{\rm I}(\rho) = \frac{\rho}{\rho_a} ,
\;\;\; c_s^2(\bar\rho) = \frac{\bar\rho}{\rho_a} =  \frac{3\lambda_4\bar\rho}{4 m^4} ,
\label{eq:quartic-Phi-I}
\ee
where in the K-essence scenario we define
\be
\rho_a = - \frac{8\Lambda^4}{3 k_2} , \;\;\; \lambda_4 = - \frac{k_2}{2} \frac{m^4}{\Lambda^4} .
\ee
}
To avoid small-scale instabilities, we require $c_s^2 \geq 0$, and hence
$\lambda_4 > 0$ or $k_2 < 0$ in the K-essence case.
More generally, $c_s^2$ is guaranteed to be positive when all the $\lambda_n\ge 0$ and
$k_n \le 0$ in the self-interaction potential and the nonstandard kinetic term.
However, this is not the most general case, and we only need the effective
potential $\Phi_{\rm I}$ to be a monotonic increasing function of $\rho$.
The quartic model gives a Jeans length that is independent of the density
\cite{Goodman:2000tg,Chavanis:2011uv}, hence of redshift,
\be
r_J = \frac{1}{\sqrt{4\pi{\cal G} \rho_a}} .
\label{eq:rJ-quartic}
\ee
To ensure that galaxies and Lyman-$\alpha$ clouds form, the Jeans length should be smaller
than about $20 \, {\rm kpc}$.
This happens to coincide with the constraint (\ref{eq:r-I-20}), which expresses that the
background pressure is negligible up to redshift $z_{\rm eq}$, see also
Eqs.(\ref{eq:lambda4-DM}) and (\ref{eq:lambda4-ra}).

Thus, in this scenario the scalar field self-interactions typically become
relevant on galactic or subgalactic scales at most, below $20$ kpc, and the
development of the cosmic web, cluster of galaxies, and large galaxies,
proceeds as in the standard CDM scenario.

{
In the K-essence scenario, the nonrelativistic effective potential reads
\be
\Phi_{\rm I}(\rho) = \frac{\rho}{\rho_a} - \frac{20 k_3}{9 k_2^2} \left( \frac{\rho}{\rho_a} \right)^2 
+ \dots 
\ee
If the coefficients $k_n$ are of order unity, that is, there is no other scale than the cutoff
$\Lambda$, we can use the leading approximation (\ref{eq:quartic-Phi-I}) for densities
$\rho \ll \rho_a$, which is satisfied as soon as $\rho \ll 10^{11} \bar\rho_0$.
Therefore, this is sufficient for all cosmological and galactic halos.
}

\section{Small-scale dynamics}
\label{sec:small-scale}

\subsection{Fluid approach}

On galactic and subgalactic scales, we neglect the expansion of the Universe and
the background density. This applies to high-density inner parts of virialized halos
or to astrophysical scales.
Then, the continuity and Euler equations (\ref{eq:continuity-cosmo}) and
(\ref{eq:Euler}) become
\ba
&& \dot \rho + \nabla_r (\rho \vec v) = 0 ,
\label{eq:continuity-small-scale} \\
&& \dot {\vec v} +  (\vec v \cdot \nabla_r) \vec v = - \nabla_r (\Phi + \Phi_{\rm I} ) ,
\label{eq:Euler-small-scale}
\ea
where we neglected the quantum potential $\Phi_{\rm Q}$ as we focus on models
with $m \gg 10^{-21} {\rm eV}$.
Here $\nabla_r$ is the gradient or divergence operator with respect to the physical coordinate
$\vec r = a \vec x$.
The Poisson equation (\ref{eq:Poisson}) becomes
\be
\nabla^2_r \Phi = 4\pi {\cal G} \rho  \;\;\; \mbox{hence} \;\;
\Phi(\vec r) = - {\cal G} \int d\vec r^{\,'} \frac{\rho(\vec r^{\,'})}{|\vec r^{\,'} - \vec r|} .
\label{eq:Poisson-small-scale}
\ee

We can check that the total energy $E$, given by the sum of the kinetic,
gravitational, and internal (i.e. self-interaction) energies,
\be
E = \int d{\vec r} \left[ \rho \frac{\vec v^{\,2}}{2} + \frac{1}{2} \rho \Phi + {\cal V}_{\rm I} \right] ,
\label{eq:E-rho-v}
\ee
is conserved by the dynamics \cite{Chavanis:2011zi}.
Indeed, taking its time derivative and using
Eqs.(\ref{eq:continuity-small-scale})-(\ref{eq:Poisson-small-scale}) we obtain
$dE/dt=0$.

In terms of the phase $S$, instead of the peculiar velocity ${\vec v}$, the equations
of motion read
\ba
&& \dot \rho +  \frac{1}{m} \nabla_r (\rho \nabla_r S) = 0 ,
\label{eq:S-continuity-small-scale} \\
&& \dot S + \frac{(\nabla S)^2}{2m} = - m (\Phi + \Phi_{\rm I} ) ,
\label{eq:S-Euler-small-scale}
\ea
where we again neglected the quantum pressure term.
The total energy now reads
\be
E = \int d{\vec r} \left[ \rho \frac{(\nabla S)^2}{2 m^2} + \frac{1}{2} \rho \Phi
+ {\cal V}_{\rm I} \right] ,
\label{eq:E-rho-S}
\ee
and it is again conserved by the equations of motion
(\ref{eq:S-continuity-small-scale})-(\ref{eq:S-Euler-small-scale}).

\subsection{Complex scalar field}

In terms of the nonrelativistic complex field $\psi$, the equation of motion
(\ref{eq:psi-NR-motion}) becomes
\ba
i  \dot\psi & = & - \frac{\nabla^2_r \psi}{2 m} + m \Phi \psi
+ \frac{\partial {\cal V}_{\rm I}}{\partial\psi^\star}  \\
& = & - \frac{\nabla^2_r \psi}{2 m} + m (\Phi + \Phi_{\rm I}) \psi ,
\label{eq:psi-NR-motion-static}
\ea
while the total energy reads \cite{Chavanis:2011zi,RindlerDaller:2011kx}
\be
E = \int d{\vec r} \left[ \frac{\nabla_r\psi \cdot \nabla_r\psi^\star}{2 m} + \frac{1}{2} m \psi
\psi^\star \Phi + {\cal V}_{\rm I} \right] .
\label{eq:E-psi}
\ee
Again, we can check that the total energy (\ref{eq:E-psi}) is conserved by the
equation of motion (\ref{eq:psi-NR-motion-static}).
However, it is actually different from the energy defined in Eqs.(\ref{eq:E-rho-v})
and (\ref{eq:E-rho-S}), as in the fluid approach above we neglected the quantum
pressure term, whereas in Eq.(\ref{eq:E-psi}) we cannot separate the terms associated
with the quantum pressure from the spatial derivative terms associated with the velocity
$\vec v = \nabla_r S/m$.

\subsection{Static soliton}
\label{sec:static}

On small scales, we can look for static equilibrium solutions with
$\vec v=0$.
From the Euler equation (\ref{eq:Euler-small-scale}), the equation of hydrostatic
equilibrium reads
\be
\nabla_r ( \Phi + \Phi_{\rm I} ) = 0 .
\label{eq:hydrostatic}
\ee
This corresponds to a soliton solution, where the Newtonian gravity is balanced by
the dark-matter self-interactions, which act as a pressure, rather than by the velocity
dispersion or the orbital angular momentum as in standard cold dark matter scenarios.
We can integrate Eq.(\ref{eq:hydrostatic}) once, which gives
\be
\Phi + \Phi_{\rm I} = \alpha ,
\label{eq:Phi-Phi-I-0}
\ee
where $\alpha$ is an integration constant.
On the other hand, taking the divergence of Eq.(\ref{eq:hydrostatic}) and using the Poisson
equation (\ref{eq:Poisson-small-scale}) yields
\be
\nabla^2_r \Phi_{\rm I} = -4\pi{\cal G} \rho .
\label{eq:Poisson-nonlinear}
\ee
The self-interaction potential $\Phi_{\rm I}$ is a function of $\rho$, which defines the
inverse function $\rho(\Phi_{\rm I})$, and we obtain the Lane-Emden equation
\be
\nabla^2_r \Phi_{\rm I} + 4\pi{\cal G} \rho(\Phi_{\rm I}) = 0  .
\ee
This governs the shape of static soliton equilibria, which may form at the center
of dark matter halos where the self-interactions become large.

In terms of the nonrelativistic complex field $\psi$, the equation of motion
was given in Eq.(\ref{eq:psi-NR-motion-static}).
We can check that the static soliton (\ref{eq:Phi-Phi-I-0}) is a solution,
with
\be
\psi = \sqrt{\frac{\rho}{m}} e^{-i\alpha m t} , \;\;\; \mbox{hence} \;\;
S = - \alpha m t ,
\ee
where $\alpha$ is the integration constant introduced in Eq.(\ref{eq:Phi-Phi-I-0}).
Depending on the choice of normalization for the Newtonian gravitational potential,
hence for $\alpha$, there is a uniform time dependent phase, which corresponds
to a vanishing velocity $\vec v=\nabla_r S/m$.
Here we used the fact that the Laplacian term in Eq.(\ref{eq:psi-NR-motion-static})
is negligible, because we focus on the large-$m$ regime (\ref{eq:no-quantum-pressure})
where the quantum pressure is negligible and we have
\be
\frac{\nabla^2_r \psi}{2 m} \sim \frac{q^2}{m^2} m \psi \ll 10^{-6} \, m\psi \lesssim m \Phi \psi ,
\ee
where $q=k/a$ is the physical wave number.
Indeed, from Eq.(\ref{eq:m-k}) we have $q/m < 10^{-5}$ for $q< 1 \, {\rm kpc}^{-1}$
and $m>10^{-21}\,{\rm eV}$.

\subsection{Power-law potentials}
\label{sec:power-law}

For power-law interaction potentials or nonlinear kinetic terms,
we have $\Phi_{\rm I} \propto \rho^{n-1}$
as in Eq.(\ref{eq:Phi-rho-power-law}), which can be inverted as
\be
\rho(\Phi_{\rm I}) = \rho_a \Phi_{\rm I}^{1/(n-1)} .
\label{eq:rho-Phi-power-law}
\ee
Defining the characteristic radius $r_a$ and the dimensionless coordinate
$\vec x = \vec{r}/r_a$, as
\be
r_a = (4\pi{\cal G} \rho_a)^{-1/2} ,
\label{eq:ra-def}
\ee
we obtain
\be
\nabla_x^2 \Phi_{\rm I} + \Phi_{\rm I}^{1/(n-1)} = 0 .
\ee
For a spherically symmetric halo, this reads
\be
\frac{d^2\Phi_{\rm I}}{dx^2} + \frac{2}{x} \frac{d\Phi_{\rm I}}{dx} + \Phi_{\rm I}^{1/(n-1)} = 0,
\label{eq:Helmoltz}
\ee
with  the boundary conditions that $\Phi_{\rm I}$ tends to zero
at infinity and $d\Phi_{\rm I}/dx$ vanishes at the origin.

We can derive the scaling laws of the soliton profiles without explicitly solving
Eq.(\ref{eq:Helmoltz}). For a collapsed object of mass $M$ and radius $R$,
the Newtonian gravitational potential is $\Phi \sim {\cal G} M/R$.
The hydrostatic equilibrium implies $\Phi_{\rm I} \sim \Phi \sim {\cal G} M/R$.
Then, using Eq.(\ref{eq:rho-Phi-power-law}) we obtain the scaling laws
\be
\frac{\rho}{\rho_a} \propto \left( \frac{M}{M_a} \right)^{2/(3n-4)} , \;\;\;
\frac{R}{r_a} \propto \left( \frac{M}{M_a} \right)^{(n-2)/(3n-4)} ,
\label{eq:rho-R-M-n}
\ee
where we defined the characteristic mass $M_a$ by
\be
M_a = \frac{4\pi}{3} \rho_a r_a^3 .
\ee
These scaling laws will be derived in a more explicit fashion below in
section~\ref{sec:power-law-profiles}.
For $n=2$, which corresponds to the quartic model, the equilibrium radius of the halos
does not depend on their mass, while it grows with $M$ for $n>2$.

\subsection{The quartic model}
\label{sec:quartic}

We now focus on the quartic model with $n=2$, described by Eq.(\ref{eq:quartic-Phi-I}).
This also gives for the characteristic radius $r_a$ defined in Eq.(\ref{eq:ra-def})
\be
r_a = \frac{1}{\sqrt{4\pi{\cal G} \rho_a}} =
\sqrt{\frac{2}{3\Omega_{\rm m}}} \frac{c_s}{H} .
\ee
In this simple case, $r_a$ is equal to the Jeans length $r_J$ obtained in
Eq.(\ref{eq:rJ-quartic}).
The soliton equilibrium profile (\ref{eq:Helmoltz})
is now given by the linear equation
\be
\frac{d^2 \Phi_{\rm I}}{dx^2} + \frac{2}{x} \frac{d \Phi_{\rm I}}{dx} + \Phi_{\rm I} = 0 ,
\label{eq:soliton-quartic}
\ee
with the solution
\cite{Riotto:2000kh,Arbey:2003sj,Boehmer:2007um,Harko:2011jy,Chavanis:2011zi},
\be
\Phi_{\rm I}(x) = \Phi_{\rm I}(0) \, \frac{\sin x}{x} , \;\;\;
\rho(x) = \rho(0) \frac{\sin x}{x} ,
\label{eq:y-sinc}
\ee
where $\Phi_{\rm I}(0)$ and $\rho(0)$ are the values of the
self-interaction potential and the density at the center.
We immediately find that the soliton has a finite radius
\be
R_s= \pi r_a ,
\label{eq:Rh-def}
\ee
which must be smaller than about 50 kpc to match observations.
In the cosmological setting, the solitons, if they exist, are not isolated.
They form in the center of virialized halos, with outer parts that may deviate from
the soliton profile, with non-negligible radial and tangential velocities, and that connect
to the infalling matter from the cosmic web. An estimate of the size of the solitons
inside cosmological halos will be given in section~\ref{sec:halos}.
However, the expected size of the solitonic core remains set by $r_a$,
as in the simplified isolated case (\ref{eq:Rh-def}).
From the constraint (\ref{eq:r-I-20}), associated with the condition of pressureless background,
we also obtain $r_a \lesssim 20 \, {\rm kpc}$.
Thus, as noticed in Eqs.(\ref{eq:lambda4-DM}) and (\ref{eq:lambda4-ra}),
these two requirements happen to give the same condition in the plane
$(m,\lambda_4)$ and on $\rho_a$.

The density at the center is given by
\be
\rho(0) = \Phi_{\rm I}(0) \rho_a = \frac{\Phi_{\rm I}(0)}{c_s^2} \bar\rho .
\ee
For $\Phi_{\rm I} \sim 10^{-6}$ and $\rho_a \gtrsim 10^{11} \bar\rho_0$, this
gives a density contrast above $10^5$ today, as found on galactic scales,
while the speed of sound is constrained to be $c_s \lesssim 10^{-6}$.
Such solitonic cores embedded in galactic halos can provide a good agreement
with the rotation curves measured in galaxies \cite{Boehmer:2007um}.
However, the fact that the characteristic radius $R_s$ is independent of the
density $\rho(0)$ differs from the observation that galactic cores show the approximate
scaling law $\rho(0) \propto 1/R_c^\beta$ with $\beta \simeq 1$ \cite{Deng:2018jjz}.
Nevertheless, one can still obtain a reasonably good agreement with observed rotation
curves \cite{Zhang:2018okg}, especially if the halo has a nonzero rotation.
On the other hand, at small radii baryonic effects are non-negligible and could
modify these results.

\subsection{Power-law models}
\label{sec:power-law-profiles}

\begin{figure}
\begin{center}
\epsfxsize=8. cm \epsfysize=5. cm {\epsfbox{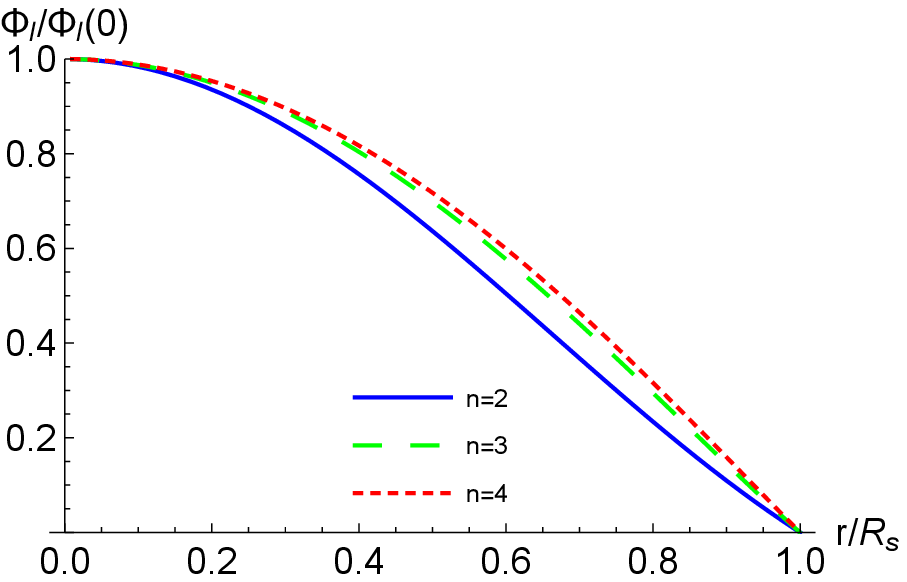}}\\
\epsfxsize=8. cm \epsfysize=5. cm {\epsfbox{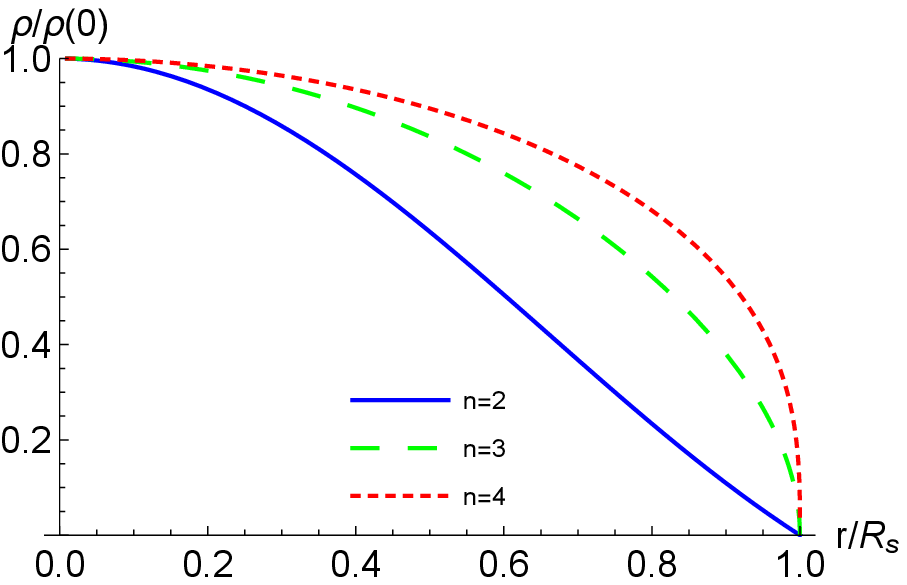}}
\end{center}
\caption{Profiles of the nonrelativistic self-interaction potential $\Phi_{\rm I}$
(upper panel) and of the density $\rho$ (lower panel) for the power-law cases
$n=2,3,$ and $4$.}
\label{fig_Phi-rho-n}
\end{figure}

For the generic power-law cases (\ref{eq:rho-Phi-power-law}), the scaling laws
(\ref{eq:rho-R-M-n}) show that it is convenient to define
the rescaled radial coordinate $u$ and self-interaction potential $y$ by
\be
u= \frac{r}{r_M} \;\;\; \mbox{with} \;\;\;
r_M = r_a \left( \frac{M}{M_a} \right)^{(n-2)/(3n-4)} ,
\ee
\be
\frac{\Phi_{\rm I}(r)}{\Phi_{\rm I}(0)} = y(u) , \;\;\; \frac{\rho(r)}{\rho(0)} = y(u)^{1/(n-1)} .
\ee
Then, the Lane-Emden equation (\ref{eq:Helmoltz}) reads
\be
\frac{d^2y}{du^2} + \frac{2}{u} \frac{dy}{du} +\alpha^{n-2} y^{1/(n-1)} = 0 ,
\label{eq:soliton-n}
\ee
where we introduced the quantity $\alpha$ defined by
\be
\alpha =  \left( \frac{M}{M_a} \right)^{2/(3n-4)} \Phi_{\rm I}(0)^{-1/(n-1)} .
\label{eq:alpha-def}
\ee
As for the case $n=2$, the soliton has a finite radius $R_s=U r_M$, where $U$ is the first
zero of the function $y(u)$. The normalization of the profile is set by the condition
$M = \int_0^R dr \, 4\pi r^2 \rho$, which reads
\be
\alpha = 3 \int_0^U du \, u^2 y^{1/(n-1)} .
\label{eq:alpha-norm}
\ee
Thus, for each index $n$, we must find the value $\alpha$ that satisfies the condition
(\ref{eq:alpha-norm}), where $y(u)$ is the $\alpha$-dependent solution of
Eq.(\ref{eq:soliton-n}) with the boundary conditions $y(0)=1$ and $y'(0)=0$.
From this fundamental solution, we obtain the profile for any mass $M$ from
Eq.(\ref{eq:alpha-def}), which gives
$\Phi_{\rm I}(0) = \alpha^{1-n} (M/M_a)^{2(n-1)/(3n-4)}$.
This gives in turn the scaling laws (\ref{eq:rho-R-M-n}).
In the case $n=2$, the explicit solution (\ref{eq:y-sinc}), $y_2(u)=\sin(u)/u$,
gives at once $U_2= \pi$ and $\alpha_2=3\pi$.
From a numerical computation, we obtain for $n=3$ the values
$U_3 \simeq 1.7$ and $\alpha_3 \simeq 2.6$, and for $n=4$ the values
$U_4 \simeq 1.4$ and $\alpha_4 \simeq 1.9$.

We compare in Fig.~\ref{fig_Phi-rho-n} the profiles of the nonrelativistic potential $\Phi_{\rm I}$
and of the density $\rho$ for the cases $n=2,3,$ and $4$, normalized to their value at the center.
The radial coordinate is normalized to the radius $R_s$ of the soliton.
We can see that the shape of the potential $\Phi_{\rm I}$ does not vary much from $n=2$
to $n=4$ but the density profile looks increasingly like a top-hat for higher $n$,
with a flatter core and a vertical slope at the boundary $R_s$ for $n>2$.

\subsection{The cosine model}
\label{sec:cosine}

\begin{figure}
\begin{center}
\epsfxsize=8. cm \epsfysize=5. cm {\epsfbox{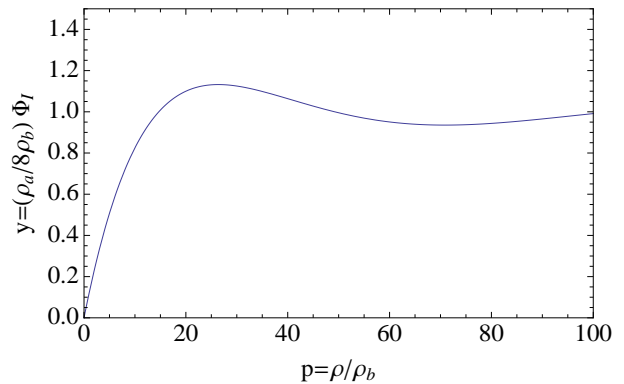}}
\end{center}
\caption{Nonrelativistic self-interaction potential $\Phi_{\rm I}(\rho)$ for a cosine
scalar field potential $V_{\rm I}(\phi)$.}
\label{fig_yp}
\end{figure}

For the cosine model described in section~\ref{sec:cosine-potential}, the nonrelativistic
potential $\Phi_{\rm I}(\rho)$ is given by Eq.(\ref{eq:Phi-I-J1}). In terms of the dimensionless
variables $p$ and $y$ defined by
\be
p= \frac{\rho}{\rho_b} , \;\;\; \Phi_{\rm I}(\rho) = \frac{8\rho_b}{\rho_a} y(p)  ,
\ee
we have
\be
y(p) = 1 - 2 J_1(\sqrt{p})/\sqrt{p} .
\ee
As shown in Fig.~\ref{fig_yp}, the function $y(p)$ behaves as $p/8$ for $p \ll 1$,
it reaches a maximum of $y_{\max}\simeq 1.13$ at $p_{\max}\simeq 26.37$,
and goes to unity at large $p$ with decreasing oscillations.
Defining again the characteristic radius $r_a=1/\sqrt{4\pi{\cal G}\rho_a}$,
and the dimensionless coordinate $x=r/r_a$, the soliton profile is given by the
nonlinear equation
\be
\frac{d^2y}{dx^2} + \frac{2}{x} \frac{dy}{dx} + \frac{p(y)}{8} = 0 .
\label{eq:soliton-cos}
\ee
At low density $\rho$ and potential $\Phi_{\rm I}$, we recover the linear equation
(\ref{eq:soliton-quartic}) of the quartic case.
At $p_{\max} \rho_b$ the potential $\Phi_{\rm I}$ becomes attractive, which gives
rise to an instability. At greater densities it shows a series of attractive and repulsive
domains but remains of finite amplitude. Therefore, it cannot support massive and
high-density halos.
Thus, a well-defined and smooth soliton profile only exists for halos with a central
density that is below the critical value $\rho_{\max} = p_{\max} \rho_b$.

\subsection{Stability}
\label{sec:stability}

Stable equilibria of isolated systems correspond to minima of the total energy
at fixed mass.
Saddle points are given by the equation $\delta E - \alpha \delta M = 0$
for the first-order variations, where $\alpha$ is the Lagrange multiplier associated
with the constraint of fixed mass \cite{Chavanis:2011zi}.
From Eq.(\ref{eq:E-rho-v}) this yields
\be
\int d{\vec r} \left[ \delta\rho \frac{{\vec v}^{\, 2}}{2}
+ \rho {\vec v} \cdot \delta {\vec v} + \delta\rho ( \Phi + \Phi_{\rm I} )
- \alpha \delta\rho \right] = 0 .
\ee
This must hold for any $\delta{\vec v}$, hence $\vec v=0$, and any $\delta\rho$,
hence $\Phi+\Phi_{\rm I} = \alpha$.
Thus, we recover the hydrostatic equilibrium (\ref{eq:Phi-Phi-I-0}), and the Lagrange
multiplier $\alpha$ is given by the integration constant of Eq.(\ref{eq:Phi-Phi-I-0}).

The second variation of the total energy (\ref{eq:E-rho-v}) reads
\be
\delta^2 E = \int d{\vec r} \left[ \rho \frac{\delta{\vec v}^{\,2}}{2} + \frac{1}{2} \delta\rho
\, \delta\Phi + \frac{1}{2} \Phi_{\rm I}' \, \delta\rho^2 \right] .
\ee
Therefore, the hydrostatic equilibrium (\ref{eq:Phi-Phi-I-0}) is stable if
\be
\delta^2 E(\delta\rho) = \frac{1}{2} \int d{\vec r} \left[ \delta\rho \,
\delta\Phi + \Phi_{\rm I}' \, \delta\rho^2 \right] > 0 ,
\ee
for all perturbations $\delta\rho$ that conserve the total mass,
$\int d{\vec r} \, \delta \rho = 0$.
This is identical to Chandrasekhar's variational principle for barotropic fluids
\cite{1987gady.book.....B}.

This condition greatly simplifies for the quartic model, where the energy
(\ref{eq:E-rho-v}) is a quadratic functional of the density and the second variation
$\delta^2 E$ does not depend on the equilibrium profile.
Using $\Phi_{\rm I}'=1/\rho_a$ from Eq.(\ref {eq:quartic-Phi-I}) and
the Poisson equation, $\nabla_r^2\delta\Phi = 4\pi {\cal G} \delta\rho$,
we obtain
\be
\delta^2 E(\delta\rho) = \frac{1}{2\rho_a}  \int d{\vec r} \left[ \frac{1}{r_a^2}
\delta\rho \cdot \nabla_r^{-2} \cdot \delta\rho + \delta\rho^2 \right] > 0 ,
\label{eq:d2E-quartic}
\ee
where the characteristic radius $r_a$ was defined in Eq.(\ref{eq:ra-def}).
Defining the Fourier space normalization as
$\delta\rho({\vec r}) = \int d{\vec q} \,e^{i{\vec q}\cdot{\vec r}} \delta\rho({\vec q})$,
this also reads as
\be
\delta^2 E = \frac{(2\pi)^3}{2\rho_a}  \int d{\vec q} \, \left| \delta\rho({\vec q}) \right|^2
\left[ 1 - \frac{1}{(q r_a)^2} \right] .
\label{eq:d2E-Fourier}
\ee
We can see that large-scale wave numbers $q<r_a^{-1}$ are unstable.
Thus, we recover the Jeans length $r_J$ obtained in Eq.(\ref{eq:rJ-quartic})
for perturbations with respect to the uniform cosmological background.
This is because for the quartic model the energy is a quadratic functional of the
density and the second variation (\ref{eq:d2E-Fourier}) does not depend on the
background profile, whether it is the homogeneous cosmological background or
the finite-size soliton in vacuum.

However, whereas plane waves can describe perturbations of the cosmological
background, they are not appropriate for the stability of the soliton profile (\ref{eq:y-sinc}),
because they extend over all space. For the study of the isolated soliton, which is a compact
object of finite radius $R_s=\pi r_a$ within vacuum, physical perturbations correspond
to continuous local matter redistributions and velocity fluctuations.
They conserve mass and do not extend far beyond the halo radius
(matter is not created in a discontinuous manner far from the object and needs
to be transported from radius $R_s$ in a continuous manner).
Therefore, to ensure dynamical stability, it is sufficient to show that the second variation
(\ref {eq:d2E-quartic}) is strictly positive for all perturbations $\delta\rho$ that conserve
mass and are restricted to a finite radius $R$, such that $R>R_s$.
If such a radius $R$ can be found and is strictly greater than $R_s$, this implies that
all local perturbations, including small increases of the halo radius, are taken into account.
The positivity of the symmetric quadratic form (\ref{eq:d2E-quartic}),
where $\int d{\vec r} \, \delta\rho \cdot \nabla_r^{-2} \cdot \delta\rho
= - \int d{\vec r}d{\vec r}^{\,'} \, \delta\rho({\vec r}) \delta\rho({\vec r}^{\,'})
/ (4\pi | {\vec r}-{\vec r}^{\,'} |)$, means that all eigenvalues
$\lambda$ of the associated eigenvector problem are strictly positive,
\be
\frac{1}{r_a^2} \nabla_r^{-2} \delta\rho + \delta\rho = \lambda \, \delta\rho , \;\;\; \lambda > 0 .
\ee
Taking the Laplacian of this equation we obtain the Helmholtz equation
\be
\nabla_r^2 \delta\rho = - \mu \, \delta\rho , \;\;\; \mu=\frac{1}{(1-\lambda) r_a^2} ,
\label{eq:Helmholtz-mu}
\ee
with the stability criterion
\be
\mbox{unstable iff} \;\;\; 0 \leq \mu \leq \frac{1}{r_a^2} .
\label{eq:mu-unstable}
\ee
The eigenfunctions of Eq.(\ref{eq:Helmholtz-mu}) that are finite at the origin and
vanish at some finite radius are the usual spherical harmonic eigenfunctions of the
Laplacian,
\be
\delta\rho({\vec r}) = j_{\ell}(\sqrt{\mu} r) Y_{\ell}^m(\theta,\varphi) , \;\;\; \mu > 0 .
\ee
Let us first consider the case $\ell \geq 1$. These eigenfunctions automatically conserve mass,
$\int d{\vec r} \, \delta \rho = 0$, through the integration over angles.
As we consider perturbations within a radius $R$, the boundary condition that sets the
discrete values of $\mu$ is $\delta\rho(R)=0$. This gives the eigenvalues
\be
\ell \geq 1: \;\;\; \mu_n^{(\ell)} = \left( \frac{x_n^{(\ell+1/2)}}{R} \right)^2 , \;\;\; n=1, 2, 3, \dots ,
\ee
where $x_n^{(\nu)}$ is the $n$-th strictly positive zero of the Bessel function of the first
kind $J_\nu$.
From Eq.(\ref{eq:mu-unstable}), stability is ensured if $\mu_n^{(\ell)} > 1/r_a^2$
for all $\ell$ and $n$. The smallest value corresponds to $\ell=1$ and $n=1$,
which gives the stability criterion
\be
\mbox{stable iff} \;\;\; \frac{x_1^{(3/2)}}{R} > \frac{1}{r_a} .
\label{eq:stable-ell-1}
\ee
Let us now consider the case $\ell=0$. We now take for boundary condition
$\delta M(R)=0$, to ensure that mass is conserved within the radius $R$.
For $\delta\rho({\vec r}) \propto j_0(\sqrt{\mu}r)$ we obtain
$\delta M(R) \propto (\sqrt{\mu} R)^{3/2} J_{3/2}(\sqrt{\mu}R)$. This yields the eigenvalues
\be
\ell = 0 : \;\;\; \mu_n^{(0)} =  \left( \frac{x_n^{(3/2)}}{R} \right)^2 , \;\;\; n=1, 2, 3, ... ,
\ee
and we obtain the same stability criterion as in Eq.(\ref{eq:stable-ell-1}).
Thus, the soliton profile is stable with respect to perturbations within radius $R$,
provided $R<x_1^{(3/2)} r_a$.
We have $x_1^{(3/2)} \simeq 4.493$, therefore we can choose for instance $R=4.4 \, r_a$.
This is strictly greater than the soliton radius $R_s=\pi r_a$ of Eq.(\ref{eq:Rh-def}).
Hence we conclude that the soliton profile (\ref{eq:y-sinc}) is dynamically stable.
As the energy (\ref{eq:E-rho-v}) is truly a quadratic functional of the density,
this goes beyond linear perturbations and the soliton is nonlinearly stable with
respect to finite perturbations, provided they correspond to finite density changes
and are restricted within radius $R$.
Of course, up to factors of order unity, we recover the result of the Fourier analysis
(\ref{eq:d2E-Fourier}), that the system is stable with respect to small-scale perturbations,
of wavelength below $2\pi r_a$ for the Fourier analysis, and of radius below
$x_1^{(3/2)} r_a$ for the local analysis.
The latter analysis provides a more accurate and appropriate criterion for the case
of the isolated soliton.

In the case of the cosine model described in section~\ref{sec:cosine}
the analysis is more intricate. In particular, because no solitons exist with central
density beyond the critical value $\rho_{\max}=p_{\max}\rho_b$,
we can expect new behaviors and nonlinear instabilities for equilibria close to this
threshold. On the other hand, for low-density equilibria, where we recover the quartic model,
we should also recover dynamical stability as well as nonlinear stability, within a radius $R$
greater than the soliton radius, but for density perturbations that remain below the
threshold $\rho_{\max}$.

\section{Cosmological halos}
\label{sec:halos}

Numerical simulations of fuzzy dark matter models
\cite{Schive:2014dra,Schwabe:2016rze,Mocz:2017wlg,Veltmaat:2018dfz},
without self-interactions but a non-negligible quantum pressure on galactic scales,
show that the cosmic web develops as in the standard CDM scenario, although
galaxy formation can be delayed.
However, inside galaxies solitonic cores form, surrounded by extended halos of
fluctuating density granules with a spherically averaged density profile that
matches the NFW profile \cite{Navarro:1995iw} observed in numerical simulations
of standard collisionless dark matter.

The case of strong attractive self-interactions and weak gravity
has also been studied in numerical simulations \cite{Guzman:2006yc,Amin:2019ums}.
Then, solitons are governed by the balance between
the quantum pressure and the attractive self-interactions and the numerical simulations
find a rich dynamics with formation, mergers and scatterings of the solitons.
At late times, only stable solitons survive.

The regime we study in this paper is different from those works, as we consider repulsive
self-interactions, which are weak on large scales, and negligible quantum pressure,
so that our solitons arise from the balance between the repulsive
self-interactions and gravity.
However, we also expect solitons to form and merge inside halos generated
by gravitational collapse and stable configurations to survive, while the outer regions
should again follow the standard NFW profile.
The case of solitonic cores inside isothermal halos was considered in the recent analytical
work \cite{Chavanis:2019faf}, which appeared during completion of this paper.
Here we do not assume thermodynamical equilibrium within an isothermal halo
and build a simple matching to the outer NFW halo.

\subsection{The quartic model}
\label{sec:quartic-NFW}

We now consider the profiles of cosmological halos, embedded in the cosmic web.
We focus on the quartic model, where $\Phi_{\rm I} = \rho/\rho_a$,
described in section~\ref{sec:quartic} above, which we normalize by
\be
\rho_a = 10^{12} \bar\rho_0 , \;\;\; r_a = 4.3 \, h^{-1} {\rm kpc} ,
\ee
to make sure that $\bar\Phi_{\rm I} \ll 1$ until the matter-radiation equality.
For $\lambda_4 \sim 1$ this corresponds to $m \sim 1 \, {\rm eV}$. { This model applies to either a quartic model with a potential
term or a K-essence theory with a quartic correction to the kinetic terms.}

At low redshifts, the size of cosmological halos such as clusters of galaxies is much greater
than $r_a$, and the dark matter force $F_{\rm I} = - \nabla\Phi_{\rm I}$ associated
with the self-interactions is negligible. Therefore, at large radii we can expect to recover the
usual NFW density profiles.
There, gravity is balanced by the velocity dispersion and orbital angular momentum
of the dark matter.
At small radii, the Newtonian gravitational potential typically decreases as
$\Phi \propto \rho r^2 \propto r$, while the self-interaction potential increases
as $\Phi_{\rm I} \propto \rho \propto r^{-1}$.
Therefore, below a radius that is mainly set by $r_a$, the new dark matter
self-interaction becomes important and can no longer be neglected.
In this regime, we expect the profile to follow the soliton solution
(\ref{eq:y-sinc}), where gravity is balanced by the self-interaction as in
Eq.(\ref{eq:hydrostatic}).
This expectation is motivated by the fact that the soliton profiles are dynamically and
nonlinearly stable as seen in section~\ref{sec:stability}.
However, numerical simulations are required to check that these equilibria can be reached
in the context of the virialized halos that arise from the cosmic web.

\begin{figure}
\begin{center}
\epsfxsize=8.5 cm \epsfysize=5.5 cm {\epsfbox{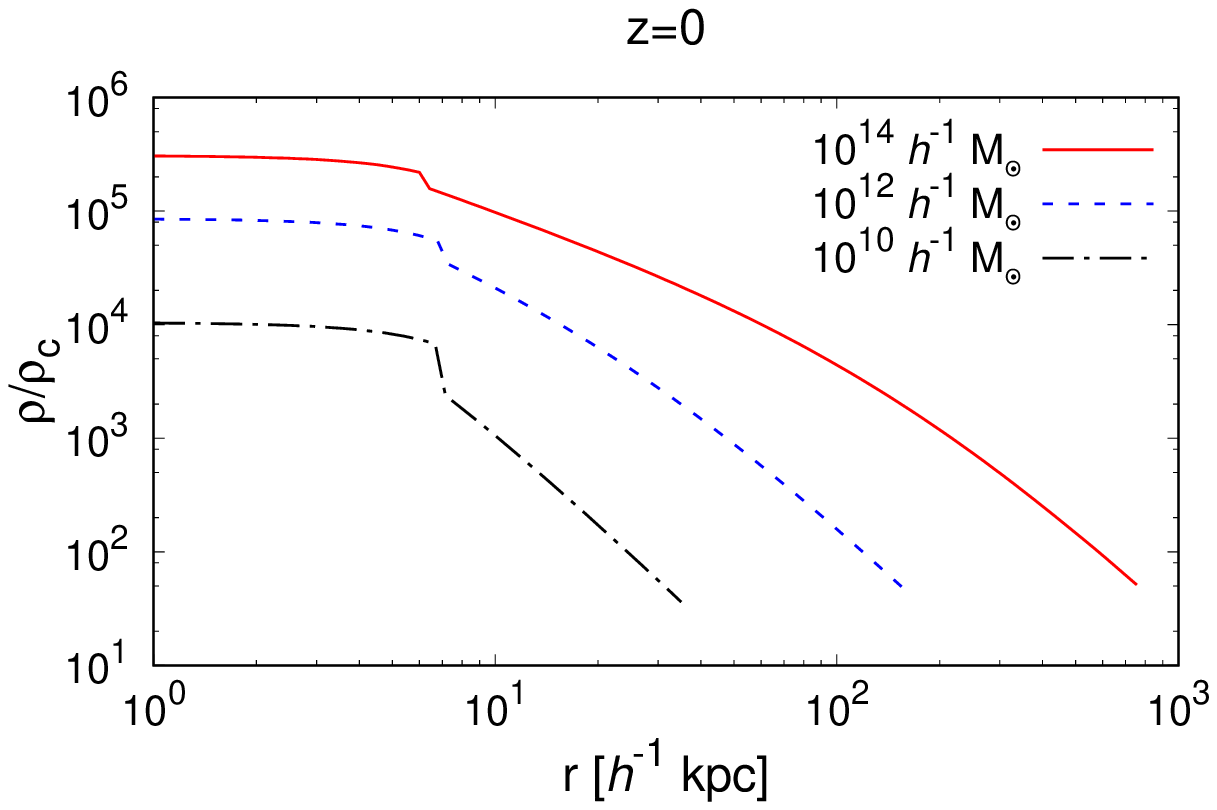}}\\
\epsfxsize=8.5 cm \epsfysize=5.5 cm {\epsfbox{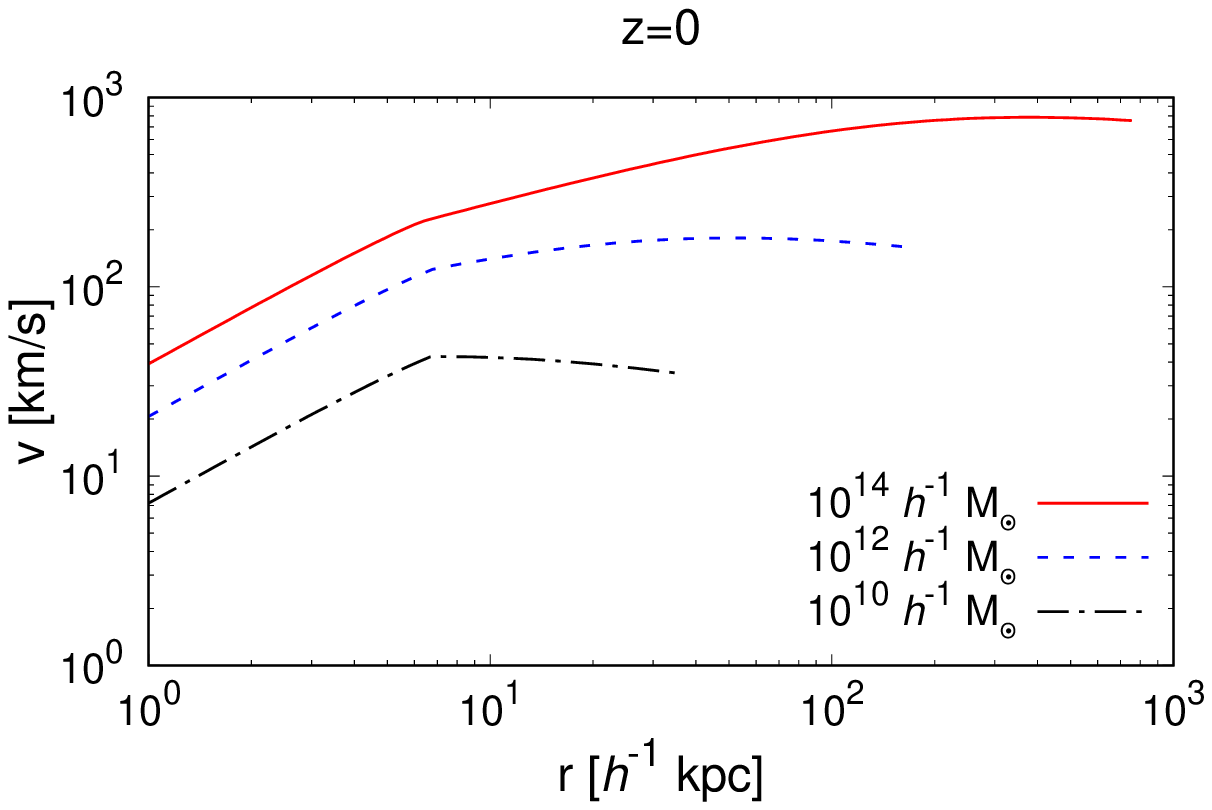}}\\
\epsfxsize=8.5 cm \epsfysize=5.5 cm {\epsfbox{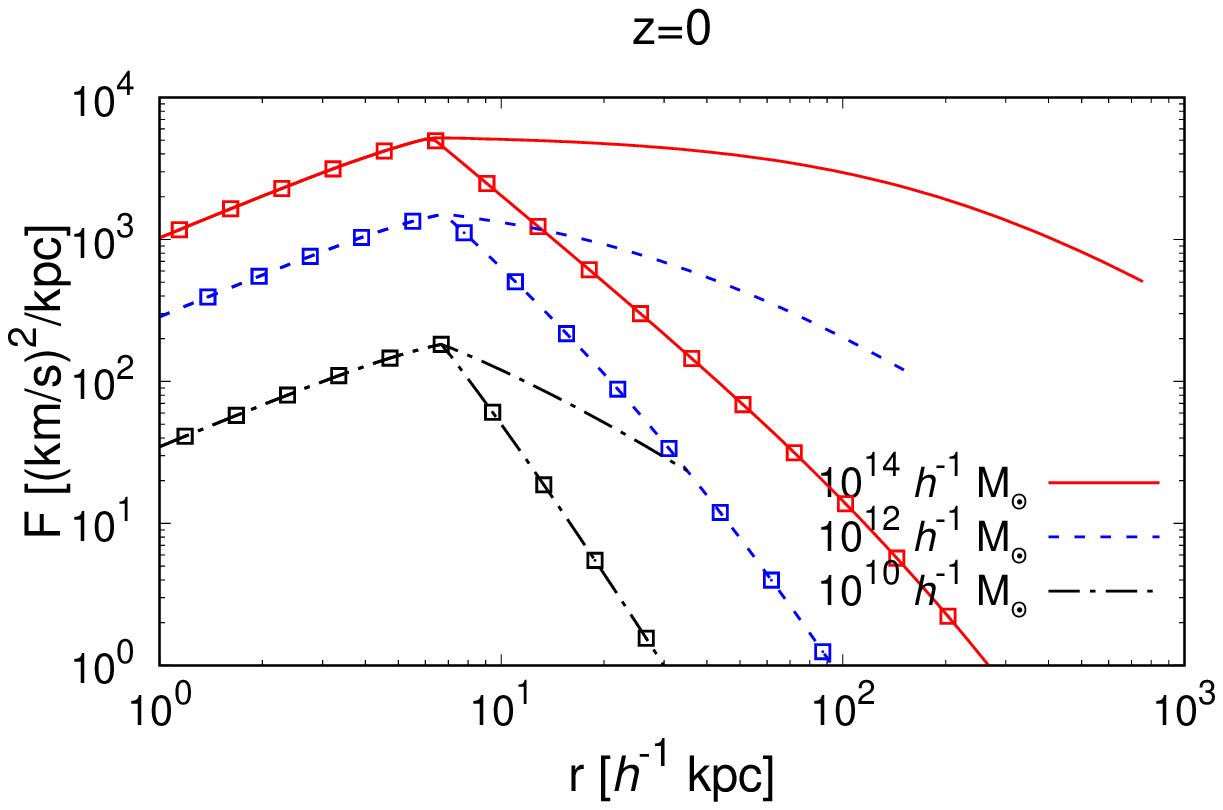}}
\end{center}
\caption{Dark matter halo profiles at redshift $z=0$ for halos of mass
$M=10^{14}$, $10^{12}$, and $10^{10} h^{-1} M_\odot$.
{\it Upper panel:} dark matter density $\rho$ in units of the critical density $\rho_c$.
{\it Middle panel:} circular velocity $v$.
{\it Lower panel:} Newtonian force $-F_{\rm N}$ and dark-matter force $F_{\rm I}$
(lines with squares), in units of ${\rm (km/s)^2/kpc}$.}
\label{fig_halo_z0}
\end{figure}

To estimate the impact of the self-interactions on the halo profiles, we adopt the simple
following model. On large radii, $r>R_\star$, we follow the standard NFW profile,
\be
R_\star < r < R : \;\;\; \rho(r) = \frac{\rho_s}{\frac{r}{r_s} \left(1+\frac{r}{r_s}\right)^2}
\ee
up to the halo radius $R$, with a concentration parameter $c=R/r_s$ from $\Lambda$-CDM
simulations \cite{Maccio:2008pcd}, and a density contrast of $200$ within radius $R$
with respect to the critical density $\rho_c(z)$.
We define the transition radius $R_\star$ as the radius where $F_{\rm I}=-F_{\rm N}$,
where $F_{\rm N}=-{\cal G}M/r^2$ is the Newtonian gravitational force.
At lower radius, we switch to the soliton profile (\ref{eq:y-sinc}),
\be
r < R_{\star} : \;\;\; \rho(r) = \rho(0) \frac{\sin(r/r_a)}{r/r_a} .
\ee
From Eq.(\ref{eq:hydrostatic}), in the soliton regime we have $F_{\rm I}=-F_{\rm N}$
at all radii $r<R_\star$.
The normalization $\rho(0)$ is set by the conservation of matter: the mass
associated with the soliton profile up to $R_\star$ is equal to the mass that would
have been associated with the NFW profile below $R_\star$,
\ba
&& M(<R_\star) =  \rho(0) 4\pi r_a^3 \left[ \sin\left(\frac{R_\star}{r_a}\right)
- \frac{R_\star}{r_a} \cos\left(\frac{R_\star}{r_a}\right) \right] \nonumber \\
&& \hspace{0.5cm} = \rho_s 4\pi r_s^3 \left[ \ln\left(1+\frac{R_\star}{r_s}\right)
- \frac{R_\star/r_s}{1+R_\star/r_s} \right] .
\label{eq:M-star-M-R-star}
\ea
In other words, we assume that below $R_\star$, where the dark matter self-interaction
is important, the mass is redistributed to converge to the static soliton solution
(\ref{eq:hydrostatic}), where gravity is balanced by the interactions instead of the
velocity dispersion or rotation.
The density $\rho$ and the self-interaction potential $\Phi_{\rm I}$ are discontinuous
at the transition $R_\star$ in this simplified treatment, because the support provided
by the velocity dispersion or rotational velocity in the NFW regime is abruptly set to zero
in the soliton regime. However, the mass and both the Newtonian and dark-matter forces
are continuous.

\begin{figure}
\begin{center}
\epsfxsize=8.5 cm \epsfysize=5.5 cm {\epsfbox{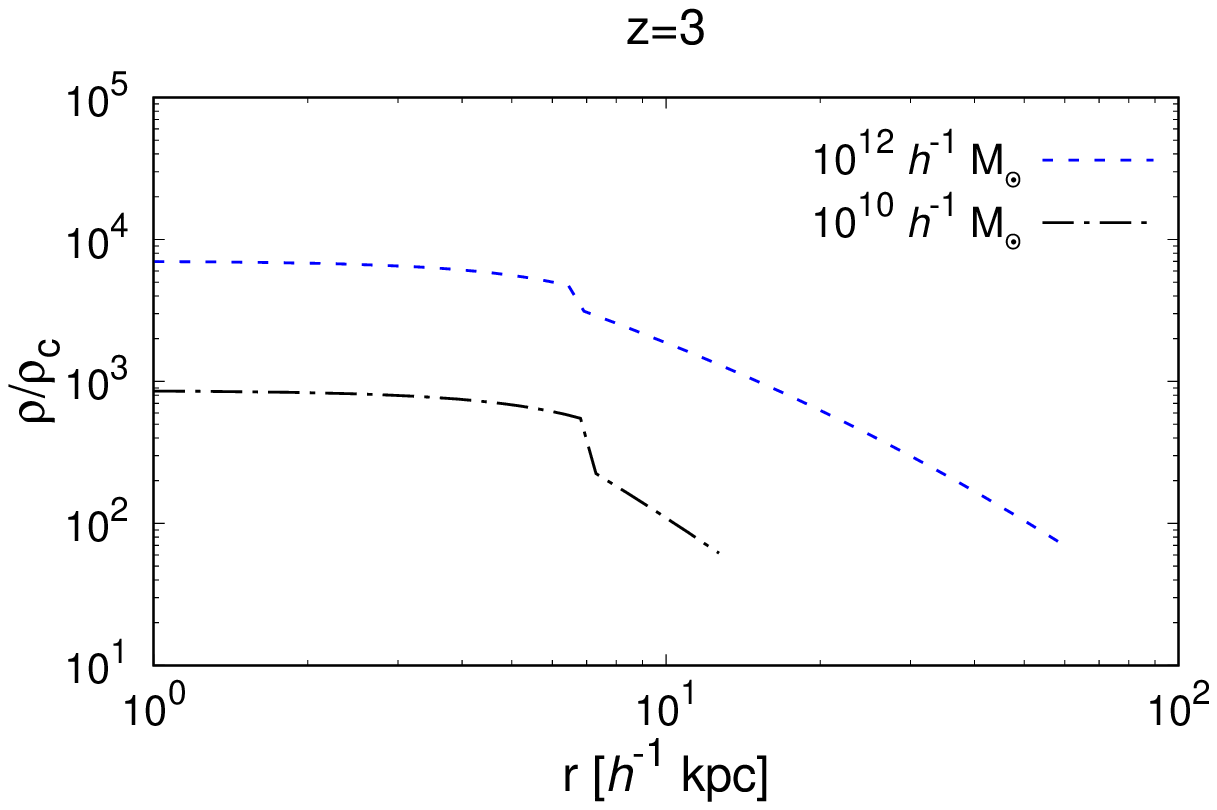}}\\
\epsfxsize=8.5 cm \epsfysize=5.5 cm {\epsfbox{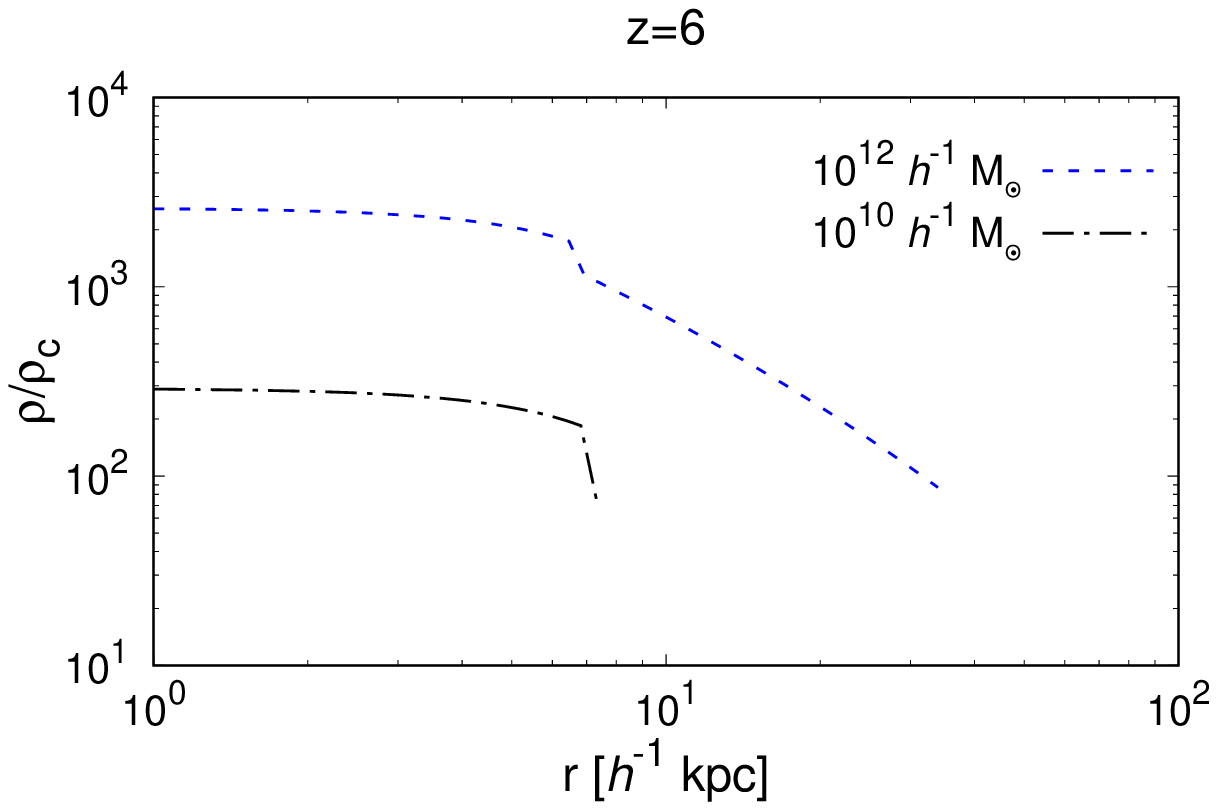}}
\end{center}
\caption{Dark matter halo density profiles at redshifts $z=3$ (upper panel)
and $z=6$ (lower panel).}
\label{fig_halo_z3-z6}
\end{figure}

\begin{figure}
\begin{center}
\epsfxsize=8.5 cm \epsfysize=5.5 cm {\epsfbox{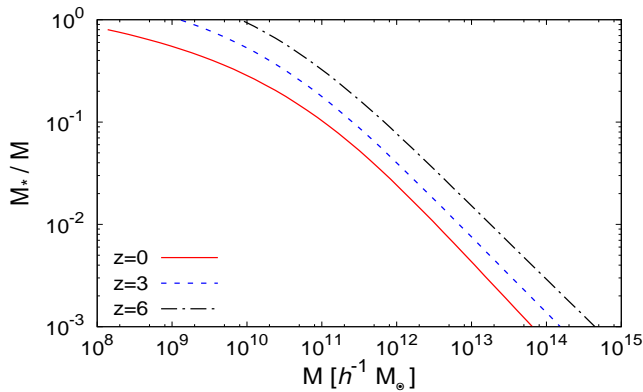}}
\end{center}
\caption{Mass ratio $M_\star/M$ of the solitonic core to the total halo mass
$M$ at redshifts $z=0$, $z=3$ and $z=6$.}
\label{fig_mass}
\end{figure}

We show our results at $z=0$ in Fig.~\ref{fig_halo_z0}.
As expected, we can see that $R_\star$ is of order $r_a$. Beyond $R_\star$
the self-interaction force $F_{\rm I}$ shows a fast decrease, following the dark matter
density $\rho$. Below $R_\star$, the soliton profile mainly gives a flat core,
defined by the scale $R_\star$ (mainly set by $r_a$) and its total mass
$M(R_\star)$.
Because of this flat density core, the circular velocity $v=\sqrt{{\cal G}M(r)/r}$
decreases as $r$ in the central region. However, on these small radii baryons are typically
more concentrated than dark matter and can have a non-negligible impact on their
velocity profile.
Although we expect these main characteristics to be robust, i.e. a flat core below $r_a$,
the details of the transition between the NFW and soliton regimes should be studied
with numerical simulations.

We show the density profiles at $z=3$ and $z=6$ in Fig.~\ref{fig_halo_z3-z6}.
We can see that for masses below $10^{10} h^{-1} M_\odot$ the soliton extends
over the whole halo. This could leave a signature on weak lensing at high $z$.
On the other hand, we can expect gas cooling to allow the gas to fall within the potential
wells and to form small galaxies as in the standard $\Lambda$-CDM scenario,
although with slightly different properties that could be investigated with numerical
simulations.

We show the ratio of the solitonic core to the total halo mass in Fig.~\ref{fig_mass}.
In agreement with Figs.~\ref{fig_halo_z0} and \ref{fig_halo_z3-z6}, it decreases
with $M$ and the solitonic core becomes negligible in terms of relative mass
for $M>10^{12} h^{-1} M_\odot$.
To understand the slope of this relation $M_\star/M$, let us simplify the problem and
consider that the solitonic core is embedded in a power-law halo profile, 
instead of the NFW profile,
\be
R_\star < r < R : \;\;\; \rho(r) \propto r^{-\alpha} , \;\;\; 
M(<r) = \left( \frac{r}{R} \right)^{3-\alpha} M .
\ee
Then, approximating the solitonic core $R_\star$ as a constant, in agreement
with Eq.(\ref{eq:Rh-def}) and Fig.~\ref{fig_halo_z0}, and using the hypothesis of 
mass conservation within $R_\star$ as in Eq.(\ref{eq:M-star-M-R-star}), 
$M_\star = M(<R_\star)$, we obtain
\be
\frac{M_\star}{M} =  \left( \frac{R_\star}{R} \right)^{3-\alpha}  \propto M^{(\alpha-3)/3} .
\label{eq:M-star-M-alpha}
\ee
Here we used that halos are defined by a constant density contrast at radius $R$, and hence
$M \propto R^3$.
For an isothermal halo, which corresponds to $\alpha=2$, we obtain
$\frac{M_\star}{M} \propto M^{-1/3}$.
This agrees with the result obtained by \cite{Chavanis:2019faf},
which appeared during completion of this paper, where the author considers thermodynamical
equilibrium of a solitonic core inside an isothermal halo of temperature $T$. 
This can also be understood from the continuity of the Newtonian potential $\Phi$
and the fact that in the isothermal halo the circular velocity $v_c$ is constant and
$\Phi(r) \propto \ln(r)$, so that $\Phi_\star \simeq \Phi(R)$ up to logarithmic corrections,
and $v_c(R_\star) = v_c(R)$.
However, in the central parts of NFW halos we have $\alpha=1$. Then, 
Eq.(\ref{eq:M-star-M-alpha}) gives the steeper slope $\frac{M_\star}{M} \propto M^{-2/3}$,
which roughly agrees with the large mass slope obtained in Fig.~\ref{fig_mass}.
Thus, our hypothesis of local relaxation within the radius $R_\star$ leads to a strong 
dependence of the mass ratio $M_\star/M$ on the slope of the halo profile.
It may happen that the relaxation process is instead a global phenomenon that involves
a redistribution of mass over the full halo extent, up to radius $R$. This more efficient
relaxation could then make $M_\star/M$ independent of the initial halo profile.
It would be interesting to study this point with numerical simulations,
to obtain the slope of the relation $M_\star/M$ and to check the extent of the relaxation process.

\subsection{The cosine model}
\label{sec:cosine-NFW}

We now consider the case of a cosine scalar-field potential, described in
section~\ref{sec:cosine}, which we normalize by
\be
\rho_a = 10^{13} \bar\rho_0 , \;\;\; \rho_b = 10^5 \bar\rho_0 ,
\ee
This implies that $\Phi_{\rm I} < 10^{-7}$ for all densities.
This also ensures that $V_{\rm I}(\phi) \ll \rho$ at all redshifts and the scalar field $\phi$
always behaves as pressureless dark matter at the background level.

\begin{figure*}
\begin{center}
\epsfxsize=8.5 cm \epsfysize=5.5 cm {\epsfbox{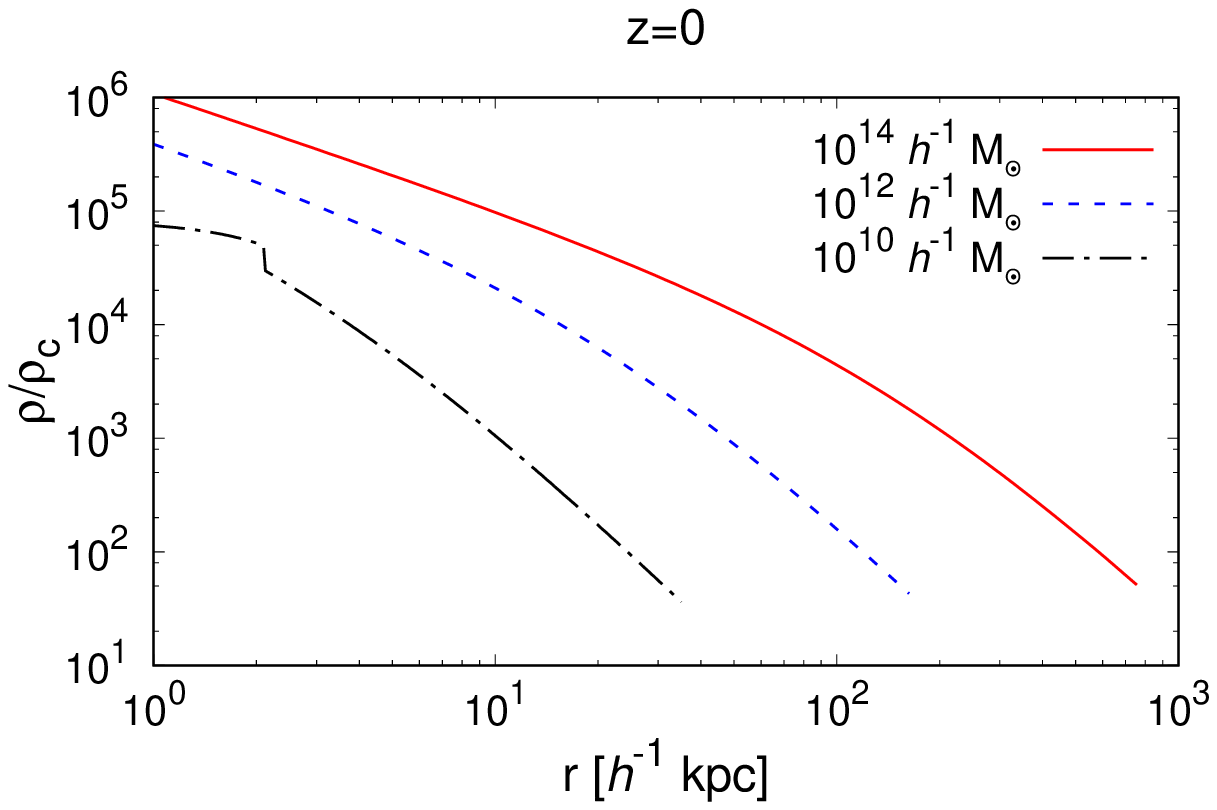}}
\epsfxsize=8.5 cm \epsfysize=5.5 cm {\epsfbox{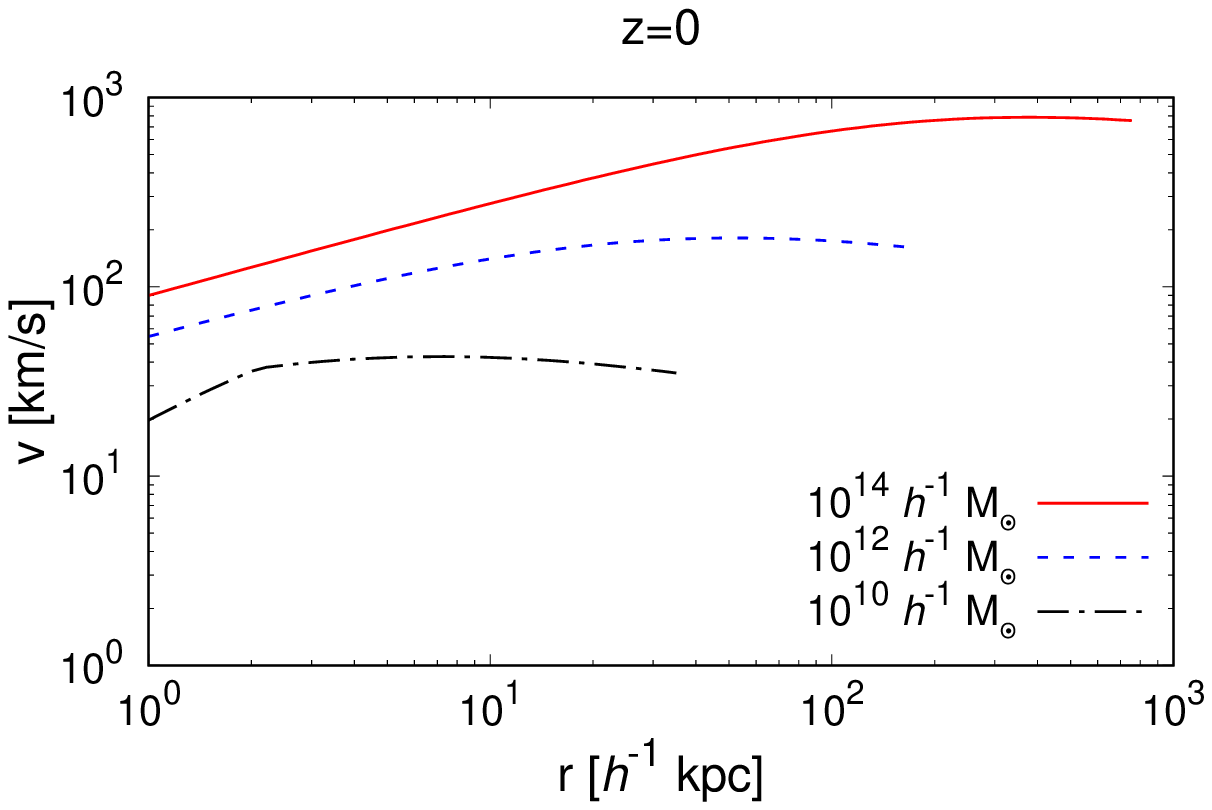}}\\
\epsfxsize=8.5 cm \epsfysize=5.5 cm {\epsfbox{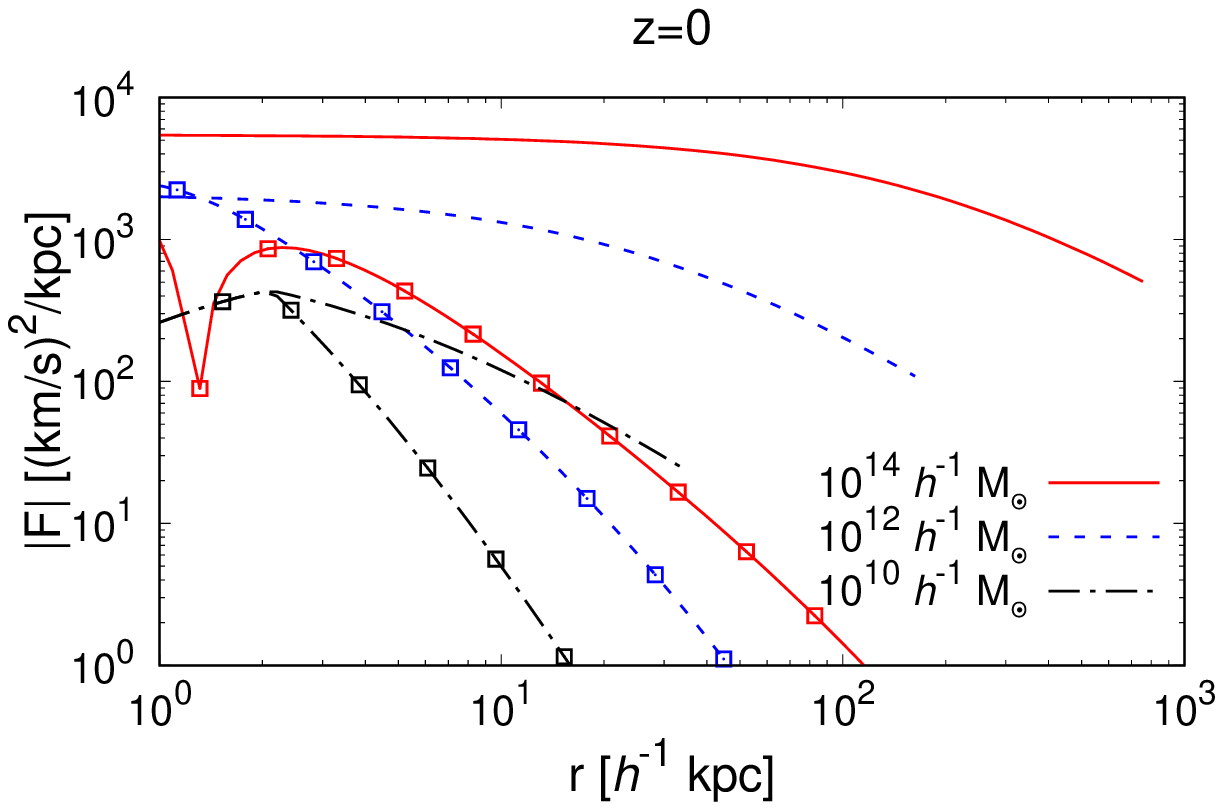}}
\epsfxsize=8.5 cm \epsfysize=5.5 cm {\epsfbox{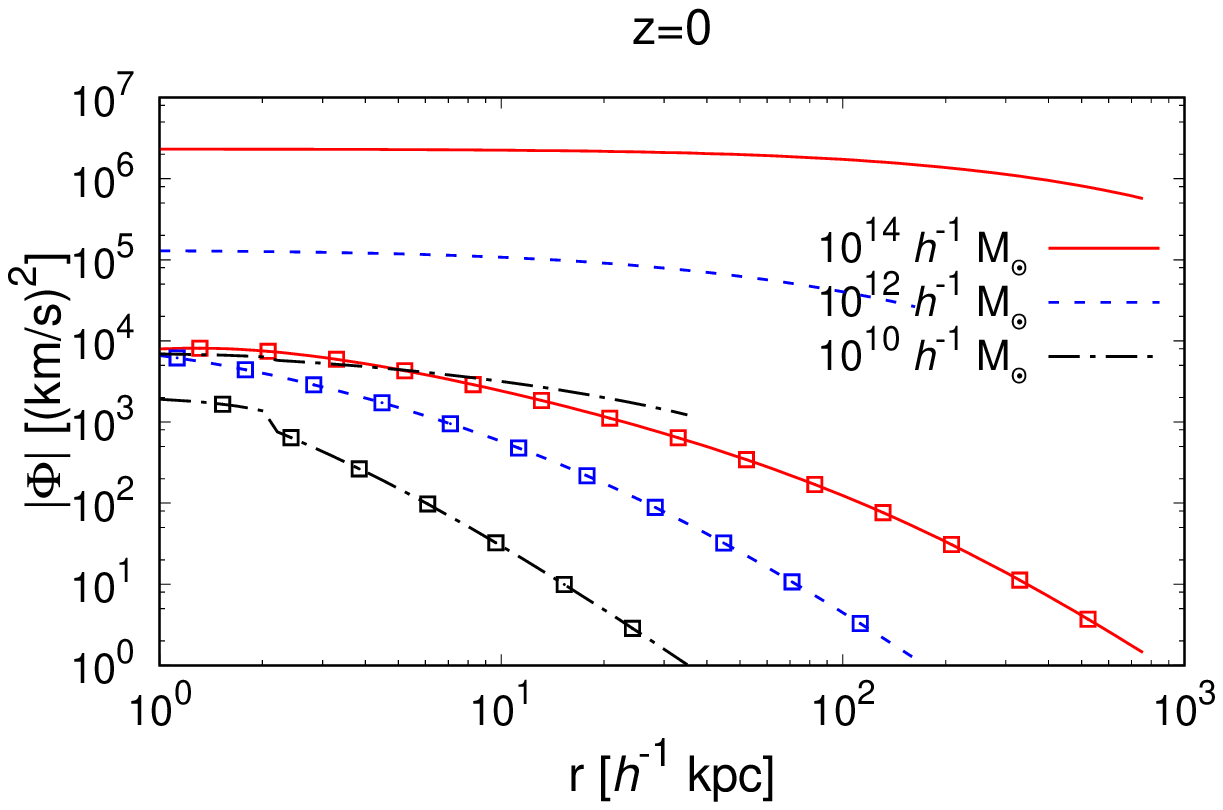}}
\end{center}
\caption{Dark matter halo profiles at redshift $z=0$ for halos of mass
$M=10^{14}$, $10^{12}$, and $10^{10} h^{-1} M_\odot$.
{\it Upper panel left:} dark matter density $\rho$ in units of the critical density $\rho_c$.
{\it Upper right panel:} circular velocity $v$.
{\it Lower left panel:} Newtonian force $-F_{\rm N}$ and dark-matter force $F_{\rm I}$
(lines with squares), in units of ${\rm (km/s)^2/kpc}$.
{\it Lower right panel:} Newtonian potential $\Phi$ and dark-matter potential $\Phi_{\rm I}$
(lines with squares).
}
\label{fig_cos_halo_z0}
\end{figure*}

We follow the same prescription as for the quartic model. On large radii, where
the dark matter force $F_{\rm I}$ is small, we follow the NFW density profile,
while on small radii we follow the soliton profile determined by Eq.(\ref{eq:soliton-cos}).
We again set the transition radius $R_\star$ as that where $F_{\rm I}=-F_{\rm N}$,
and we ensure mass conservation within radius $R$.
Within the soliton regime, we obtain at once from Eq.(\ref{eq:hydrostatic})
\be
r \leq R_\star : \;\;\; \frac{d\Phi_{\rm I}}{dr} = - \frac{d\Phi}{dr} = - \frac{{\cal G} M(r)}{r^2} .
\ee
Therefore, the soliton profile is defined by Eq.(\ref {eq:soliton-cos}) with the two
boundary conditions $d\Phi_{\rm I}/dr=0$ at $r=0$ and
$d\Phi_{\rm I}/dr=-{\cal G} M(R_\star)/R_\star^2$ at $r=R_\star$.

In massive halos, we find that no smooth
soliton profile exists with these boundary conditions.
Then, the NFW profile extends down to the center of the halo and there is no solitonic core.
In the NFW regime, the dark matter force reads from Eq.(\ref{eq:Phi-I-J1}) as
\be
r \geq R_\star: \;\;\; F_{\rm I} = \frac{8\rho_b}{\rho_a r} J_2(\sqrt{\rho/\rho_b})
\frac{1+3r/r_s}{1+r/r_s} .
\ee
Since $\rho \propto r^{-1}$ at small radii in the NFW profile, we have
$F_{\rm I} \sim r^{-3/4} \cos(r^{-1/2})$, which grows to infinity with an infinite number
of changes of sign, whereas the Newtonian force goes to a constant.
On the other hand, the self-interaction potential behaves as
$\Phi_{\rm I} \sim 1 + r^{3/4} \cos(r^{-1/2})$.
Thus, in the massive halos the large and oscillating dark matter force is due to the
fast oscillations of the dark matter potential $\Phi_{\rm I}$, but the magnitude of the latter
remains small and goes to zero at the center.
Therefore, the large value of $F_{\rm I}$ has no strong effect on the dynamics,
as particles move through the potential well at constant energy $E$ with
$E=v^2/2+\Phi+\Phi_{\rm I}$, and $\Phi_{\rm I}$ only induces small and fast
oscillations of their velocity $v$.
Then, we expect to recover the NFW profile down to the center of these massive halos.

\begin{figure}
\begin{center}
\epsfxsize=8.5 cm \epsfysize=5.5 cm {\epsfbox{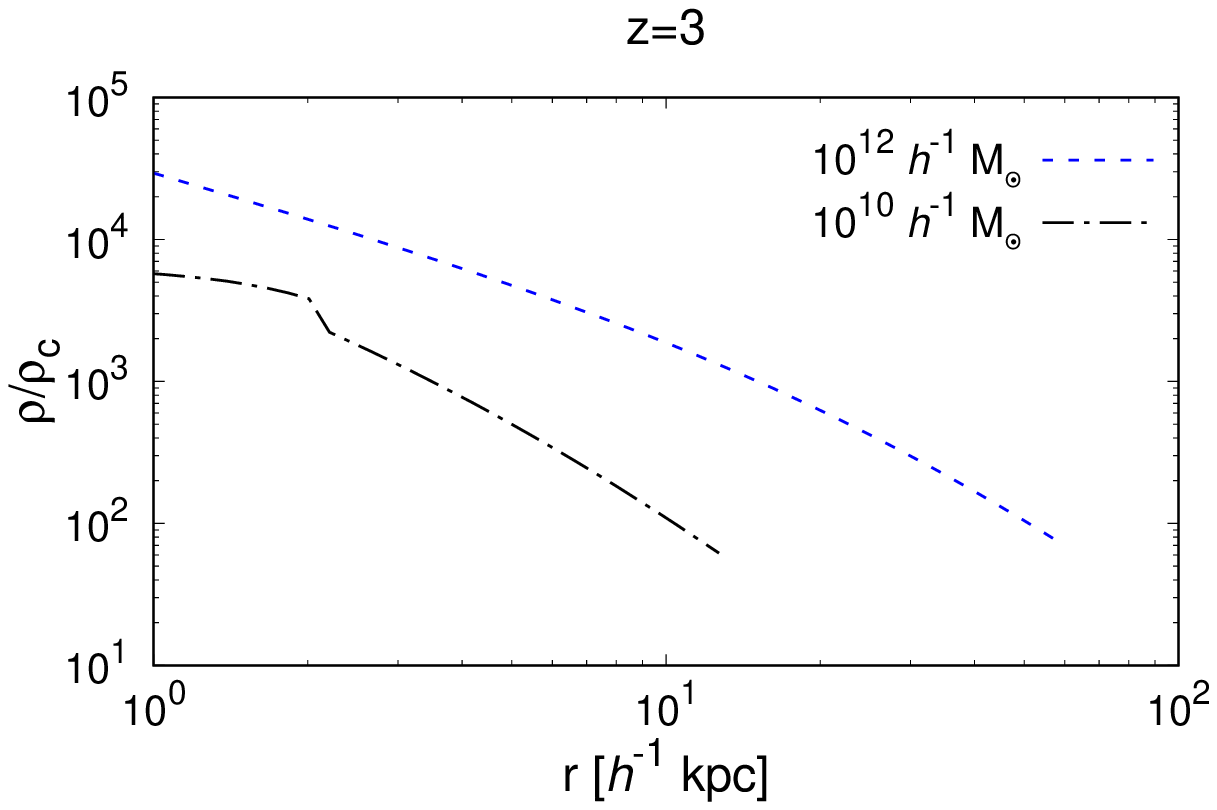}}\\
\epsfxsize=8.5 cm \epsfysize=5.5 cm {\epsfbox{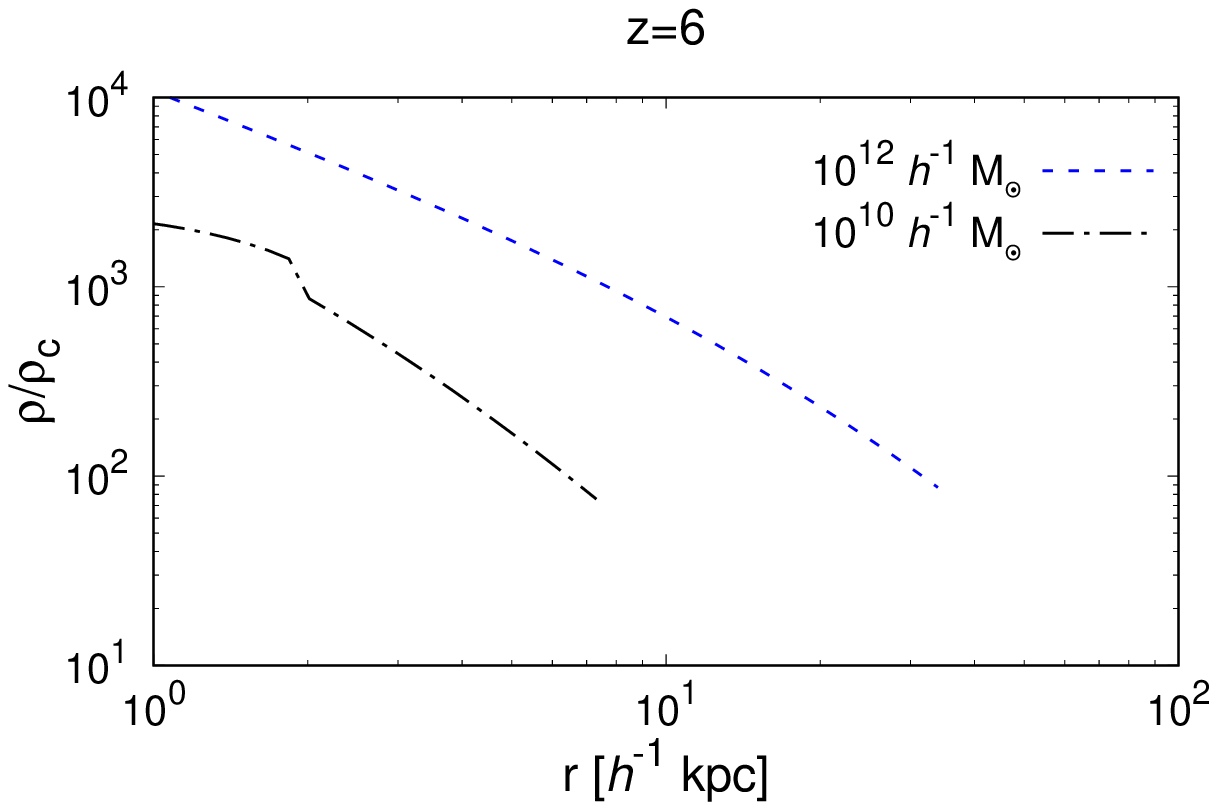}}
\end{center}
\caption{Dark matter halo density profiles at redshifts $z=3$ (upper panel)
and $z=6$ (lower panel).}
\label{fig_cos_halo_z3-z6}
\end{figure}

We show our results at $z=0$ in Fig.~\ref{fig_cos_halo_z0}.
We can see that for massive halos, $M \gtrsim 10^{12} h^{-1} M_{\odot}$, there is
no solitonic core and we follow the NFW density profile down to the center.
This is most clearly seen in the case of the most massive case,
$M = 10^{14} h^{-1} M_{\odot}$. We can see in the lower right panel that the
self-interaction potential converges at small radii to a value that is 100 times smaller
than the Newtonian potential and should have no effect on the dynamics.
On the other hand, the small oscillations of $\Phi_{\rm I}$ around its asymptotic value
give rise to the first change of sign of $F_{\rm I}$ at $r \simeq 1.5 h^{-1} {\rm kpc}$
in the lower left panel.
In contrast, for the low-mass case, $M = 10^{10} h^{-1} M_{\odot}$, we have a solitonic
core as for the quartic model presented in section~\ref {sec:quartic-NFW}.
Indeed, the density remains below the threshold
$\rho_{\max} \simeq 2.6 \times 10^6 \bar\rho_0$ down to the center and the
self-interaction potential is well described by its low-density regime,
$\Phi_{\rm I} \simeq \rho/\rho_a$, which is identical to the quartic model.

\begin{figure}
\begin{center}
\epsfxsize=8.5 cm \epsfysize=5.5 cm {\epsfbox{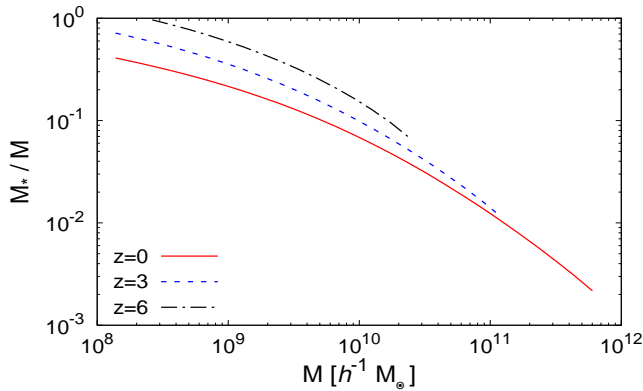}}
\end{center}
\caption{Mass ratio $M_\star/M$ of the solitonic core to the total halo mass
$M$ at redshifts $z=0$, $z=3$ and $z=6$.}
\label{fig_cos_mass}
\end{figure}

We show the density profiles at $z=3$ and $z=6$ in Fig.~\ref{fig_cos_halo_z3-z6}.
We can see that again a solitonic core only develops for the low-mass halos,
$M \lesssim 10^{10} h^{-1} M_\odot$.
For $M = 10^{10} h^{-1} M_\odot$ the soliton core is smaller than for
the quartic model shown in Fig.~\ref{fig_halo_z3-z6}.
However, for smaller masses, $M \lesssim 10^{8} h^{-1} M_\odot$,
the soliton would again extend over the full halo.

We show the ratio of the solitonic core to the total halo mass in Fig.~\ref{fig_cos_mass}.
It decreases with $M$, as for the quartic model, but with a smaller ratio
at fixed mass and redshift. The solitonic core now becomes
negligible in terms of relative mass for $M>10^{11} h^{-1} M_\odot$
and it actually disappears beyond a finite halo mass. This mass threshold
decreases at higher redshifts.

\section{Conclusion}
\label{sec:conclusion}

In this work, we have studied coherent DM scalar models provided by a
self-interaction that is able to impact the structures on galactic scales. 
This effect is produced not only by non-derivative
self-interactions, but also by derivative terms. At low energies, these models also
behave as typical DM but with a deviation provided by the new terms. In particular, the derivative interactions are well motivated for light scalar fields associated with pseudo-Goldstone models. We have shown that the nonrelativistic limit of such scalar models admits an effective description as a hydrodynamic fluid governed not only by quantum pressure but also by a new potential term in the Euler equation. This potential takes into account both the derivative and non-derivative self-interactions, in the regime where they are subleading with respect to the canonical free term.

We have studied in detail models with masses much higher than  $10^{-21} \, {\rm eV}$,
where we can neglect the quantum pressure with respect to the self-interactions on galactic scales. In fuzzy dark matter scenarios, the quantum pressure prevents or delays
the gravitational collapse of small scales below the Jeans length.
This allows one to constrain these models through Lyman-$\alpha$ statistics
\cite{Nori:2018pka},
which probe the matter power spectrum at redshift $z \sim 3$ and scales
of order $1\; h^{-1} {\rm Mpc}$. In these models of fuzzy dark matter, the quantum pressure plays a crucial role. In our case, the repulsive self-interaction plays
a similar role to  the quantum pressure and also leads to a non-negligible Jeans length. 

In the repulsive Landau-Ginzburg models that we consider, whose best example is a canonical massive field with a $(\partial \phi)^4$ interaction, we find that cores of virialized structures of sizes several kpc's exist. In these cases, as the Jean's length is of the same order of magnitude, the
 Lyman-$\alpha$ statistics may also constrain these scenarios and
the amplitude of the self-interactions. This is left for future works.
We also consider the case of a cosine
self-interaction, as an example of bounded nonpolynomial self-interaction. This gives
similar results in low-mass and low-density halos whereas solitonic cores are shown to be
absent in massive halos.
In addition, this helps to lessen cosmological constraints associated with the
BBN and the radiation era, as the upper bound on the potential ensures that
the scalar field energy density is negligible at high redshifts before matter-radiation
equality.

\section{Acknowledgments}

  This work is
supported in part by the EU Horizon 2020 research and innovation
program under the Marie-Sklodowska Grant No. 690575. This article is
based upon work related to the COST Action CA15117 (CANTATA) supported
by COST (European Cooperation in Science and Technology).
The work by J.A.R.C. is partially supported by the MINECO (Spain)
projects FIS2014-52837-P, FIS2016-78859-P(AEI/FEDER, UE),
 and Consolider-Ingenio MULTIDARK CSD2009-00064.

\appendix

\section{Weak gravity}
\label{app:weak-gravity}

\subsection{Einstein-Hilbert action}
\label{app:EH}

Up to quadratic order over the metric potentials $\Phi$ and $\Psi$, the Einstein-Hilbert action
reads
\ba
&& S_{\rm EH} = M_{\rm Pl}^2 \int d^4x \; a^3 \left[ 6 H^2 +3 \dot{H}
+ 3 H^2 \Phi-9H^2\Psi \right.  \nonumber \\
&& -6\dot{H}\Psi - \frac{9}{2} H^2 \Phi^2 - 9 H^2 \Phi \Psi + \frac{9}{2} H^2 \Psi^2
+ 3 \dot{H} \Psi^2 \nonumber \\
&& \left. - 6 H \Phi \dot\Psi - 3 \dot\Psi^2 + a^{-2} \left( -2 \nabla\Psi\cdot\nabla\Phi+
(\nabla\Psi)^2 \right) \right] ,
\label{eq:S-EH-2}
\ea
while the Einstein tensor $G^\mu_\nu$ writes, up to linear order over $\Phi$ and $\Psi$,
\be
G^0_0 = - 3 H^2 + 6 H^2 \Phi + 6 H \dot\Psi - 2 a^{-2} \nabla^2 \Psi ,
\label{eq:G00}
\ee
\be
G^0_i = - 2 \partial_i ( H \Phi + \dot\Psi ) ,
\label{eq:G0i}
\ee
\ba
&& G^i_j = \delta^i_j \left[ - 3 H^2 - 2 \dot{H} + 2 \ddot\Psi + 6 H \dot\Psi + 2 H \dot\Phi
\right. \nonumber \\
&& \left. + (6H^2+4\dot{H}) \Phi + a^{-2} \nabla^2 (\Phi-\Psi)  \right]
+ a^{-2} \partial_i\partial_j (\Psi-\Phi) . \nonumber \\
&&
\label{eq:Gij}
\ea

\subsection{Complex scalar field $\psi$}
\label{app:complex-scalar}

Substituting the expression (\ref{eq:psi-def}) into the scalar-field action (\ref{eq:S-phi-2})
gives
\ba
&& S_\phi = \int d^4x \; a^3 \biggl \lbrace e^{-2imt} \left[ \frac{1-\Phi-3\Psi}{4m}
(\dot\psi^2 - 2 i m \dot\psi \psi) \right. \nonumber \\
&& \left. - \frac{1+\Phi-\Psi}{4 m a^2} (\nabla\psi)^2
- \frac{1-3\Psi}{2} m \psi^2 \right]
+ e^{2imt} [ {\rm c.c.} ] \nonumber \\
&& + \frac{1-\Phi-3\Psi}{2m} ( \dot\psi \dot\psi^\star + i m \dot\psi \psi^\star - i m \psi \dot\psi^\star)
\nonumber \\
&& - \frac{1+\Phi-\Psi}{2 m a^2} (\nabla\psi)\cdot(\nabla\psi^\star) - m \Phi \psi \psi^\star
+ K_{\rm I}(X) - V_{\rm I}(\phi) \biggl \rbrace  \nonumber \\
&&
\label{eq:S-psi-full-2}
\ea
where the bracket associated with $e^{2imt}$ is the complex conjugate of the first bracket
associated with $e^{-2imt}$.

\subsection{Fluid picture $(\rho,S)$}
\label{app:fluid}

Substituting the expression (\ref{eq:rho-S-def}) into the action (\ref{eq:S-psi-full-2}) gives
\ba
&& S_\phi = \int d^4x \; a^3 \biggl \lbrace e^{-2imt+2iS} \biggl[ \frac{1-\Phi-3\Psi}{4m^2}
\biggl( \frac{\dot\rho^2}{4\rho} - \rho \dot{S}^2 \nonumber \\
&& +i \dot\rho \dot{S} - i m \dot\rho + 2 m \rho \dot{S} \biggl) - \frac{1+\Phi-\Psi}{4 m^2 a^2}
\biggl( \frac{(\nabla\rho)^2}{4\rho} - \rho (\nabla S)^2 \nonumber \\
&& + i \nabla\rho \cdot \nabla S \biggl) - \frac{1-3\Psi}{2} \rho \biggl]
+ e^{2imt-2iS} [ {\rm c.c.} ] \nonumber \\
&& + \frac{1-\Phi-3\Psi}{2m^2} \biggl( \frac{\dot\rho^2}{4\rho} + \rho \dot{S}^2 - 2 m \rho \dot{S}
\biggl) \nonumber \\
&& - \frac{1+\Phi-\Psi}{2 m^2 a^2} \biggl( \frac{(\nabla\rho)^2}{4\rho} + \rho (\nabla S)^2 \biggl)
- \rho \Phi + K_{\rm I}(\phi) - V_{\rm I}(\phi) \biggl \rbrace  \nonumber \\
&&
\label{eq:S-rho-full-2}
\ea
where the bracket associated with $e^{2imt-2iS}$ is the complex conjugate of the first bracket
associated with $e^{-2imt+2iS}$.

\vspace{-.3cm}

\bibliography{ref1}

\begin{thebibliography}{74}
\expandafter\ifx\csname natexlab\endcsname\relax\def\natexlab#1{#1}\fi
\expandafter\ifx\csname bibnamefont\endcsname\relax
  \def\bibnamefont#1{#1}\fi
\expandafter\ifx\csname bibfnamefont\endcsname\relax
  \def\bibfnamefont#1{#1}\fi
\expandafter\ifx\csname citenamefont\endcsname\relax
  \def\citenamefont#1{#1}\fi
\expandafter\ifx\csname url\endcsname\relax
  \def\url#1{\texttt{#1}}\fi
\expandafter\ifx\csname urlprefix\endcsname\relax\def\urlprefix{URL }\fi
\providecommand{\bibinfo}[2]{#2}
\providecommand{\eprint}[2][]{\url{#2}}

\bibitem[{\citenamefont{Ostriker and Steinhardt}(2003)}]{Ostriker:2003qj}
\bibinfo{author}{\bibfnamefont{J.~P.} \bibnamefont{Ostriker}} \bibnamefont{and}
  \bibinfo{author}{\bibfnamefont{P.~J.} \bibnamefont{Steinhardt}},
  \bibinfo{journal}{Science} \textbf{\bibinfo{volume}{300}},
  \bibinfo{pages}{1909} (\bibinfo{year}{2003}), \eprint{astro-ph/0306402}.

\bibitem[{\citenamefont{Weinberg et~al.}(2015)\citenamefont{Weinberg, Bullock,
  Governato, Kuzio~de Naray, and Peter}}]{Weinberg:2013aya}
\bibinfo{author}{\bibfnamefont{D.~H.} \bibnamefont{Weinberg}},
  \bibinfo{author}{\bibfnamefont{J.~S.} \bibnamefont{Bullock}},
  \bibinfo{author}{\bibfnamefont{F.}~\bibnamefont{Governato}},
  \bibinfo{author}{\bibfnamefont{R.}~\bibnamefont{Kuzio~de Naray}},
  \bibnamefont{and} \bibinfo{author}{\bibfnamefont{A.~H.~G.}
  \bibnamefont{Peter}}, \bibinfo{journal}{Proc. Nat. Acad. Sci.}
  \textbf{\bibinfo{volume}{112}}, \bibinfo{pages}{12249}
  (\bibinfo{year}{2015}), \bibinfo{note}{[Proc. Nat. Acad.
  Sci.112,2249(2015)]}, \eprint{1306.0913}.

\bibitem[{\citenamefont{Pontzen and Governato}(2014)}]{Pontzen:2014lma}
\bibinfo{author}{\bibfnamefont{A.}~\bibnamefont{Pontzen}} \bibnamefont{and}
  \bibinfo{author}{\bibfnamefont{F.}~\bibnamefont{Governato}},
  \bibinfo{journal}{Nature} \textbf{\bibinfo{volume}{506}},
  \bibinfo{pages}{171} (\bibinfo{year}{2014}), \eprint{1402.1764}.

\bibitem[{\citenamefont{Boylan-Kolchin
  et~al.}(2011)\citenamefont{Boylan-Kolchin, Bullock, and
  Kaplinghat}}]{BoylanKolchin:2011de}
\bibinfo{author}{\bibfnamefont{M.}~\bibnamefont{Boylan-Kolchin}},
  \bibinfo{author}{\bibfnamefont{J.~S.} \bibnamefont{Bullock}},
  \bibnamefont{and}
  \bibinfo{author}{\bibfnamefont{M.}~\bibnamefont{Kaplinghat}},
  \bibinfo{journal}{Mon. Not. Roy. Astron. Soc.}
  \textbf{\bibinfo{volume}{415}}, \bibinfo{pages}{L40} (\bibinfo{year}{2011}),
  \eprint{1103.0007}.

\bibitem[{\citenamefont{Moore et~al.}(1999)\citenamefont{Moore, Ghigna,
  Governato, Lake, Quinn, Stadel, and Tozzi}}]{Moore:1999nt}
\bibinfo{author}{\bibfnamefont{B.}~\bibnamefont{Moore}},
  \bibinfo{author}{\bibfnamefont{S.}~\bibnamefont{Ghigna}},
  \bibinfo{author}{\bibfnamefont{F.}~\bibnamefont{Governato}},
  \bibinfo{author}{\bibfnamefont{G.}~\bibnamefont{Lake}},
  \bibinfo{author}{\bibfnamefont{T.~R.} \bibnamefont{Quinn}},
  \bibinfo{author}{\bibfnamefont{J.}~\bibnamefont{Stadel}}, \bibnamefont{and}
  \bibinfo{author}{\bibfnamefont{P.}~\bibnamefont{Tozzi}},
  \bibinfo{journal}{Astrophys. J.} \textbf{\bibinfo{volume}{524}},
  \bibinfo{pages}{L19} (\bibinfo{year}{1999}), \eprint{astro-ph/9907411}.

\bibitem[{\citenamefont{de~Blok}(2010)}]{deBlok:2009sp}
\bibinfo{author}{\bibfnamefont{W.~J.~G.} \bibnamefont{de~Blok}},
  \bibinfo{journal}{Adv. Astron.} \textbf{\bibinfo{volume}{2010}},
  \bibinfo{pages}{789293} (\bibinfo{year}{2010}), \eprint{0910.3538}.

\bibitem[{\citenamefont{Hu et~al.}(2000)\citenamefont{Hu, Barkana, and
  Gruzinov}}]{Hu:2000ke}
\bibinfo{author}{\bibfnamefont{W.}~\bibnamefont{Hu}},
  \bibinfo{author}{\bibfnamefont{R.}~\bibnamefont{Barkana}}, \bibnamefont{and}
  \bibinfo{author}{\bibfnamefont{A.}~\bibnamefont{Gruzinov}},
  \bibinfo{journal}{Phys. Rev. Lett.} \textbf{\bibinfo{volume}{85}},
  \bibinfo{pages}{1158} (\bibinfo{year}{2000}), \eprint{astro-ph/0003365}.

\bibitem[{\citenamefont{Turner}(1983)}]{Turner:1983he}
\bibinfo{author}{\bibfnamefont{M.~S.} \bibnamefont{Turner}},
  \bibinfo{journal}{Phys. Rev.} \textbf{\bibinfo{volume}{D28}},
  \bibinfo{pages}{1243} (\bibinfo{year}{1983}).

\bibitem[{\citenamefont{Sahni and Wang}(2000)}]{Sahni:1999qe}
\bibinfo{author}{\bibfnamefont{V.}~\bibnamefont{Sahni}} \bibnamefont{and}
  \bibinfo{author}{\bibfnamefont{L.-M.} \bibnamefont{Wang}},
  \bibinfo{journal}{Phys. Rev.} \textbf{\bibinfo{volume}{D62}},
  \bibinfo{pages}{103517} (\bibinfo{year}{2000}), \eprint{astro-ph/9910097}.

\bibitem[{\citenamefont{Johnson and Kamionkowski}(2008)}]{Johnson:2008se}
\bibinfo{author}{\bibfnamefont{M.~C.} \bibnamefont{Johnson}} \bibnamefont{and}
  \bibinfo{author}{\bibfnamefont{M.}~\bibnamefont{Kamionkowski}},
  \bibinfo{journal}{Phys. Rev.} \textbf{\bibinfo{volume}{D78}},
  \bibinfo{pages}{063010} (\bibinfo{year}{2008}), \eprint{0805.1748}.

\bibitem[{\citenamefont{Cembranos et~al.}(2012)\citenamefont{Cembranos,
  Hallabrin, Maroto, and Jare{\~n}o}}]{Cembranos:2012kk}
\bibinfo{author}{\bibfnamefont{J.~A.~R.} \bibnamefont{Cembranos}},
  \bibinfo{author}{\bibfnamefont{C.}~\bibnamefont{Hallabrin}},
  \bibinfo{author}{\bibfnamefont{A.~L.} \bibnamefont{Maroto}},
  \bibnamefont{and} \bibinfo{author}{\bibfnamefont{S.~J.~N.}
  \bibnamefont{Jare{\~n}o}}, \bibinfo{journal}{Phys. Rev.}
  \textbf{\bibinfo{volume}{D86}}, \bibinfo{pages}{021301}
  (\bibinfo{year}{2012}), \eprint{1203.6221}.

\bibitem[{\citenamefont{Cembranos et~al.}(2013)\citenamefont{Cembranos, Maroto,
  and N{\'u}{\~n}ez~Jare{\~n}o}}]{Cembranos:2012ng}
\bibinfo{author}{\bibfnamefont{J.~A.~R.} \bibnamefont{Cembranos}},
  \bibinfo{author}{\bibfnamefont{A.~L.} \bibnamefont{Maroto}},
  \bibnamefont{and} \bibinfo{author}{\bibfnamefont{S.~J.}
  \bibnamefont{N{\'u}{\~n}ez~Jare{\~n}o}}, \bibinfo{journal}{Phys. Rev.}
  \textbf{\bibinfo{volume}{D87}}, \bibinfo{pages}{043523}
  (\bibinfo{year}{2013}), \eprint{1212.3201}.

\bibitem[{\citenamefont{Cembranos et~al.}(2014)\citenamefont{Cembranos, Maroto,
  and N{\'u}{\~n}ez~Jare{\~n}o}}]{Cembranos:2013cba}
\bibinfo{author}{\bibfnamefont{J.~A.~R.} \bibnamefont{Cembranos}},
  \bibinfo{author}{\bibfnamefont{A.~L.} \bibnamefont{Maroto}},
  \bibnamefont{and} \bibinfo{author}{\bibfnamefont{S.~J.}
  \bibnamefont{N{\'u}{\~n}ez~Jare{\~n}o}}, \bibinfo{journal}{JCAP}
  \textbf{\bibinfo{volume}{1403}}, \bibinfo{pages}{042} (\bibinfo{year}{2014}),
  \eprint{1311.1402}.

\bibitem[{\citenamefont{Hwang and Noh}(2009)}]{Hwang:2009js}
\bibinfo{author}{\bibfnamefont{J.-c.} \bibnamefont{Hwang}} \bibnamefont{and}
  \bibinfo{author}{\bibfnamefont{H.}~\bibnamefont{Noh}},
  \bibinfo{journal}{Phys. Lett.} \textbf{\bibinfo{volume}{B680}},
  \bibinfo{pages}{1} (\bibinfo{year}{2009}), \eprint{0902.4738}.

\bibitem[{\citenamefont{Park et~al.}(2012)\citenamefont{Park, Hwang, and
  Noh}}]{Park:2012ru}
\bibinfo{author}{\bibfnamefont{C.-G.} \bibnamefont{Park}},
  \bibinfo{author}{\bibfnamefont{J.-c.} \bibnamefont{Hwang}}, \bibnamefont{and}
  \bibinfo{author}{\bibfnamefont{H.}~\bibnamefont{Noh}},
  \bibinfo{journal}{Phys. Rev.} \textbf{\bibinfo{volume}{D86}},
  \bibinfo{pages}{083535} (\bibinfo{year}{2012}), \eprint{1207.3124}.

\bibitem[{\citenamefont{Hlozek et~al.}(2015)\citenamefont{Hlozek, Grin, Marsh,
  and Ferreira}}]{Hlozek:2014lca}
\bibinfo{author}{\bibfnamefont{R.}~\bibnamefont{Hlozek}},
  \bibinfo{author}{\bibfnamefont{D.}~\bibnamefont{Grin}},
  \bibinfo{author}{\bibfnamefont{D.~J.~E.} \bibnamefont{Marsh}},
  \bibnamefont{and} \bibinfo{author}{\bibfnamefont{P.~G.}
  \bibnamefont{Ferreira}}, \bibinfo{journal}{Phys. Rev.}
  \textbf{\bibinfo{volume}{D91}}, \bibinfo{pages}{103512}
  (\bibinfo{year}{2015}), \eprint{1410.2896}.

\bibitem[{\citenamefont{Cembranos et~al.}(2016)\citenamefont{Cembranos, Maroto,
  and N{\'u}{\~n}ez~Jare{\~n}o}}]{Cembranos:2015oya}
\bibinfo{author}{\bibfnamefont{J.~A.~R.} \bibnamefont{Cembranos}},
  \bibinfo{author}{\bibfnamefont{A.~L.} \bibnamefont{Maroto}},
  \bibnamefont{and} \bibinfo{author}{\bibfnamefont{S.~J.}
  \bibnamefont{N{\'u}{\~n}ez~Jare{\~n}o}}, \bibinfo{journal}{JHEP}
  \textbf{\bibinfo{volume}{03}}, \bibinfo{pages}{013} (\bibinfo{year}{2016}),
  \eprint{1509.08819}.

\bibitem[{\citenamefont{Cembranos et~al.}(2017)\citenamefont{Cembranos, Maroto,
  and N{\'u}{\~n}ez~Jare{\~n}o}}]{Cembranos:2016ugq}
\bibinfo{author}{\bibfnamefont{J.~A.~R.} \bibnamefont{Cembranos}},
  \bibinfo{author}{\bibfnamefont{A.~L.} \bibnamefont{Maroto}},
  \bibnamefont{and} \bibinfo{author}{\bibfnamefont{S.~J.}
  \bibnamefont{N{\'u}{\~n}ez~Jare{\~n}o}}, \bibinfo{journal}{JHEP}
  \textbf{\bibinfo{volume}{02}}, \bibinfo{pages}{064} (\bibinfo{year}{2017}),
  \eprint{1611.03793}.

\bibitem[{\citenamefont{Schive et~al.}(2014)\citenamefont{Schive, Chiueh, and
  Broadhurst}}]{Schive:2014dra}
\bibinfo{author}{\bibfnamefont{H.-Y.} \bibnamefont{Schive}},
  \bibinfo{author}{\bibfnamefont{T.}~\bibnamefont{Chiueh}}, \bibnamefont{and}
  \bibinfo{author}{\bibfnamefont{T.}~\bibnamefont{Broadhurst}},
  \bibinfo{journal}{Nature Phys.} \textbf{\bibinfo{volume}{10}},
  \bibinfo{pages}{496} (\bibinfo{year}{2014}), \eprint{1406.6586}.

\bibitem[{\citenamefont{Peccei and Quinn}(1977)}]{Peccei:1977hh}
\bibinfo{author}{\bibfnamefont{R.~D.} \bibnamefont{Peccei}} \bibnamefont{and}
  \bibinfo{author}{\bibfnamefont{H.~R.} \bibnamefont{Quinn}},
  \bibinfo{journal}{Phys. Rev. Lett.} \textbf{\bibinfo{volume}{38}},
  \bibinfo{pages}{1440} (\bibinfo{year}{1977}).

\bibitem[{\citenamefont{Wilczek}(1978)}]{Wilczek:1977pj}
\bibinfo{author}{\bibfnamefont{F.}~\bibnamefont{Wilczek}},
  \bibinfo{journal}{Phys. Rev. Lett.} \textbf{\bibinfo{volume}{40}},
  \bibinfo{pages}{279} (\bibinfo{year}{1978}).

\bibitem[{\citenamefont{Weinberg}(1978)}]{Weinberg:1977ma}
\bibinfo{author}{\bibfnamefont{S.}~\bibnamefont{Weinberg}},
  \bibinfo{journal}{Phys. Rev. Lett.} \textbf{\bibinfo{volume}{40}},
  \bibinfo{pages}{223} (\bibinfo{year}{1978}).

\bibitem[{\citenamefont{Marsh}(2016)}]{Marsh:2015xka}
\bibinfo{author}{\bibfnamefont{D.~J.~E.} \bibnamefont{Marsh}},
  \bibinfo{journal}{Phys. Rept.} \textbf{\bibinfo{volume}{643}},
  \bibinfo{pages}{1} (\bibinfo{year}{2016}), \eprint{1510.07633}.

\bibitem[{\citenamefont{Hui et~al.}(2017)\citenamefont{Hui, Ostriker, Tremaine,
  and Witten}}]{Hui:2016ltb}
\bibinfo{author}{\bibfnamefont{L.}~\bibnamefont{Hui}},
  \bibinfo{author}{\bibfnamefont{J.~P.} \bibnamefont{Ostriker}},
  \bibinfo{author}{\bibfnamefont{S.}~\bibnamefont{Tremaine}}, \bibnamefont{and}
  \bibinfo{author}{\bibfnamefont{E.}~\bibnamefont{Witten}},
  \bibinfo{journal}{Phys. Rev.} \textbf{\bibinfo{volume}{D95}},
  \bibinfo{pages}{043541} (\bibinfo{year}{2017}), \eprint{1610.08297}.

\bibitem[{\citenamefont{Khlopov et~al.}(1985)\citenamefont{Khlopov, Malomed,
  and Zeldovich}}]{Khlopov:1985jw}
\bibinfo{author}{\bibfnamefont{M.}~\bibnamefont{Khlopov}},
  \bibinfo{author}{\bibfnamefont{B.~A.} \bibnamefont{Malomed}},
  \bibnamefont{and} \bibinfo{author}{\bibfnamefont{I.~B.}
  \bibnamefont{Zeldovich}}, \bibinfo{journal}{Mon. Not. Roy. Astron. Soc.}
  \textbf{\bibinfo{volume}{215}}, \bibinfo{pages}{575} (\bibinfo{year}{1985}).

\bibitem[{\citenamefont{Sarkar et~al.}(2016)\citenamefont{Sarkar, Mondal, Das,
  Sethi, Bharadwaj, and Marsh}}]{Sarkar:2015dib}
\bibinfo{author}{\bibfnamefont{A.}~\bibnamefont{Sarkar}},
  \bibinfo{author}{\bibfnamefont{R.}~\bibnamefont{Mondal}},
  \bibinfo{author}{\bibfnamefont{S.}~\bibnamefont{Das}},
  \bibinfo{author}{\bibfnamefont{S.}~\bibnamefont{Sethi}},
  \bibinfo{author}{\bibfnamefont{S.}~\bibnamefont{Bharadwaj}},
  \bibnamefont{and} \bibinfo{author}{\bibfnamefont{D.~J.~E.}
  \bibnamefont{Marsh}}, \bibinfo{journal}{JCAP}
  \textbf{\bibinfo{volume}{1604}}, \bibinfo{pages}{012} (\bibinfo{year}{2016}),
  \eprint{1512.03325}.

\bibitem[{\citenamefont{Kobayashi et~al.}(2017)\citenamefont{Kobayashi, Murgia,
  De~Simone, Ir{\v s}i{\v c}, and Viel}}]{Kobayashi:2017jcf}
\bibinfo{author}{\bibfnamefont{T.}~\bibnamefont{Kobayashi}},
  \bibinfo{author}{\bibfnamefont{R.}~\bibnamefont{Murgia}},
  \bibinfo{author}{\bibfnamefont{A.}~\bibnamefont{De~Simone}},
  \bibinfo{author}{\bibfnamefont{V.}~\bibnamefont{Ir{\v s}i{\v c}}},
  \bibnamefont{and} \bibinfo{author}{\bibfnamefont{M.}~\bibnamefont{Viel}},
  \bibinfo{journal}{Phys. Rev.} \textbf{\bibinfo{volume}{D96}},
  \bibinfo{pages}{123514} (\bibinfo{year}{2017}), \eprint{1708.00015}.

\bibitem[{\citenamefont{Abel et~al.}(2017)}]{Abel:2017rtm}
\bibinfo{author}{\bibfnamefont{C.}~\bibnamefont{Abel}} \bibnamefont{et~al.},
  \bibinfo{journal}{Phys. Rev.} \textbf{\bibinfo{volume}{X7}},
  \bibinfo{pages}{041034} (\bibinfo{year}{2017}), \eprint{1708.06367}.

\bibitem[{\citenamefont{Banik et~al.}(2017)\citenamefont{Banik, Christopherson,
  Sikivie, and Todarello}}]{Banik:2017ygz}
\bibinfo{author}{\bibfnamefont{N.}~\bibnamefont{Banik}},
  \bibinfo{author}{\bibfnamefont{A.~J.} \bibnamefont{Christopherson}},
  \bibinfo{author}{\bibfnamefont{P.}~\bibnamefont{Sikivie}}, \bibnamefont{and}
  \bibinfo{author}{\bibfnamefont{E.~M.} \bibnamefont{Todarello}},
  \bibinfo{journal}{Phys. Rev.} \textbf{\bibinfo{volume}{D95}},
  \bibinfo{pages}{043542} (\bibinfo{year}{2017}), \eprint{1701.04573}.

\bibitem[{\citenamefont{Hirano et~al.}(2018)\citenamefont{Hirano, Sullivan, and
  Bromm}}]{Hirano:2017bnu}
\bibinfo{author}{\bibfnamefont{S.}~\bibnamefont{Hirano}},
  \bibinfo{author}{\bibfnamefont{J.~M.} \bibnamefont{Sullivan}},
  \bibnamefont{and} \bibinfo{author}{\bibfnamefont{V.}~\bibnamefont{Bromm}},
  \bibinfo{journal}{Mon. Not. Roy. Astron. Soc.}
  \textbf{\bibinfo{volume}{473}}, \bibinfo{pages}{L6} (\bibinfo{year}{2018}),
  \eprint{1706.00435}.

\bibitem[{\citenamefont{Conlon et~al.}(2018)\citenamefont{Conlon, Day,
  Jennings, Krippendorf, and Muia}}]{Conlon:2017ofb}
\bibinfo{author}{\bibfnamefont{J.~P.} \bibnamefont{Conlon}},
  \bibinfo{author}{\bibfnamefont{F.}~\bibnamefont{Day}},
  \bibinfo{author}{\bibfnamefont{N.}~\bibnamefont{Jennings}},
  \bibinfo{author}{\bibfnamefont{S.}~\bibnamefont{Krippendorf}},
  \bibnamefont{and} \bibinfo{author}{\bibfnamefont{F.}~\bibnamefont{Muia}},
  \bibinfo{journal}{Mon. Not. Roy. Astron. Soc.}
  \textbf{\bibinfo{volume}{473}}, \bibinfo{pages}{4932} (\bibinfo{year}{2018}),
  \eprint{1707.00176}.

\bibitem[{\citenamefont{Brito et~al.}(2017{\natexlab{a}})\citenamefont{Brito,
  Ghosh, Barausse, Berti, Cardoso, Dvorkin, Klein, and Pani}}]{Brito:2017wnc}
\bibinfo{author}{\bibfnamefont{R.}~\bibnamefont{Brito}},
  \bibinfo{author}{\bibfnamefont{S.}~\bibnamefont{Ghosh}},
  \bibinfo{author}{\bibfnamefont{E.}~\bibnamefont{Barausse}},
  \bibinfo{author}{\bibfnamefont{E.}~\bibnamefont{Berti}},
  \bibinfo{author}{\bibfnamefont{V.}~\bibnamefont{Cardoso}},
  \bibinfo{author}{\bibfnamefont{I.}~\bibnamefont{Dvorkin}},
  \bibinfo{author}{\bibfnamefont{A.}~\bibnamefont{Klein}}, \bibnamefont{and}
  \bibinfo{author}{\bibfnamefont{P.}~\bibnamefont{Pani}},
  \bibinfo{journal}{Phys. Rev. Lett.} \textbf{\bibinfo{volume}{119}},
  \bibinfo{pages}{131101} (\bibinfo{year}{2017}{\natexlab{a}}),
  \eprint{1706.05097}.

\bibitem[{\citenamefont{Brito et~al.}(2017{\natexlab{b}})\citenamefont{Brito,
  Ghosh, Barausse, Berti, Cardoso, Dvorkin, Klein, and Pani}}]{Brito:2017zvb}
\bibinfo{author}{\bibfnamefont{R.}~\bibnamefont{Brito}},
  \bibinfo{author}{\bibfnamefont{S.}~\bibnamefont{Ghosh}},
  \bibinfo{author}{\bibfnamefont{E.}~\bibnamefont{Barausse}},
  \bibinfo{author}{\bibfnamefont{E.}~\bibnamefont{Berti}},
  \bibinfo{author}{\bibfnamefont{V.}~\bibnamefont{Cardoso}},
  \bibinfo{author}{\bibfnamefont{I.}~\bibnamefont{Dvorkin}},
  \bibinfo{author}{\bibfnamefont{A.}~\bibnamefont{Klein}}, \bibnamefont{and}
  \bibinfo{author}{\bibfnamefont{P.}~\bibnamefont{Pani}},
  \bibinfo{journal}{Phys. Rev.} \textbf{\bibinfo{volume}{D96}},
  \bibinfo{pages}{064050} (\bibinfo{year}{2017}{\natexlab{b}}),
  \eprint{1706.06311}.

\bibitem[{\citenamefont{Sarkar et~al.}(2017)\citenamefont{Sarkar, Sethi, and
  Das}}]{Sarkar:2017vls}
\bibinfo{author}{\bibfnamefont{A.}~\bibnamefont{Sarkar}},
  \bibinfo{author}{\bibfnamefont{S.~K.} \bibnamefont{Sethi}}, \bibnamefont{and}
  \bibinfo{author}{\bibfnamefont{S.}~\bibnamefont{Das}},
  \bibinfo{journal}{JCAP} \textbf{\bibinfo{volume}{1707}}, \bibinfo{pages}{012}
  (\bibinfo{year}{2017}), \eprint{1701.07273}.

\bibitem[{\citenamefont{Diacoumis and Wong}(2017)}]{Diacoumis:2017hff}
\bibinfo{author}{\bibfnamefont{J.~A.~D.} \bibnamefont{Diacoumis}}
  \bibnamefont{and} \bibinfo{author}{\bibfnamefont{Y.~Y.~Y.}
  \bibnamefont{Wong}}, \bibinfo{journal}{JCAP} \textbf{\bibinfo{volume}{1709}},
  \bibinfo{pages}{011} (\bibinfo{year}{2017}), \eprint{1707.07050}.

\bibitem[{\citenamefont{Broadhurst et~al.}(2018)\citenamefont{Broadhurst, Luu,
  and Tye}}]{Broadhurst:2018fei}
\bibinfo{author}{\bibfnamefont{T.}~\bibnamefont{Broadhurst}},
  \bibinfo{author}{\bibfnamefont{H.~N.} \bibnamefont{Luu}}, \bibnamefont{and}
  \bibinfo{author}{\bibfnamefont{S.~H.~H.} \bibnamefont{Tye}}
  (\bibinfo{year}{2018}), \eprint{1811.03771}.

\bibitem[{\citenamefont{Chavanis}(2011)}]{Chavanis:2011zi}
\bibinfo{author}{\bibfnamefont{P.-H.} \bibnamefont{Chavanis}},
  \bibinfo{journal}{Phys. Rev.} \textbf{\bibinfo{volume}{D84}},
  \bibinfo{pages}{043531} (\bibinfo{year}{2011}), \eprint{1103.2050}.

\bibitem[{\citenamefont{Cede{\~n}o et~al.}(2017)\citenamefont{Cede{\~n}o,
  Gonz{\'a}lez-Morales, and Ure{\~n}a-L{\'o}pez}}]{Cedeno:2017sou}
\bibinfo{author}{\bibfnamefont{F.~X.~L.} \bibnamefont{Cede{\~n}o}},
  \bibinfo{author}{\bibfnamefont{A.~X.} \bibnamefont{Gonz{\'a}lez-Morales}},
  \bibnamefont{and} \bibinfo{author}{\bibfnamefont{L.~A.}
  \bibnamefont{Ure{\~n}a-L{\'o}pez}}, \bibinfo{journal}{Phys. Rev.}
  \textbf{\bibinfo{volume}{D96}}, \bibinfo{pages}{061301}
  (\bibinfo{year}{2017}), \eprint{1703.10180}.

\bibitem[{\citenamefont{Desjacques et~al.}(2018)\citenamefont{Desjacques,
  Kehagias, and Riotto}}]{Desjacques:2017fmf}
\bibinfo{author}{\bibfnamefont{V.}~\bibnamefont{Desjacques}},
  \bibinfo{author}{\bibfnamefont{A.}~\bibnamefont{Kehagias}}, \bibnamefont{and}
  \bibinfo{author}{\bibfnamefont{A.}~\bibnamefont{Riotto}},
  \bibinfo{journal}{Phys. Rev.} \textbf{\bibinfo{volume}{D97}},
  \bibinfo{pages}{023529} (\bibinfo{year}{2018}), \eprint{1709.07946}.

\bibitem[{\citenamefont{Goodman}(2000)}]{Goodman:2000tg}
\bibinfo{author}{\bibfnamefont{J.}~\bibnamefont{Goodman}},
  \bibinfo{journal}{New Astron.} \textbf{\bibinfo{volume}{5}},
  \bibinfo{pages}{103} (\bibinfo{year}{2000}), \eprint{astro-ph/0003018}.

\bibitem[{\citenamefont{Li et~al.}(2014)\citenamefont{Li, Rindler-Daller, and
  Shapiro}}]{Li:2013nal}
\bibinfo{author}{\bibfnamefont{B.}~\bibnamefont{Li}},
  \bibinfo{author}{\bibfnamefont{T.}~\bibnamefont{Rindler-Daller}},
  \bibnamefont{and} \bibinfo{author}{\bibfnamefont{P.~R.}
  \bibnamefont{Shapiro}}, \bibinfo{journal}{Phys. Rev.}
  \textbf{\bibinfo{volume}{D89}}, \bibinfo{pages}{083536}
  (\bibinfo{year}{2014}), \eprint{1310.6061}.

\bibitem[{\citenamefont{Su{\'a}rez and Chavanis}(2017)}]{Suarez:2016eez}
\bibinfo{author}{\bibfnamefont{A.}~\bibnamefont{Su{\'a}rez}} \bibnamefont{and}
  \bibinfo{author}{\bibfnamefont{P.-H.} \bibnamefont{Chavanis}},
  \bibinfo{journal}{Phys. Rev.} \textbf{\bibinfo{volume}{D95}},
  \bibinfo{pages}{063515} (\bibinfo{year}{2017}), \eprint{1608.08624}.

\bibitem[{\citenamefont{Su{\'a}rez and Chavanis}(2015)}]{Suarez:2015fga}
\bibinfo{author}{\bibfnamefont{A.}~\bibnamefont{Su{\'a}rez}} \bibnamefont{and}
  \bibinfo{author}{\bibfnamefont{P.-H.} \bibnamefont{Chavanis}},
  \bibinfo{journal}{Phys. Rev.} \textbf{\bibinfo{volume}{D92}},
  \bibinfo{pages}{023510} (\bibinfo{year}{2015}), \eprint{1503.07437}.

\bibitem[{\citenamefont{Su{\'a}rez and Chavanis}(2018)}]{Suarez:2017mav}
\bibinfo{author}{\bibfnamefont{A.}~\bibnamefont{Su{\'a}rez}} \bibnamefont{and}
  \bibinfo{author}{\bibfnamefont{P.-H.} \bibnamefont{Chavanis}},
  \bibinfo{journal}{Phys. Rev.} \textbf{\bibinfo{volume}{D98}},
  \bibinfo{pages}{083529} (\bibinfo{year}{2018}), \eprint{1710.10486}.

\bibitem[{\citenamefont{Chavanis}(2018)}]{Chavanis:2018pkx}
\bibinfo{author}{\bibfnamefont{P.-H.} \bibnamefont{Chavanis}}
  (\bibinfo{year}{2018}), \eprint{1810.08948}.

\bibitem[{\citenamefont{Fan}(2016)}]{Fan:2016rda}
\bibinfo{author}{\bibfnamefont{J.}~\bibnamefont{Fan}}, \bibinfo{journal}{Phys.
  Dark Univ.} \textbf{\bibinfo{volume}{14}}, \bibinfo{pages}{84}
  (\bibinfo{year}{2016}), \eprint{1603.06580}.

\bibitem[{\citenamefont{Rindler-Daller and
  Shapiro}(2012)}]{RindlerDaller:2011kx}
\bibinfo{author}{\bibfnamefont{T.}~\bibnamefont{Rindler-Daller}}
  \bibnamefont{and} \bibinfo{author}{\bibfnamefont{P.~R.}
  \bibnamefont{Shapiro}}, \bibinfo{journal}{Mon. Not. Roy. Astron. Soc.}
  \textbf{\bibinfo{volume}{422}}, \bibinfo{pages}{135} (\bibinfo{year}{2012}),
  \eprint{1106.1256}.

\bibitem[{\citenamefont{Cembranos et~al.}(2018)\citenamefont{Cembranos, Maroto,
  N{\'u}{\~n}ez~Jare{\~n}o, and Villarrubia-Rojo}}]{Cembranos:2018ulm}
\bibinfo{author}{\bibfnamefont{J.~A.~R.} \bibnamefont{Cembranos}},
  \bibinfo{author}{\bibfnamefont{A.~L.} \bibnamefont{Maroto}},
  \bibinfo{author}{\bibfnamefont{S.~J.}
  \bibnamefont{N{\'u}{\~n}ez~Jare{\~n}o}}, \bibnamefont{and}
  \bibinfo{author}{\bibfnamefont{H.}~\bibnamefont{Villarrubia-Rojo}},
  \bibinfo{journal}{JHEP} \textbf{\bibinfo{volume}{08}}, \bibinfo{pages}{073}
  (\bibinfo{year}{2018}), \eprint{1805.08112}.

\bibitem[{\citenamefont{Dev et~al.}(2017)\citenamefont{Dev, Lindner, and
  Ohmer}}]{Dev:2016hxv}
\bibinfo{author}{\bibfnamefont{P.~S.~B.} \bibnamefont{Dev}},
  \bibinfo{author}{\bibfnamefont{M.}~\bibnamefont{Lindner}}, \bibnamefont{and}
  \bibinfo{author}{\bibfnamefont{S.}~\bibnamefont{Ohmer}},
  \bibinfo{journal}{Phys. Lett.} \textbf{\bibinfo{volume}{B773}},
  \bibinfo{pages}{219} (\bibinfo{year}{2017}), \eprint{1609.03939}.

\bibitem[{\citenamefont{Li et~al.}(2017)\citenamefont{Li, Shapiro, and
  Rindler-Daller}}]{Li:2016mmc}
\bibinfo{author}{\bibfnamefont{B.}~\bibnamefont{Li}},
  \bibinfo{author}{\bibfnamefont{P.~R.} \bibnamefont{Shapiro}},
  \bibnamefont{and}
  \bibinfo{author}{\bibfnamefont{T.}~\bibnamefont{Rindler-Daller}},
  \bibinfo{journal}{Phys. Rev.} \textbf{\bibinfo{volume}{D96}},
  \bibinfo{pages}{063505} (\bibinfo{year}{2017}), \eprint{1611.07961}.

\bibitem[{\citenamefont{Tulin and Yu}(2018)}]{Tulin:2017ara}
\bibinfo{author}{\bibfnamefont{S.}~\bibnamefont{Tulin}} \bibnamefont{and}
  \bibinfo{author}{\bibfnamefont{H.-B.} \bibnamefont{Yu}},
  \bibinfo{journal}{Phys. Rept.} \textbf{\bibinfo{volume}{730}},
  \bibinfo{pages}{1} (\bibinfo{year}{2018}), \eprint{1705.02358}.

\bibitem[{\citenamefont{Randall et~al.}(2008)\citenamefont{Randall, Markevitch,
  Clowe, Gonzalez, and Bradac}}]{Randall:2007ph}
\bibinfo{author}{\bibfnamefont{S.~W.} \bibnamefont{Randall}},
  \bibinfo{author}{\bibfnamefont{M.}~\bibnamefont{Markevitch}},
  \bibinfo{author}{\bibfnamefont{D.}~\bibnamefont{Clowe}},
  \bibinfo{author}{\bibfnamefont{A.~H.} \bibnamefont{Gonzalez}},
  \bibnamefont{and} \bibinfo{author}{\bibfnamefont{M.}~\bibnamefont{Bradac}},
  \bibinfo{journal}{Astrophys. J.} \textbf{\bibinfo{volume}{679}},
  \bibinfo{pages}{1173} (\bibinfo{year}{2008}), \eprint{0704.0261}.

\bibitem[{\citenamefont{Arbey et~al.}(2003)\citenamefont{Arbey, Lesgourgues,
  and Salati}}]{Arbey:2003sj}
\bibinfo{author}{\bibfnamefont{A.}~\bibnamefont{Arbey}},
  \bibinfo{author}{\bibfnamefont{J.}~\bibnamefont{Lesgourgues}},
  \bibnamefont{and} \bibinfo{author}{\bibfnamefont{P.}~\bibnamefont{Salati}},
  \bibinfo{journal}{Phys. Rev.} \textbf{\bibinfo{volume}{D68}},
  \bibinfo{pages}{023511} (\bibinfo{year}{2003}), \eprint{astro-ph/0301533}.

\bibitem[{\citenamefont{Peebles}(2000)}]{Peebles:2000yy}
\bibinfo{author}{\bibfnamefont{P.~J.~E.} \bibnamefont{Peebles}},
  \bibinfo{journal}{Astrophys. J.} \textbf{\bibinfo{volume}{534}},
  \bibinfo{pages}{L127} (\bibinfo{year}{2000}), \eprint{astro-ph/0002495}.

\bibitem[{\citenamefont{Madelung}(1927)}]{Madelung_1927}
\bibinfo{author}{\bibfnamefont{E.}~\bibnamefont{Madelung}},
  \bibinfo{journal}{Zeitschrift f�r Physik} \textbf{\bibinfo{volume}{40}},
  \bibinfo{pages}{322} (\bibinfo{year}{1927}), ISSN \bibinfo{issn}{1434-601X},
  \urlprefix\url{http://dx.doi.org/10.1007/BF01400372}.

\bibitem[{\citenamefont{Gradshteyn and Ryzhik}(1965)}]{Gradshteyn1965}
\bibinfo{author}{\bibfnamefont{I.~S.} \bibnamefont{Gradshteyn}}
  \bibnamefont{and} \bibinfo{author}{\bibfnamefont{I.~M.}
  \bibnamefont{Ryzhik}}, \emph{\bibinfo{title}{Table of integrals, series, and
  products}} (\bibinfo{publisher}{New York Academic Press},
  \bibinfo{year}{1965}), \bibinfo{edition}{4th} ed.,
  \urlprefix\url{http://openlibrary.org/books/OL5955048M}.

\bibitem[{\citenamefont{Verhulst}(2005)}]{Verhulst2005}
\bibinfo{author}{\bibfnamefont{F.}~\bibnamefont{Verhulst}},
  \emph{\bibinfo{title}{Methods and applications of singular perturbations.
  Boundary layers and multiple timescale dynamics}}, vol.~\bibinfo{volume}{50}
  of \emph{\bibinfo{series}{Texts in Applied Mathematics}}
  (\bibinfo{publisher}{Springer-Verlag New York}, \bibinfo{year}{2005}).

\bibitem[{\citenamefont{Marsh}(2015)}]{Marsh:2015daa}
\bibinfo{author}{\bibfnamefont{D.~J.~E.} \bibnamefont{Marsh}},
  \bibinfo{journal}{Phys. Rev.} \textbf{\bibinfo{volume}{D91}},
  \bibinfo{pages}{123520} (\bibinfo{year}{2015}), \eprint{1504.00308}.

\bibitem[{\citenamefont{Chavanis}(2012)}]{Chavanis:2011uv}
\bibinfo{author}{\bibfnamefont{P.-H.} \bibnamefont{Chavanis}},
  \bibinfo{journal}{Astron. Astrophys.} \textbf{\bibinfo{volume}{537}},
  \bibinfo{pages}{A127} (\bibinfo{year}{2012}), \eprint{1103.2698}.

\bibitem[{\citenamefont{Riotto and Tkachev}(2000)}]{Riotto:2000kh}
\bibinfo{author}{\bibfnamefont{A.}~\bibnamefont{Riotto}} \bibnamefont{and}
  \bibinfo{author}{\bibfnamefont{I.}~\bibnamefont{Tkachev}},
  \bibinfo{journal}{Phys. Lett.} \textbf{\bibinfo{volume}{B484}},
  \bibinfo{pages}{177} (\bibinfo{year}{2000}), \eprint{astro-ph/0003388}.

\bibitem[{\citenamefont{Boehmer and Harko}(2007)}]{Boehmer:2007um}
\bibinfo{author}{\bibfnamefont{C.~G.} \bibnamefont{Boehmer}} \bibnamefont{and}
  \bibinfo{author}{\bibfnamefont{T.}~\bibnamefont{Harko}},
  \bibinfo{journal}{JCAP} \textbf{\bibinfo{volume}{0706}}, \bibinfo{pages}{025}
  (\bibinfo{year}{2007}), \eprint{0705.4158}.

\bibitem[{\citenamefont{Harko}(2011)}]{Harko:2011jy}
\bibinfo{author}{\bibfnamefont{T.}~\bibnamefont{Harko}}, \bibinfo{journal}{Mon.
  Not. Roy. Astron. Soc.} \textbf{\bibinfo{volume}{413}}, \bibinfo{pages}{3095}
  (\bibinfo{year}{2011}), \eprint{1101.3655}.

\bibitem[{\citenamefont{Deng et~al.}(2018)\citenamefont{Deng, Hertzberg,
  Namjoo, and Masoumi}}]{Deng:2018jjz}
\bibinfo{author}{\bibfnamefont{H.}~\bibnamefont{Deng}},
  \bibinfo{author}{\bibfnamefont{M.~P.} \bibnamefont{Hertzberg}},
  \bibinfo{author}{\bibfnamefont{M.~H.} \bibnamefont{Namjoo}},
  \bibnamefont{and} \bibinfo{author}{\bibfnamefont{A.}~\bibnamefont{Masoumi}},
  \bibinfo{journal}{Phys. Rev.} \textbf{\bibinfo{volume}{D98}},
  \bibinfo{pages}{023513} (\bibinfo{year}{2018}), \eprint{1804.05921}.

\bibitem[{\citenamefont{Zhang et~al.}(2018)\citenamefont{Zhang, Chan, Harko,
  Liang, and Leung}}]{Zhang:2018okg}
\bibinfo{author}{\bibfnamefont{X.}~\bibnamefont{Zhang}},
  \bibinfo{author}{\bibfnamefont{M.~H.} \bibnamefont{Chan}},
  \bibinfo{author}{\bibfnamefont{T.}~\bibnamefont{Harko}},
  \bibinfo{author}{\bibfnamefont{S.-D.} \bibnamefont{Liang}}, \bibnamefont{and}
  \bibinfo{author}{\bibfnamefont{C.~S.} \bibnamefont{Leung}},
  \bibinfo{journal}{Eur. Phys. J.} \textbf{\bibinfo{volume}{C78}},
  \bibinfo{pages}{346} (\bibinfo{year}{2018}), \eprint{1804.08079}.

\bibitem[{\citenamefont{{Binney} and {Tremaine}}(1987)}]{1987gady.book.....B}
\bibinfo{author}{\bibfnamefont{J.}~\bibnamefont{{Binney}}} \bibnamefont{and}
  \bibinfo{author}{\bibfnamefont{S.}~\bibnamefont{{Tremaine}}},
  \emph{\bibinfo{title}{{Galactic dynamics}}} (\bibinfo{publisher}{Princeton
  University Press}, \bibinfo{year}{1987}).

\bibitem[{\citenamefont{Schwabe et~al.}(2016)\citenamefont{Schwabe, Niemeyer,
  and Engels}}]{Schwabe:2016rze}
\bibinfo{author}{\bibfnamefont{B.}~\bibnamefont{Schwabe}},
  \bibinfo{author}{\bibfnamefont{J.~C.} \bibnamefont{Niemeyer}},
  \bibnamefont{and} \bibinfo{author}{\bibfnamefont{J.~F.}
  \bibnamefont{Engels}}, \bibinfo{journal}{Phys. Rev.}
  \textbf{\bibinfo{volume}{D94}}, \bibinfo{pages}{043513}
  (\bibinfo{year}{2016}), \eprint{1606.05151}.

\bibitem[{\citenamefont{Mocz et~al.}(2017)\citenamefont{Mocz, Vogelsberger,
  Robles, Zavala, Boylan-Kolchin, Fialkov, and Hernquist}}]{Mocz:2017wlg}
\bibinfo{author}{\bibfnamefont{P.}~\bibnamefont{Mocz}},
  \bibinfo{author}{\bibfnamefont{M.}~\bibnamefont{Vogelsberger}},
  \bibinfo{author}{\bibfnamefont{V.~H.} \bibnamefont{Robles}},
  \bibinfo{author}{\bibfnamefont{J.}~\bibnamefont{Zavala}},
  \bibinfo{author}{\bibfnamefont{M.}~\bibnamefont{Boylan-Kolchin}},
  \bibinfo{author}{\bibfnamefont{A.}~\bibnamefont{Fialkov}}, \bibnamefont{and}
  \bibinfo{author}{\bibfnamefont{L.}~\bibnamefont{Hernquist}},
  \bibinfo{journal}{Mon. Not. Roy. Astron. Soc.}
  \textbf{\bibinfo{volume}{471}}, \bibinfo{pages}{4559} (\bibinfo{year}{2017}),
  \eprint{1705.05845}.

\bibitem[{\citenamefont{Veltmaat et~al.}(2018)\citenamefont{Veltmaat, Niemeyer,
  and Schwabe}}]{Veltmaat:2018dfz}
\bibinfo{author}{\bibfnamefont{J.}~\bibnamefont{Veltmaat}},
  \bibinfo{author}{\bibfnamefont{J.~C.} \bibnamefont{Niemeyer}},
  \bibnamefont{and} \bibinfo{author}{\bibfnamefont{B.}~\bibnamefont{Schwabe}},
  \bibinfo{journal}{Phys. Rev.} \textbf{\bibinfo{volume}{D98}},
  \bibinfo{pages}{043509} (\bibinfo{year}{2018}), \eprint{1804.09647}.

\bibitem[{\citenamefont{Navarro et~al.}(1996)\citenamefont{Navarro, Frenk, and
  White}}]{Navarro:1995iw}
\bibinfo{author}{\bibfnamefont{J.~F.} \bibnamefont{Navarro}},
  \bibinfo{author}{\bibfnamefont{C.~S.} \bibnamefont{Frenk}}, \bibnamefont{and}
  \bibinfo{author}{\bibfnamefont{S.~D.~M.} \bibnamefont{White}},
  \bibinfo{journal}{Astrophys. J.} \textbf{\bibinfo{volume}{462}},
  \bibinfo{pages}{563} (\bibinfo{year}{1996}), \eprint{astro-ph/9508025}.

\bibitem[{\citenamefont{Guzman and Urena-Lopez}(2006)}]{Guzman:2006yc}
\bibinfo{author}{\bibfnamefont{F.~S.} \bibnamefont{Guzman}} \bibnamefont{and}
  \bibinfo{author}{\bibfnamefont{L.~A.} \bibnamefont{Urena-Lopez}},
  \bibinfo{journal}{Astrophys. J.} \textbf{\bibinfo{volume}{645}},
  \bibinfo{pages}{814} (\bibinfo{year}{2006}), \eprint{astro-ph/0603613}.

\bibitem[{\citenamefont{Amin and Mocz}(2019)}]{Amin:2019ums}
\bibinfo{author}{\bibfnamefont{M.~A.} \bibnamefont{Amin}} \bibnamefont{and}
  \bibinfo{author}{\bibfnamefont{P.}~\bibnamefont{Mocz}}
  (\bibinfo{year}{2019}), \eprint{1902.07261}.

\bibitem[{\citenamefont{Chavanis}(2019)}]{Chavanis:2019faf}
\bibinfo{author}{\bibfnamefont{P.-H.} \bibnamefont{Chavanis}}
  (\bibinfo{year}{2019}), \eprint{1905.08137}.

\bibitem[{\citenamefont{Maccio' et~al.}(2008)\citenamefont{Maccio', Dutton, and
  Bosch}}]{Maccio:2008pcd}
\bibinfo{author}{\bibfnamefont{A.~V.} \bibnamefont{Maccio'}},
  \bibinfo{author}{\bibfnamefont{A.~A.} \bibnamefont{Dutton}},
  \bibnamefont{and} \bibinfo{author}{\bibfnamefont{F.~C. v.~d.}
  \bibnamefont{Bosch}}, \bibinfo{journal}{Mon. Not. Roy. Astron. Soc.}
  \textbf{\bibinfo{volume}{391}}, \bibinfo{pages}{1940} (\bibinfo{year}{2008}),
  \eprint{0805.1926}.

\bibitem[{\citenamefont{Nori et~al.}(2019)\citenamefont{Nori, Murgia, Ir{\v
  s}i{\v c}, Baldi, and Viel}}]{Nori:2018pka}
\bibinfo{author}{\bibfnamefont{M.}~\bibnamefont{Nori}},
  \bibinfo{author}{\bibfnamefont{R.}~\bibnamefont{Murgia}},
  \bibinfo{author}{\bibfnamefont{V.}~\bibnamefont{Ir{\v s}i{\v c}}},
  \bibinfo{author}{\bibfnamefont{M.}~\bibnamefont{Baldi}}, \bibnamefont{and}
  \bibinfo{author}{\bibfnamefont{M.}~\bibnamefont{Viel}},
  \bibinfo{journal}{Mon. Not. Roy. Astron. Soc.}
  \textbf{\bibinfo{volume}{482}}, \bibinfo{pages}{3227} (\bibinfo{year}{2019}),
  \eprint{1809.09619}.

\end{thebibliography}


\end{document}